%% file: Bayes_arXiv_v1.tex
\documentclass[12pt]{article}
\PassOptionsToPackage{hyphens}{url}
\usepackage[pagebackref=true,colorlinks]{hyperref}
\usepackage{amsmath,amssymb,amsthm,mathtools,amsrefs}
\usepackage{pifont}% http://ctan.org/pkg/pifont
\usepackage{mathpazo} % add possibly `sc` and `osf` options
\usepackage{eulervm}
\usepackage{etex} %allows xRightarrow
\usepackage[table]{xcolor}%This is used to add colors to rows and columns in arrays (which I use for matrices) WARNING: Must load before tikz packages (i believe arrows.meta uses xcolor so there is a clash). or another option is to add "table" as a global option to the document class, ex. \documentclass[12pt,table]{article}
\usepackage{tikz}
\usetikzlibrary{arrows.meta}
\usetikzlibrary{decorations.pathmorphing,shapes}
\usepackage{adjustbox}
\usepackage{scalerel,stackengine}%Used for reallywidehat
\usepackage{stmaryrd}
\usepackage{fancyhdr}
\usepackage{soul}
\usepackage{dsfont}%this is for mathbb{1} but you write mathds{1}
\usepackage[normalem]{ulem}
\usepackage[bottom]{footmisc}%this forces footnotes to appear on the bottom of page
\usepackage{caption}
\usepackage{subfig} %this is to plot figures vertically
\usepackage{pictexwd, dcpic}
\usepackage{float}
\usepackage{enumerate}
\usepackage[all]{xy} %This is to make higher categorical stuff
\usetikzlibrary{circuits.ee.IEC}
\hypersetup{
	colorlinks=true,%set true if you want colored links
    linktoc=all,%set to all if you want both sections and subsections linked
    linkcolor=blue,  %choose some color if you want links to stand out
    citecolor=blue,
    }
\makeatletter
\newcommand*{\shifttext}[2]{%
  \settowidth{\@tempdima}{#2}%
  \makebox[\@tempdima]{\hspace*{#1}#2}%
}
\makeatother
\makeatletter
\renewcommand*\env@matrix[1][\arraystretch]{%
  \edef\arraystretch{#1}%
  \hskip -\arraycolsep
  \let\@ifnextchar\new@ifnextchar
  \array{*\c@MaxMatrixCols c}}
\makeatother

\stackMath
\newcommand\reallywidehat[1]{%
\savestack{\tmpbox}{\stretchto{%
  \scaleto{%
    \scalerel*[\widthof{\ensuremath{#1}}]{\kern.1pt\mathchar"0362\kern.1pt}%
    {\rule{0ex}{\textheight}}%WIDTH-LIMITED CIRCUMFLEX
  }{\textheight}% 
}{2.4ex}}%
\stackon[-6.9pt]{#1}{\tmpbox}%
}
\theoremstyle{theorem}
\newtheorem{theorem}[equation]{Theorem}
\newtheorem{lemma}[equation]{Lemma}
\newtheorem{proposition}[equation]{Proposition}
\newtheorem{corollary}[equation]{Corollary}
\theoremstyle{definition}
\newtheorem{definition}[equation]{Definition}
\newtheorem{construction}[equation]{Construction}
\newtheorem{question}[equation]{Question}
\newtheorem{problem}[equation]{Problem}
\newtheorem{example}[equation]{Example}
\newtheorem{exercise}[equation]{Exercise}
\newtheorem*{answer}{Answer}
\newtheorem*{solution}{Solution}
\newtheorem{remark}[equation]{Remark}

\newtheorem{notation}[equation]{Notation}

\newcommand\define[1]{\emph{\textbf{#1}}}%italicize and bold-face %this seems like a good alternative

\numberwithin{equation}{section}
\let\a=\alpha \let\b=\beta \let\g=\gamma \let\de=\delta 
\let\z=\zeta    
\let\l=\lambda 
\let\s=\sigma \let\t=\tau    
\let\w=\omega       \let\D=\Delta  \let\L=\Lambda
 \let\P=\Pi    
\let\C=\Chi \let\W=\Omega
\def\vf{\varphi}

\newcommand{\be}{\begin{equation}}
\newcommand{\ee}{\end{equation}}
\def\ba{\begin{align}} %previously this was ``array''
\def\ea{\end{align}}
\newcommand{\bea}{\begin{eqnarray}}
\newcommand{\eea}{\end{eqnarray}}
\newcommand{\bx}{\begin{example}}
\newcommand{\ex}{\end{example}}
\newcommand{\bex}{\begin{exercise}}
\newcommand{\eex}{\end{exercise}}
\newcommand{\ban}{\begin{answer}}
\newcommand{\ean}{\end{answer}}
\newcommand{\bt}{\begin{theorem}}
\newcommand{\et}{\end{theorem}}
\newcommand{\bc}{\begin{corollary}}
\newcommand{\ec}{\end{corollary}}
\newcommand{\blem}{\begin{lemma}}
\newcommand{\elem}{\end{lemma}}
\newcommand{\bp}{\begin{problem}}
\newcommand{\ep}{\end{problem}}
\newcommand{\bn}{\begin{proposition}}
\newcommand{\en}{\end{proposition}}
\newcommand{\bd}{\begin{definition}}
\newcommand{\ed}{\end{definition}}
\newcommand{\bcon}{\begin{construction}}
\newcommand{\econ}{\end{construction}}
\newcommand{\bq}{\begin{question}}
\newcommand{\eq}{\end{question}}
\newcommand{\bprf}{\begin{proof}}
\newcommand{\eprf}{\end{proof}}
\newcommand{\br}{\begin{remark}}
\newcommand{\er}{\end{remark}}
\newcommand{\bs}{\begin{solution}}
\newcommand{\es}{\end{solution}}
\newcommand{\beqs}{\begin{eqnarray}}
\newcommand{\eeqs}{\end{eqnarray}}

 \let\ov=\overline

\newcommand{\<}{\langle}
\renewcommand{\>}{\rangle}

\newcommand{\id}{\mathrm{id}}

\newcommand{\mC}{\mathcal{C}}
\newcommand{\mM}{\mathcal{M}}
\newcommand{\mD}{\mathcal{D}}

\newcommand{\tr}{{\rm tr} }

\def\C{{{\mathbb C}}}

\def\N{{{\mathbb N}}}

\def\Hi{{{\mathcal{H}}}}

\def\mA{{{\mathcal{A}}}}

\def\mB{{{\mathcal{B}}}}

\newcommand{\FinSet}{\mathbf{FinSet}}
\newcommand{\FinStoch}{\mathbf{FinStoch}}

\newcommand{\Choi}{\mathrm{Choi}}

\newcommand{\stoch}{\;\xy0;/r.25pc/:(-3,0)*{}="1";(3,0)*{}="2";{\ar@{~>}"1";"2"|(1.06){\hole}};\endxy\!}
\newcounter{sarrow}
\newcommand\xstoch[1]{%
\stepcounter{sarrow}%
\mathrel{\begin{tikzpicture}[baseline= {( $ (current bounding box.south) + (0,-0.1ex) $ )}]
\node[inner sep=.5ex] (\thesarrow) {\;$\scriptstyle #1$\;};
\path[draw,{<[scale=1.5,width=3,length=2]}-,decorate,
  decoration={snake,amplitude=0.3mm,segment length=2.1mm,pre=lineto,pre length=1pt}] 
    (\thesarrow.south east) -- (\thesarrow.south west);
\end{tikzpicture}}%
}

\newcommand{\ben}{\renewcommand{\theenumi}{\alph{enumi}} 
\renewcommand{\labelenumi}{(\theenumi)}\begin{enumerate}}
\newcommand{\een}{\end{enumerate}}

\newcommand\blfootnote[1]{%
  \begingroup
  \renewcommand\thefootnote{}\footnote{#1}%
  \addtocounter{footnote}{-1}%
  \endgroup
}

\input{tikzdefs}
\title{A non-commutative Bayes' theorem}
\author{Arthur J.~Parzygnat
and Benjamin P.~Russo}
\date{\today}

\newcommand{\Addresses}{{% additional braces for segregating \footnotesize
  \bigskip
  \footnotesize

  A.~Parzygnat, \textsc{Institut des Hautes \'Etudes Scientifiques, 35 Routes de Chartres 91440, Bures-sur-Yvette, France}\par\nopagebreak
  \textit{E-mail address}, A.~Parzygnat: \texttt{parzygnat@ihes.fr}

  \medskip

  B.~Russo, \textsc{Department of Mathematics, Farmingdale State College SUNY,
    Farmingdale, New York 11735}\par\nopagebreak
  \textit{E-mail address}, B.~Russo: \texttt{russobp@farmingdale.edu}

}}

\begin{document}
\maketitle
\vspace{-9mm} 

\begin{abstract}
Using a diagrammatic reformulation of Bayes' theorem, we provide a necessary and sufficient condition for the existence of Bayesian inference in the setting of finite-dimensional $C^*$-algebras. In other words, we prove an analogue of Bayes' theorem in the joint classical and quantum context. Our analogue is justified by recent advances in categorical probability theory, which have provided an abstract formulation of the classical Bayes' theorem. In the process, we further develop non-commutative almost everywhere equivalence and illustrate its important role in non-commutative Bayesian inversion. The construction of such Bayesian inverses, when they exist, involves solving a positive semidefinite matrix completion problem for the Choi matrix. This gives a solution to the open problem of constructing Bayesian inversion for completely positive unital maps acting on density matrices that do not have full support. We illustrate how the procedure works for several examples relevant to quantum information theory. 

\blfootnote{\emph{2020 Mathematics Subject Classification.} 
46L53, %Non-commutative probability and statistics
81P45 (Primary); %Quantum information, communication, networks (quantum-theoretic aspects)
%60B05, %Probability theory: Probability measures on topological spaces
15A83, %Matrix completion problems
%46J10, %Banach algebras of continuous functions, function algebras
46M99, %Functional analysis: methods of category theory in functional analysis
47B65, %Operator Theory: Positive operators and order-bounded operators
%18A40 %Category theory: Adjoint functors
81R15 (Secondary) %operator algebraic methods in quantum mechanics
%15B48, %Positive matrices and their generalizations; cones of matrices
}
\blfootnote{
\emph{Key words and phrases.} 
Quantum probability;
quantum information;
%density matrix;
optimal hypothesis;
%von Neumann algebra;
%completely positive;
%$C^*$-algebra;
%non-commutative measure theory; 
%non-commutative probability theory; 
disintegration;
regular conditional probability;
Bayesian inference;
%conditional probability;
%conditional expectation;
%Markov kernel;
%transition kernel;
%stochastic map;
probability monad;
%Kleisli category;
%quantum operation;
%reversible quantum channel;
%probabilistic programming;
%quantum programming;
%pre-Hilbert module;
%categorical quantum mechanics;
positive semidefinite matrix completion
%pseudoinverse; 
%Choi matrix;
%Schur complement
%measurement;
%L\"uders projection; 
}
\end{abstract}

\vspace{-11mm}
\tableofcontents

%\pagebreak
%%%%%%%%%%%%%%%%%%%%%%%%%%%%%%%%%%%%%

%%%%%%%%%%%%%%%%%%%%%%%%%%%%%%%%%%%%%
\section{Introduction and outline} 
%%%%%%%%%%%%%%%%%%%%%%%%%%%%%%%%%%%%%

Bayesian inference is an act of inferring likelihoods based on information that becomes available to us. It can be used to help understand how we make decisions and as a consequence assist in constructing forms of artificial intelligence. Bayesian inference has been a key ingredient in data analysis and machine learning algorithms~\cite{Bi95,Ti04}. 
Occasionally, one needs to manage and analyze enormous amounts of data for artificial intelligence. Since certain tasks have algorithms that can be performed on quantum computers more quickly than currently known algorithms on classical computers~\cite{Sh94}, these two aspects suggest the importance of a suitable quantum analogue of Bayesian inference~\cite{Re92}.
Although there have been numerous approaches to formulating quantum Bayesian inference~\cite{Ja18EPTCS,Va18,CoSp12,FaKo12,Le06,Le07,Ac74,AcCe82}, each one has its pros and cons and there does not yet seem to be a standard method. 
More importantly, the case of density matrices without full support (or more generally non-faithful states) is, in our opinion, a subtle, yet crucial, aspect to quantum Bayesian inference that has not been addressed adequately in the literature. The present paper aims to fill that gap.

We propose an inherently process-theoretic and diagrammatic formulation of quantum Bayesian inference~\cite{PaBayes}. 
The approach is based on category theory, which has been providing an interesting perspective on the foundations of probability theory~\cite{La62,Gi82,Pa99,Ja11,BFL,Fo12,CuSt14,CDDG17,GaPa18,Ja18,ChJa18,Fr19,Ja19}, and more recently quantum probability~\cite{Fr02,CoSp12,PaBayes,Ja18EPTCS} and the information-theoretic foundations of quantum mechanics~\cite{CDP10}. 
The categorical perspective provides a formulation of Bayes' theorem regarding the existence of a certain morphism satisfying a condition equivalent to Bayes' rule in the category of stochastic maps (morphisms are interpreted as a conditional probabilities in this context). This formulation can be used to \emph{define} Bayesian inference in quantum or classical Markov categories, where many of the essential features of probabilistic concepts have been categorified~\cite{ChJa18,Fr19,PaBayes}. In the present paper, we analyze the existence and almost everywhere uniqueness properties of Bayesian inference in the category of finite-dimensional $C^*$-algebras and completely positive unital (CPU) maps. This category contains both classical and quantum-mechanical systems. It includes a general class of evolutions of open quantum systems, and it includes positive-operator valued measures and instruments (which can be formulated as certain CPU maps between $C^*$-algebras that are not necessarily matrix factors but direct sums of matrix algebras). When restricted to the commutative algebras, the notion reproduces the standard Bayes' theorem. 

We state here the definition of a Bayesian inverse to highlight our main theorem afterwards. 
\bd
Let $\mB\xstoch{F}\mA$ be a CPU map between finite-dimensional $C^*$-algebras, let $\mA\xstoch{\w}\C$ be a state, and set $\xi:=\w\circ F$. A \define{Bayesian inverse} of $(F,\w)$ is a CPU map $\mA\xstoch{G}\mB$ such that%
%footnote
\footnote{The equals sign in this diagram indicates that the diagram commutes.}
%end footnote
\be
\xy0;/r.25pc/:
(0,-7.5)*+{\C}="1";
(-25,-7.5)*+{\mB}="B";
(25,-7.5)*+{\mA}="A";
(-25,7.5)*+{\mB\otimes\mB}="BB";
(25,7.5)*+{\mA\otimes\mA}="AA";
(0,7.5)*+{\mA\otimes\mB}="AB";
{\ar@{~>}"A";"1"^{\w}};
{\ar@{~>}"B";"1"_{\xi}};
{\ar@{~>}"BB";"B"_{\mu_{\mB}}};
{\ar@{~>}"AA";"A"^{\mu_{\mA}}};
{\ar@{~>}"AB";"BB"_{G\otimes\id_{\mB}}};
{\ar@{~>}"AB";"AA"^{\id_{\mA}\otimes F}};
(0,0)*{=\joinrel=\joinrel=};
\endxy
,
\ee
i.e.
\be
\label{eq:bayesintro}
\xi\big(G(A)B\big)=\w\big(AF(B)\big)
\ee
for all $A\in\mA$ and $B\in\mB$. 
\ed

Key features of classical Bayesian inference are its almost everywhere (a.e.) uniqueness, its compositionality for iterated processes, and its reversibility. Non-commutative a.e.\ equivalence was recently introduced and studied in~\cite{PaRu19}, while a categorical notion of a.e.\ equivalence was introduced in~\cite{ChJa18}. The two were shown to coincide for \emph{all} $C^*$-algebras algebras in~\cite{PaBayes}. In the present paper, we continue this development of non-commutative a.e.\ equivalence and indicate its subtle aspects for constructing non-commutative Bayesian inference. 

Our main result is Theorem~\ref{thm:Bayesianinversionmatrixalgebracase}, which provides a necessary and sufficient condition for $(F,\w)$ to have a CPU Bayesian inverse when $\mA=\mathcal{M}_{m}(\C)$ and $\mB=\mathcal{M}_{n}(\C)$ are matrix algebras (Theorem~\ref{thm:Bayesianinversiongeneralcase} generalizes this to arbitrary finite-dimensional $C^*$-algebras). The condition is expressed in terms of a certain Choi matrix and also provides a construction of a Bayesian inverse. Proposition~\ref{prop:qBayesianinverseformula} provides a formula for any \emph{linear} map $G$ satisfying~(\ref{eq:bayesintro}) on the support of the state $\xi:=\w\circ F$ with no additional assumptions whatsoever. However, this formula alone does not specify $G$ on the full algebra and one must extend it to guarantee unitality (unitality is analogous to probability preservation). Requiring $G$ to be $*$-preserving provides several additional constraints that are strong enough to demand complete positivity on the supported corner (cf.\ Definition~\ref{defn:supportedcorner}) of $\xi$ and this (along with several other equivalent conditions) is proved in Proposition~\ref{prop:selfadjointimpliesCP}. Corollary~\ref{cor:KrausdecompBayesmatrixcase} provides a Kraus decomposition for Bayesian inverses on the supported corner and provides a direct relationship between our Bayesian inverses and those of Leifer~\cite{Le06,Le07,CoSp12}. However, this formula only specifies the Bayesian inverse on the supported corner. It turns out to be a non-trivial requirement for a CPU extension to the full algebra. This requirement is explained in terms of a positive semidefinite matrix completion problem whose solution is provided in Theorem~\ref{thm:Bayesianinversionmatrixalgebracase}. 

The result is a bit technical to state here in the introduction, but the rough idea is the following. The Choi matrix of the Bayesian inverse breaks up into four parts based on the support of $\xi$. Let $\mathfrak{A}$ be the Choi matrix on the supported corner of $\xi$, $\hat{\mathfrak{A}}$ is its pseudoinverse, and $\mathfrak{B}$ the part of the Choi matrix that is on the support of $\xi$ but off the supported corner. These are both completely determined by the Bayes condition (and require no additional assumption for their existence). Provided that $\mathfrak{A}^{\dag}=\mathfrak{A}$ ($\mathfrak{A}$ is self-adjoint), a completion exists if and only if the partial trace of $\mathfrak{B}^{\dag}\hat{\mathfrak{A}}\mathfrak{B}$ (which is necessarily a positive semidefinite matrix) is bounded from above by the orthogonal complement of the support of $\xi$. Once this completion problem condition is satisfied, several CPU Bayesian inverses may exist, all of them being a.e.\ equivalent to one another. In many cases, such as the case of disintegrations (when $F$ is a $^*$-homomorphism) or wave collapse, the extension is unique. However, they do not always exist and one may find that the partial trace of $\mathfrak{B}^{\dag}\hat{\mathfrak{A}}\mathfrak{B}$ is too large. This provides a subtle aspect to our understanding of Bayesian inference in the quantum setting. Further investigation and comparison to physical phenomena is necessary. 

The outline of the paper is as follows. Section~\ref{sec:classicalBayes} reviews stochastic maps, disintegrations, diagrammatic Bayesian inversion, and the standard Bayes' theorem from a categorical perspective. Section~\ref{sec:measurezero} reviews non-commutative analogues of these notions for finite-dimensional $C^*$-algebras. Section~\ref{sec:examples} goes over several physically relevant examples. In particular, we find a surprising connection to Kribs' interaction algebra and fixed point set theorem~\cite{Kr03} in the context of finding Bayesian inverses for positive operator-valued measures. Section~\ref{sec:quantumBayes} is the main section of this paper and contains our main theorem for finding necessary and sufficient conditions for Bayesian inversion on \emph{matrix algebras}, the important results leading to it, and many consequences. Furthermore, we illustrate in several examples how to construct Bayesian inverses via a matrix completion problem, which we solve. We use this construction to reprove our disintegration theorem~\cite{PaRu19}, which we know must happen due to the fact that disintegrations are special kinds of Bayesian inverses~\cite{PaBayes}. Section~\ref{sec:quantumBayesgeneral} generalizes our version of Bayes' theorem to finite-dimensional $C^*$-algebras (direct sums of matrix algebras). Appendix~\ref{appendix:permutation} contains a lemma for permuting rows and columns of a partially defined matrix in a way that allows us to apply a theorem for positive semidefinite matrix completion. 

%%%%%%%%%%%%%%%%%%%%%%%%%%%%%%%%%%%%%
\section{Classical disintegration and Bayes' theorem}
\label{sec:classicalBayes}
%%%%%%%%%%%%%%%%%%%%%%%%%%%%%%%%%%%%%

What should quantum Bayesian inference look like? We will answer this question by first studying the classical case from the category-theoretic perspective. 
We recall the category $\FinStoch$ of finite sets and stochastic maps~\cite{BaFr14,Fr19,PaRu19}. The objects are finite sets and the morphisms are stochastic maps (also known as stochastic matrices or Markov kernels, though the latter name is usually reserved when working on sets of infinite cardinality).
In more detail, if $X$ and $Y$ are finite sets, a \define{stochastic map} 
$X\xstoch{f}Y$ associating a probability measure $f_{x}$ on $Y$ to each $x\in X$ (all finite sets will be equipped with the discrete $\s$-algebra). The value of this probability measure on $y\in Y$ will be denoted by $f_{yx}$. Stochastic maps are drawn with squiggly arrows to distinguish them from deterministic maps, which are stochastic maps assigning Dirac delta measures to each point in the domain.%
%footnote
\footnote{The Dirac delta measure $\de_{y}$ at $y\in Y$ is given by $\de_{y}(E)=1$ if $y\in E$ and $0$ otherwise for all measurable subsets $E\subseteq Y$. Since $Y$ is finite, $\de_{y}(\{y'\})$ will be written as $\de_{y'y}$.}
%end footnote
 The latter are drawn with straight arrows $\rightarrow$ and correspond to functions (the relationship is described thoroughly in~\cite{Pa17}). A single element set will be denoted by $\{\bullet\}$. A stochastic map $\{\bullet\}\xstoch{p}X$ encodes a probability measure on $X$. Stochastic maps $X\xstoch{f}Y\xstoch{g}Z$ can be composed via the Chapman--Kolmogorov equation 
\be
(g\circ f)_{zx}:=\sum_{y\in Y}g_{zy}f_{yx}. 
\ee
The category $\FinStoch$ also has the following monoidal structure. 
Let $X\xstoch{f}X'$ and $Y\xstoch{g}Y'$ be two stochastic maps. 
Their \define{product} is the stochastic map $X\times Y\xstoch{f\times g}X'\times Y'$ defined by 
\be
\label{eq:productstochasticmaps}
(X,Y)\ni(x,y)\mapsto\Big((X',Y')\ni(x',y')\mapsto(f\times g)_{(x',y')(x,y)}:=f_{x'x}g_{y'y}\Big).
\ee
This product reproduces the usual product of functions as the following example shows. 
\bx
Let $X\xrightarrow{f}X'$ and $Y\xrightarrow{g}Y'$ be functions viewed as stochastic maps (the measure associated to each point is a Dirac delta measure). Their product evaluated on elements is given by 
\be
(f\times g)_{(x',y')(x,y)}:=f_{x'x}g_{y'y}=\de_{x'f(x)}\de_{y'g(y)}=\de_{(x',y')(f(x),g(y))}=\de_{(x',y')(f\times g)(x,y)}, 
\ee
where the last expression $f\times g$ is the usual Cartesian product of functions. 
\ex

\br
The cartesian product of finite sets and the product of stochastic maps turns $\FinStoch$ into a symmetric monoidal category. The functor $\FinSet\hookrightarrow\FinStoch$ preserves this monoidal structure. Here, $\FinSet$ is the category of finite sets and functions with the cartesian product as the monoidal structure. 
\er

Due to this symmetric monoidal structure, a graphical calculus can be used~\cite{ChJa18,Fo12}. We will only use these string diagrams on occasion (mainly in remarks and proofs), and it will not be essential that the reader is familiar with the notation. Strings parallel to each other represent taking the product as in~(\ref{eq:productstochasticmaps}). We will occasionally keep the string-diagram notation next to the usual (functional) diagrammatic notation as to keep our work more accessible. We will also occasionally label wires if there is a potential for confusion.

There are two important stochastic maps (functions, in fact) associated to every finite set $X$. These are 
\be
\xy0;/r.25pc/:
(0,-5)*+{X}="X";
(0,5)*+{X\times X}="XX";
{\ar"X";"XX"_{\D_{X}}};
\endxy
\quad
\xy0;/r.25pc/:
(0,-5)*+{x}="X";
(0,5)*+{(x,x)}="XX";
{\ar@{|->}"X";"XX"};
\endxy
%\begin{matrix}
%X\xrightarrow{\;\D_{X}\;}X\times X\\
%x\xmapsto{\;\quad\;}(x,x)
%\end{matrix}
\quad
\vcenter{\hbox{
\begin{tikzpicture}%[scale=1.5]
\newcommand\dist{0.6};
\node at (0,0) {$\bullet$};
\draw (0,0) to [out=180,in=270] (-\dist,\dist);
\draw (0,0) to [out=0,in=270] (\dist,\dist);
\draw (0,0) to [out=270,in=90] (0,-\dist);
\end{tikzpicture}
}}
\qquad\text{ and }\qquad
\xy0;/r.25pc/:
(0,-5)*+{X}="X";
(0,5)*+{\{\bullet\}}="I";
{\ar"X";"I"_(0.45){!_{X}}};
\endxy
\quad
\xy0;/r.25pc/:
(0,-5)*+{x}="X";
(0,5)*+{\bullet}="I";
{\ar@{|->}"X";"I"};
\endxy
\quad
\vcenter{\hbox{
\begin{tikzpicture}[scale=1.0,circuit ee IEC]
\newcommand\dist{0.5};
\node[ground,align=center,rotate=90] (g) at (0,\dist) {};
\node at (0,{1.2*\dist}) {};
\draw(g) to (0,-\dist);
\end{tikzpicture}
}}
.
\ee
These maps are called \define{copy} and \define{discard}, respectively. 
When multiple sets are in use, these string diagrams may be labelled. 
Discard can be used to define projections \cite{ChJa18}. Namely, 
\be
\xy0;/r.25pc/:
(0,-10)*+{X\times Y}="XY";
(0,10)*+{X}="X";
{\ar"XY";"X"_{\pi_{X}}};
\endxy
\quad
:=
\quad
\xy0;/r.25pc/:
(0,-10)*+{X\times Y}="XY";
(0,0)*+{X\times\{\bullet\}}="XI";
(0,10)*+{X}="X";
{\ar"XY";"XI"_{\id_{X}\times!_{X}}};
{\ar"XI";"X"_{\cong}};
\endxy
\quad
\vcenter{\hbox{
\begin{tikzpicture}[scale=1.0,circuit ee IEC]
\newcommand\dist{1.0};
\node[ground,align=center,rotate=90] (g) at ({0.5*\dist},{0.5*\dist}) {};
\draw(g) to node[right]{\footnotesize$Y$} ({0.5*\dist},-\dist);
\draw ({-0.5*\dist},\dist) to node[left]{\footnotesize$X$} ({-0.5*\dist},-\dist);
\end{tikzpicture}
}}
, 
\ee
where the isomorphism is the one given by $(x,\bullet)\mapsto x$. 
A similar projection $X\times Y\xrightarrow{\pi_{Y}}Y$ can be defined onto $Y$ using $!_{X}\times\id_{Y}$. 
Copy and discard also satisfy many important relations including commutativity of
\be
\xy0;/r.25pc/:
(0,-10)*+{X}="Xb";
(-20,-10)*+{X\times X}="XXL";
(-20,10)*+{\{\bullet\}\times X}="pX";
(20,-10)*+{X\times X}="XXR";
(20,10)*+{X\times\{\bullet\}}="Xp";
(0,10)*+{X}="Xt";
{\ar"Xb";"XXL"|-{\D_{X}}};
{\ar"Xb";"XXR"|-{\D_{X}}};
{\ar"XXL";"pX"|-{!_{X}\times\id_{X}}};
{\ar"XXR";"Xp"|-{\id_{X}\times!_{X}}};
{\ar"Xb";"Xt"|-{\id_{X}}};
{\ar"XXL";"Xt"|-{\pi_{X}}};
{\ar"XXR";"Xt"|-{\pi_{X}}};
{\ar"pX";"Xt"^(0.6){\cong}};
{\ar"Xp";"Xt"_(0.6){\cong}};
\endxy
\qquad\text{ and }\qquad
\xy0;/r.25pc/:
(-20,-10)*+{X}="X";
(-20,10)*+{X\times X}="XX1";
(20,-10)*+{X\times X}="XX2";
(20,10)*+{X\times X\times X}="XXX";
{\ar"X";"XX1"|-{\D_{X}}};
{\ar"X";"XX2"|-{\D_{X}}};
{\ar"XX1";"XXX"|-{\id_{X}\times\D_{X}}};
{\ar"XX2";"XXX"|-{\D_{X}\times\id_{X}}};
\endxy
,
\ee
which, in the string diagram notation, translate to  
\be
\vcenter{\hbox{%
\begin{tikzpicture}[font=\small]
\node[copier] (c) at (0,0) {};
\coordinate (x1) at (-0.3,0.3);
\node[discarder] (d) at (x1) {};
\coordinate (x2) at (0.3,0.5);
\draw (c) to[out=165,in=-90] (x1);
\draw (c) to[out=15,in=-90] (x2);
\draw (c) to (0,-0.3);
\end{tikzpicture}}}
\quad=\quad
\vcenter{\hbox{%
\begin{tikzpicture}[font=\small]
\draw (0,0) to (0,0.8);
\end{tikzpicture}}}
\quad=\quad
\vcenter{\hbox{%
\begin{tikzpicture}[font=\small]
\node[copier] (c) at (0,0) {};
\coordinate (x1) at (-0.3,0.5);
\coordinate (x2) at (0.3,0.3);
\node[discarder] (d) at (x2) {};
\draw (c) to[out=165,in=-90] (x1);
\draw (c) to[out=15,in=-90] (x2);
\draw (c) to (0,-0.3);
\end{tikzpicture}}}
\qquad\text{ and }\qquad
\vcenter{\hbox{%
\begin{tikzpicture}[font=\small]
\node[copier] (c2) at (0,0) {};
\node[copier] (c1) at (-0.3,0.3) {};
\draw (c2) to[out=165,in=-90] (c1);
\draw (c2) to[out=15,in=-90] (0.4,0.6);
\draw (c1) to[out=165,in=-90] (-0.6,0.6);
\draw (c1) to[out=15,in=-90] (0,0.6);
\draw (c2) to (0,-0.3);
\end{tikzpicture}}}
\quad=\quad
\vcenter{\hbox{%
\begin{tikzpicture}[font=\small]
\node[copier] (c2) at (-0.5,0) {};
\node[copier] (c1) at (-0.2,0.3) {};
\draw (c2) to[out=15,in=-90] (c1);
\draw (c2) to[out=165,in=-90] (-1,0.6);
\draw (c1) to[out=165,in=-90] (-0.5,0.6);
\draw (c1) to[out=15,in=-90] (0.1,0.6);
\draw (c2) to (-0.5,-0.3);
\end{tikzpicture}}}
,
\ee
respectively. 
$\FinStoch$ also satisfies 
\be
\label{eq:markovcatidentitiesII}
\vcenter{\hbox{%
\begin{tikzpicture}[font=\small]
\node[discarder] (d) at (0,0) {};
\draw (d) to +(0,-0.5);
\node at (0.6,-0.3) {$X\times Y$};
\end{tikzpicture}}}
=
\vcenter{\hbox{%
\begin{tikzpicture}[font=\small]
\node[discarder] (d) at (0,0) {};
\node[discarder] (d2) at (0.6,0) {};
\draw (d) to +(0,-0.5);
\draw (d2) to +(0,-0.5);
\node at (0.2,-0.3) {$X$};
\node at (0.8,-0.3) {$Y$};
\end{tikzpicture}}}
,
\quad
\vcenter{\hbox{%
\begin{tikzpicture}[font=\small]
\node[discarder] (d) at (0,0) {};
\draw (d) to +(0,-0.5);
\node at (0.2,-0.3) {$I$};
\end{tikzpicture}}}
=\;
\vcenter{\hbox{%
\begin{tikzpicture}[font=\small]
\draw [gray,dashed] (0,0) rectangle (0.45,0.65);
\end{tikzpicture}}}
\;\;,
\quad
\vcenter{\hbox{%
\begin{tikzpicture}[font=\small]
\node[copier] (c) at (0,0.4) {};
\draw (c)
to[out=15,in=-90] (0.25,0.7);
\draw (c)
to[out=165,in=-90] (-0.25,0.7);
\draw (c) to (0,0);
\node at (0.6,0.1) {$X\times Y$};
\end{tikzpicture}}}
=
\vcenter{\hbox{%
\begin{tikzpicture}[font=\small]
\node[copier] (c) at (0,0) {};
\node[copier] (c2) at (0.4,0) {};
\draw (c) to[out=15,in=-90] +(0.5,0.45);
\draw (c) to[out=165,in=-90] +(-0.4,0.45);
\draw (c) to +(0,-0.4);
\draw (c2) to[out=15,in=-90] +(0.4,0.45);
\draw (c2) to[out=165,in=-90] +(-0.5,0.45);
\draw (c2) to +(0,-0.4);
\node at (-0.2,-0.3) {$X$};
\node at (0.6,-0.3) {$Y$};
\end{tikzpicture}}}
,
\quad
\vcenter{\hbox{%
\begin{tikzpicture}[font=\small]
\node[copier] (c) at (0,0.4) {};
\draw (c)
to[out=15,in=-90] (0.25,0.7);
\draw (c)
to[out=165,in=-90] (-0.25,0.7);
\draw (c) to (0,0);
\node at (0.2,0.1) {$I$};
\end{tikzpicture}}}
=\;
\vcenter{\hbox{%
\begin{tikzpicture}[font=\small]
\draw [gray,dashed] (0,0) rectangle (0.45,0.65);
\end{tikzpicture}}}
\;\;,\quad
\vcenter{\hbox{%
\begin{tikzpicture}[font=\small]
\node[copier] (c) at (0,0.4) {};
\draw (c)
to[out=15,in=-90] (0.25,0.65)
to[out=90,in=-90] (-0.25,1.2);
\draw (c)
to[out=165,in=-90] (-0.25,0.65)
to[out=90,in=-90] (0.25,1.2);
\draw (c) to (0,0.1);
\end{tikzpicture}}}
\;=\;
\vcenter{\hbox{%
\begin{tikzpicture}[font=\small]
\node[copier] (c) at (0,0.4) {};
\draw (c)
to[out=15,in=-90] (0.25,0.7);
\draw (c)
to[out=165,in=-90] (-0.25,0.7);
\draw (c) to (0,0.1);
\end{tikzpicture}}}
,
\ee
where $I:=\{\bullet\}$. 
These diagrammatic relations have been used to \emph{define} sufficient categorical structure that enables one to make sense of disintegration and Bayesian inference in the classical (as opposed to the quantum) setting~\cite{ChJa18,Fr19}. Categories with such structure are called (classical) Markov categories~\cite{Fr19}. Quantum Markov categories drop the last condition in~(\ref{eq:markovcatidentitiesII}) and replace it with a suitable condition that reflects the relation $(ab)^*=b^*a^*$ in $C^*$-algebras~\cite{PaBayes}. As such, many (though not all) of the diagrammatic manipulations from the theory of Markov categories can also be done in this more general setting. We will not assume the reader is familiar with these recent developments. 

Furthermore, notice that every stochastic map $X\xstoch{f}Y$ satisfies
\be
\xy0;/r.20pc/:
(-15,-10)*+{X}="X";
(15,-10)*+{Y}="Y";
(0,10)*+{\{\bullet\}}="0";
{\ar@{~>}"X";"Y"_{f}};
{\ar"X";"0"^{!_{X}}};
{\ar"Y";"0"_{!_{Y}}};
{\ar@{=}(6,-5);(-3,-1)};
\endxy
,
\qquad
\text{i.e.}
\quad
\vcenter{\hbox{
\begin{tikzpicture}[scale=1.0,circuit ee IEC]
\newcommand\dist{0.5};
\node[ground,align=center,rotate=90] (g) at (0,{1.75*\dist}) {};
\node[draw,align=center,inner sep=5pt] (f) at (0,0){$f$};
\draw (g) to node[left]{\footnotesize$Y$} (f);
\draw (f) to node[left]{\footnotesize$X$} (0,{-2*\dist});
\end{tikzpicture}
}}
\;\;=\;\;
\vcenter{\hbox{
\begin{tikzpicture}[scale=1.0,circuit ee IEC]
\newcommand\dist{0.5};
\node[ground,align=center,rotate=90] (g) at (0,{1.75*\dist}) {};
\draw(g) to node[left]{\footnotesize$X$} (0,{-2*\dist});
\end{tikzpicture}
}}
,
\qquad
\text{i.e.}
\quad
\sum_{y\in Y}f_{yx}=1\quad\forall\;x\in X.
\ee
This condition is called \emph{causality} in the literature~\cite{KiUi19,ChJa18,SSC18}. Hence, such morphisms will be called \define{causal}. Causality of a morphism in this case means that it is probability-preserving. 

An important notion in probability theory is that of almost everywhere equivalence. This has recently been defined in the non-commutative setting~\cite{PaRu19} and in the general categorical setting~\cite{ChJa18}. The notion will be especially crucial in specifying the \emph{uniqueness} of Bayesian inference in the non-commutative setting. 

\bd
\label{defn:classicalaeequivalence}
Let $X$ and $Y$ be finite sets, let $\{\bullet\}\xstoch{p}X$ be a probability measure on $X$, and let $f,g:X\stoch Y$ be stochastic maps. Then $f$ is said to be \define{$p$-a.e.\ equivalent to} $g$ iff
\be
\label{eq:aeequivalencediagram}
\xy;/r.25pc/:
(-30,0)*+{\{\bullet\}}="0";
(-15,7.5)*+{X}="Xt";
(-15,-7.5)*+{X}="Xb";
(7.5,7.5)*+{X\times X}="XXt";
(7.5,-7.5)*+{X\times X}="XXb";
(30,0)*+{X\times Y}="XY";
{\ar@{~>}"0";"Xt"^{p}};
{\ar@{~>}"0";"Xb"_{p}};
{\ar"Xt";"XXt"^(0.45){\D_{X}}};
{\ar"Xb";"XXb"_(0.45){\D_{X}}};
{\ar@{~>}"XXt";"XY"^{\id_{X}\times f}};
{\ar@{~>}"XXb";"XY"_{\id_{X}\times g}};
{\ar@{=}(-4,3);(-4,-3)};
\endxy
,\qquad\text{i.e.}\qquad
\vcenter{\hbox{%
\begin{tikzpicture}[font=\small]
\node[state] (p) at (0,0) {$p$};
\node[copier] (copier) at (0,0.3) {};
\node[arrow box] (f) at (0.5,0.95) {$f$};
\coordinate (X) at (-0.5,1.5);
\coordinate (Y) at (0.5,1.5);
\draw (p) to (copier);
\draw (copier) to[out=150,in=-90] (X);
\draw (copier) to[out=15,in=-90] (f);
\draw (f) to (Y);
\path[scriptstyle]
node at (-0.7,1.45) {$X$}
node at (0.7,1.45) {$Y$};
\end{tikzpicture}}}
\quad=\quad
\vcenter{\hbox{
\begin{tikzpicture}[font=\small]
\node[state] (p) at (0,0) {$p$};
\node[copier] (copier) at (0,0.3) {};
\node[arrow box] (g) at (0.5,0.95) {$g$};
\coordinate (X) at (-0.5,1.5);
\coordinate (Y) at (0.5,1.5);
\draw (p) to (copier);
\draw (copier) to[out=150,in=-90] (X);
\draw (copier) to[out=15,in=-90] (g);
\draw (g) to (Y);
\path[scriptstyle]
node at (-0.7,1.45) {$X$}
node at (0.7,1.45) {$Y$};
\end{tikzpicture}}}
.
\ee
In this case, the notation $f\underset{\raisebox{0.3ex}[0pt][0pt]{\scriptsize$p$}}{=}g$ will be used. As an equation, $f\underset{\raisebox{.3ex}[0pt][0pt]{\scriptsize$p$}}{=}g$ says $f_{yx}p_{x}=g_{yx}p_{x}$ for all $x\in X$ and $y\in Y$. 
\ed

\br
\label{rmk:aequivalenceleft}
Although condition~(\ref{eq:aeequivalencediagram}) looks asymmetric, one can show (\ref{eq:aeequivalencediagram}) commutes if and only if 
\be
\label{eq:leftaediagram}
\xy;/r.25pc/:
(-30,0)*+{\{\bullet\}}="0";
(-15,7.5)*+{X}="Xt";
(-15,-7.5)*+{X}="Xb";
(7.5,7.5)*+{X\times X}="XXt";
(7.5,-7.5)*+{X\times X}="XXb";
(30,0)*+{Y\times X}="XY";
{\ar@{~>}"0";"Xt"^{p}};
{\ar@{~>}"0";"Xb"_{p}};
{\ar"Xt";"XXt"^(0.45){\D_{X}}};
{\ar"Xb";"XXb"_(0.45){\D_{X}}};
{\ar@{~>}"XXt";"XY"^{f\times\id_{X}}};
{\ar@{~>}"XXb";"XY"_{g\times\id_{X}}};
{\ar@{=}(-4,3);(-4,-3)};
\endxy
,\qquad\text{i.e.}\qquad
\vcenter{\hbox{
\begin{tikzpicture}[font=\small]
\node[state] (p) at (0,0) {$p$};
\node[copier] (copier) at (0,0.3) {};
\node[arrow box] (f) at (-0.5,0.95) {$f$};
\coordinate (X) at (0.5,1.5);
\coordinate (Y) at (-0.5,1.5);
\draw (p) to (copier);
\draw (copier) to[out=30,in=-90] (X);
\draw (copier) to[out=165,in=-90] (f);
\draw (f) to (Y);
\path[scriptstyle]
node at (-0.7,1.45) {$Y$}
node at (0.7,1.45) {$X$};
\end{tikzpicture}}}
\quad=\quad
\vcenter{\hbox{
\begin{tikzpicture}[font=\small]
\node[state] (p) at (0,0) {$p$};
\node[copier] (copier) at (0,0.3) {};
\node[arrow box] (g) at (-0.5,0.95) {$g$};
\coordinate (X) at (0.5,1.5);
\coordinate (Y) at (-0.5,1.5);
\draw (p) to (copier);
\draw (copier) to[out=30,in=-90] (X);
\draw (copier) to[out=165,in=-90] (g);
\draw (g) to (Y);
\path[scriptstyle]
node at (-0.7,1.45) {$Y$}
node at (0.7,1.45) {$X$};
\end{tikzpicture}}}
.
\ee 
\er

\br
\label{eq:twonotionsofaeequivalence}
That the notion of a.e.\ equivalence agrees with the usual definition from probability theory will be described in Section \ref{sec:quantumBayes}. In fact, Definition~\ref{defn:classicalaeequivalence} agrees with \emph{both} the classical (commutative) and quantum (non-commutative) notions of a.e.\ equivalence, the latter of which was introduced recently and independently in \cite{PaRu19}. The equivalence between these two definitions in the quantum context is actually quite subtle. If one restricts to positive (or more generally $*$-preserving) unital maps, then the notions of a.e.\ equivalence are all equivalent. However, if one uses linear maps (that are not necessarily $*$-preserving), then the conditions (\ref{eq:aeequivalencediagram}) and (\ref{eq:leftaediagram}) are not equivalent in general~\cite[Remark~5.10]{PaBayes}. We will see later that Bayesian inverses always exist as linear unital maps, but \emph{not} necessarily as  \emph{$*$-preserving} linear maps, so choosing an appropriate definition of a.e.\ equivalence 
will be important.
\er

\bd
Let $(X,p)$ and $(Y,q)$ be probability spaces and let
$f:X\to Y$ be a function such that $q=f\circ p$.
Such an $f$ is said to be \define{measure/probability-preserving}. A \define{disintegration} of 
$(f,p,q)$ 
is a stochastic map
$Y\xstoch{r} X$ such that 
\be
\label{eq:disintconditions}
\xy0;/r.25pc/:
(0,7.5)*+{\{\bullet\}}="o";
(-10,-7.5)*+{X}="X";
(10,-7.5)*+{Y}="Y";
{\ar@{~>}"o";"X"_{p}};
{\ar@{~>}"o";"Y"^{q}};
{\ar@{~>}"Y";"X"^{r}};
{\ar@{=}(-3,0);(5,-4.5)};
\endxy
\qquad\text{and}\qquad
\xy0;/r.25pc/:
(0,7.5)*+{X}="X";
(10,-7.5)*+{Y}="Y1";
(-10,-7.5)*+{Y}="Y2";
{\ar@{~>}"Y1";"X"_{r}};
{\ar"X";"Y2"_{f}};
{\ar"Y1";"Y2"^{\mathrm{id}_{Y}}};
{\ar@{=}(-3,0);(5,-4.5)_{q}};
\endxy
\;,
\ee
the latter diagram signifying commutativity $q$-a.e. A disintegration is also called a \define{regular conditional probability} and an \define{optimal hypothesis}. 
\ed

The conditions of a disintegration from~(\ref{eq:disintconditions}) can be drawn using string diagrams as
\be
\vcenter{\hbox{%
\begin{tikzpicture}[font=\small]
\node[state] (q) at (0,0) {$q$};
\node[arrow box] (r) at (0,0.75) {$r$};
\draw (q) to (r);
\draw (r) to (0,1.55);
\end{tikzpicture}}}
\quad
=
\quad
\vcenter{\hbox{%
\begin{tikzpicture}[font=\small]
\node[state] (p) at (0,0) {$p$};
\node (X) at (0,1.0) {};
\draw (p) to (X);
\end{tikzpicture}}}
\qquad
\text{and}
\qquad
\vcenter{\hbox{%
\begin{tikzpicture}[font=\small]
\node[state] (q) at (0,0) {$q$};
\node[copier] (copier) at (0,0.3) {};
\coordinate (f) at (-0.5,0.91) {};
\node[arrow box] (r) at (0.5,0.95) {$r$};
\node[arrow box] (e) at (0.5,1.75) {$f$};
\coordinate (X) at (-0.5,2.3);
\coordinate (Y) at (0.5,2.3);
\draw (q) to (copier);
\draw (copier) to[out=165,in=-90] (f);
\draw (f) to (X);
\draw (copier) to[out=15,in=-90] (r);
\draw (r) to (e);
\draw (e) to (Y);
\end{tikzpicture}}}
\quad
=
\quad
\vcenter{\hbox{%
\begin{tikzpicture}[font=\small]
\node[state] (q) at (0,-0.4) {$q$};
\node[copier] (copier) at (0,0.3) {};
\coordinate (X) at (-0.5,1.31);
\coordinate (X2) at (0.5,1.31);
\draw (q) to (copier);
\draw (copier) to[out=165,in=-90] (X);
\draw (copier) to[out=15,in=-90] (X2);
\end{tikzpicture}}}
\quad,
\ee
respectively. 

\br
Our definition of a disintegration is not \emph{exactly} the same as the one of Cho and Jacobs~\cite{ChJa18}. Our definition is based on Appendix~A in our work on non-commutative disintegrations~\cite{PaRu19}, which was meant to be an exact diagrammatic formulation of the standard notion of disintegration from measure theory~\cite[Section~452]{FrV4}. 
\er

\bt
\label{thm:classicaldisintegrationtheorem}
Let $(X,p)$ and $(Y,q)$ be finite sets equipped with 
probability measures $p$ and $q.$ 
Let $f:X\to Y$ be a measure-preserving function.
Then there exists a ($q$-a.e.) unique disintegration $r:Y\stoch X$ of $(f,p,q)$.
\et

A formula for the disintegration is
\be
r_{xy}:=
\begin{cases}
p_{x}\de_{yf(x)}/q_{y}&\mbox{ if } q_{y}>0\\
1/|X|&\mbox{ if $q_{y}=0$}
\end{cases}
\ee
In fact, one can use any probability measure for $r_{y}$ when $q_{y}=0$ (the uniform probability measure is not required). 

\bprf
[Proof of Theorem~\ref{thm:classicaldisintegrationtheorem}]
This is a standard result. The present formulation is proved in~\cite[Section~2.2]{PaRu19}.
\eprf

Disintegrations are special kinds of Bayesian inverses, which will be explained later (see also~\cite{PaBayes}, which goes into detail regarding their relationship to each other). The following theorem is a categorical reformulation of Bayes' theorem. 

\bt[Bayes' theorem]
\label{thm:classicalBayestheorem}
Let $X$ and $Y$ be finite sets. 
Given a probability measure $\{\bullet\}\xstoch{p}X$ and a stochastic map
$X\xstoch{f}Y$, there exists a stochastic map $Y\xstoch{g}X$ 
such that 
\be
\label{eq:Bayesdiagrams}
\xy0;/r.25pc/:
(0,7.5)*+{\{\bullet\}}="1";
(-25,7.5)*+{Y}="Y";
(25,7.5)*+{X}="X";
(-25,-7.5)*+{Y\times Y}="YY";
(25,-7.5)*+{X\times X}="XX";
(0,-7.5)*+{X\times Y}="XY";
{\ar@{~>}"1";"X"^{p}};
{\ar@{~>}"1";"Y"_{q}};
{\ar"Y";"YY"_{\Delta_{Y}}};
{\ar"X";"XX"^{\Delta_{X}}};
{\ar@{~>}"YY";"XY"_{g\times\id_{Y}}};
{\ar@{~>}"XX";"XY"^{\id_{X}\times f}};
(0,0)*{=\joinrel=\joinrel=};
\endxy
,
\ee
where $q:=f\circ p$ is the pushforward of the probability measure $p$ along $f$. 
Furthermore, for any other $g'$ satisfying 
this condition, 
$g\underset{\raisebox{.3ex}[0pt][0pt]{\scriptsize$q$}}{=}g'.$
\et

Before we provide the short proof of Bayes' theorem, we first justify why we call this Bayes' theorem in Remark~\ref{rmk:whyBayes} after establishing some terminology in Definition~\ref{defn:Bayesianinverse}.
Furthermore, our formulation of Bayes' theorem slightly differs from other theorems with a similar name in the literature on categorical probability theory, which focus on the relationship between joint probability distributions and conditionals. We will explain more about this distinction in Remark~\ref{rmk:whyBayesclassical}. However, the true justification for why we use this version of Bayes' theorem will only become more apparent when we study the quantum analogue, where our formulation seems to become more suitable for reasons that will be explained in Remark~\ref{rmk:whyBayesquantum}.

\bd
\label{defn:Bayesianinverse}
Let $f$ and $p$ be as in Theorem \ref{thm:classicalBayestheorem}, a stochastic map $g$ satisfying commutativity of (\ref{eq:Bayesdiagrams})
is called a \define{Bayesian inverse of $(f,p)$} or a \define{Bayesian inference for $(f,p)$}. 
The diagram in~(\ref{eq:Bayesdiagrams}) is referred to as \define{Bayes' diagram} or the \define{Bayes condition}.
In the language of Bayesian statistics, the measure $p$ is sometimes called the \define{prior probability}, the stochastic map $f$ is called the \define{likelihood}, the measure $q$ is called the \define{marginal likelihood}, and the stochastic map $g$ is called the \define{weighted likelihood}/\define{posterior probability}. 
It is helpful to summarize this diagrammatically as 
\be
\xy0;/r.35pc/:
(0,7.5)*+{\{\bullet\}}="o";
(-10,-7.5)*+{X}="X";
(10,-7.5)*+{Y}="Y";
{\ar@{~>}"o";"X"_{\text{prior}=p}};
{\ar@{~>}"o";"Y"^{q=\text{marginal likelihood}}};
{\ar@{~>}@/^1.0pc/"Y";"X"_{g}^{\text{posterior}}};
{\ar@{~>}@/^1.0pc/"X";"Y"_{f}^{\text{likelihood}}};
\endxy
.
\ee
If one obtains \emph{new} \define{evidence} in the form of a probability measure $\{\bullet\}\xstoch{q'}Y$, then the \define{Bayesian update}%
%footnote
\footnote{Technically, this is called \define{Jeffrey conditioning} in this context since the marginal likelihood is a probability measure in general and not a Dirac measure~\cite{Ja19}.}
%end footnote
 is the probability measure on $X$ obtained from the composite $\{\bullet\}\xstoch{q'}Y\xstoch{g}X$. 
\ed

\br
In string diagram notation, the Bayes condition is expressed as
\be
\label{eq:bayesconditionstring}
\vcenter{\hbox{%
\begin{tikzpicture}[font=\small]
\node[state] (q) at (0,0) {$q$};
\node[copier] (copier) at (0,0.3) {};
\node[arrow box] (g) at (-0.5,0.95) {$g$};
\coordinate (X) at (-0.5,1.5);
\coordinate (Y) at (0.5,1.5);
\draw (q) to (copier);
\draw (copier) to[out=165,in=-90] (g);
\draw (copier) to[out=30,in=-90] (Y);
\draw (g) to (X);
\path[scriptstyle]
node at (-0.7,1.45) {$X$}
node at (0.7,1.45) {$Y$};
\end{tikzpicture}}}
\qquad
=
\qquad
\vcenter{\hbox{%
\begin{tikzpicture}[font=\small]
\node[state] (p) at (0,0) {$p$};
\node[copier] (copier) at (0,0.3) {};
\node[arrow box] (f) at (0.5,0.95) {$f$};
\coordinate (X) at (-0.5,1.5);
\coordinate (Y) at (0.5,1.5);
\draw (p) to (copier);
\draw (copier) to[out=150,in=-90] (X);
\draw (copier) to[out=15,in=-90] (f);
\draw (f) to (Y);
\path[scriptstyle]
node at (-0.7,1.45) {$X$}
node at (0.7,1.45) {$Y$};
\end{tikzpicture}}}
\quad.
\ee
\er

\br
\label{rmk:whyBayes}
Before we prove Bayes' theorem, we should at least explain how this diagrammatic formulation is equivalent to the usual formulation of Bayes' theorem, which is commonly written as 
\be
\label{eq:Bayesonsubsets}
P(A|B)P(B)=P(B|A)P(A)
\ee
with $A\subseteq X$ and $B\subseteq Y$~\cite[Chapter~1 Section~4]{Ko33}. 
First, we will show how (\ref{eq:Bayesdiagrams}) reproduces
\be
\label{eq:Bayesonpoints}
P(x|y)P(y)=P(y|x)P(x), 
\ee
where $P(x)$ is the probability of $x\in X$ (and similarly for $P(y)$) and $P(x|y)$ is the conditional probability, namely the probability of $x$ occurring given that $y$ has occurred (and similarly for $P(y|x)$). For us, the latter is precisely a stochastic map. To reconstruct this equation, 
let $\{\bullet\}\xstoch{q:=f\circ p}Y$ be the pushforward probability measure on $Y$. This means 
\be
q_{y}=\sum_{x\in X}f_{yx}p_{x}\qquad\forall\;y\in Y. 
\ee
Working out the right-hand-side of the 
 diagram in~(\ref{eq:Bayesdiagrams}) gives a probability measure on $X\times Y.$ 
Evaluating this measure on a `rectangular' subset of the form $A\times B$ gives (we are using the associativity of the composition here freely)
\be
\label{eq:BayesonAB}
\begin{split}
\Big((\id_{X}\times f)\circ(\Delta_{X}\circ p)\Big)
(A\times B)
&=\sum_{x',x''\in X}\big((\id_{X}\times f)_{(x',x'')}(A\times B)\big)(\Delta_{X}\circ p)_{(x',x'')}\\
&=\sum_{x',x''\in X}\chi_{A}(x')f_{x''}(B)\sum_{x\in X}\de_{(x',x'')\D_{X}(x)}p_{x}\\
&=\sum_{x,x',x''\in X}\chi_{A}(x')f_{x''}(B)\de_{x'x}\de_{x''x}p_{x}\\
&=\sum_{x\in A}f_{x}(B)p_{x}.
\end{split}
\ee
Thus, commutativity of~(\ref{eq:Bayesdiagrams}) says 
\be
\sum_{y\in B}g_{y}(A)q_{y}=\sum_{x\in A}f_{x}(B)p_{x}.
\ee
In the case that $A:=\{x\}$ and $B:=\{y\}$ are singletons, this becomes
\be
g_{xy}q_{y}=f_{yx}p_{x}\qquad\forall\;(x,y)\in X\times Y. 
\ee
If we make the definitions
\be
P(x|y):=g_{xy},\qquad
P(y):=q_{y},\qquad
P(y|x):=f_{yx},\quad\text{and}\quad
P(x):=p_{x}, 
\ee
this indeed reproduces (\ref{eq:Bayesonpoints}). 
To reproduce (\ref{eq:Bayesonsubsets}), we need to define 
all the expressions. Normally, $P(B|A)$, conditioning on an event, is defined as~\cite[Chapter~1 Section~4]{Ko33}
\be
P(B|A):=\frac{P(A\cap B)}{P(A)}
\ee
assuming $P(A)\ne 0$, 
and similarly for $P(A|B)$. However, this does not make sense for our more general setup where $A$ and $B$ are subsets of difference spaces. We therefore \emph{define} the symbol
\be
P(A\sqcap B):=\sum_{x\in A}f_{x}(B)p_{x}\equiv\sum_{y\in B}f_{yx}p_{x}. 
\ee
The meaning of $P(A\sqcap B)$ is almost the same as $P(A\cap B)$.
The only difference is that now the sets $A$ and $B$ are subsets of different sets, where the conditional probability $f$ has a codomain different from its domain. 
First, one uses $p$ to describe the probability of $A$ occurring and uses it to weight the probability that $B$ occurs as made possible by $f$, 
which is motivated by the following sketch
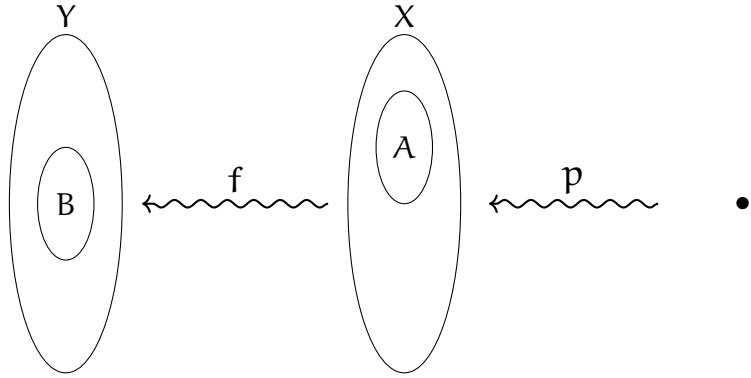
\[
\begin{tikzpicture}[xscale=-0.75,yscale=0.75]%[decoration=snake]
%\filldraw[white,even odd rule, inner color=gray, outer color=white,fill opacity=0.75] (0,1) ellipse (0.5cm and 1cm);
\draw (0,1) ellipse (0.5cm and 1cm);
\node at (-6,0) {$\bullet$};
\node at (0,1) {$A$};
\draw (0,0) ellipse (1cm and 3cm) node[above,yshift=2.2cm]{$X$};
\draw (6,0) ellipse (1cm and 3cm) node[above,yshift=2.2cm]{$Y$};
\draw (6,0) ellipse (0.5cm and 1cm) node{$B$};
\draw[thick,->,decorate,decoration={snake, amplitude=0.5mm}] (-4.5,0) -- node[above]{{\large$p$}}(-1.5,0);
\draw[thick,->,decorate,decoration={snake, amplitude=0.5mm}] (1.35,0) -- node[above]{{\large$f$}}(4.65,0);
\end{tikzpicture}
%\vspace{-4mm}
%,
\]
We therefore define 
\be
P(B|A):=\frac{P(A\sqcap B)}{P(A)}\equiv \sum_{x\in A}\frac{p_{x}}{p(A)}f_{x}(B)
\ee
provided $P(A)\ne 0,$ 
which shows how the conditional probability $P(B|A)$ is a convex combination of the $f_{x}(B)$ weighted by the normalized probability distribution $\frac{p_{x}}{p(A)}$ on $A$. A similar situation holds for $q$, $g,$ and $B$. 
By these definitions and (\ref{eq:BayesonAB}), we obtain 
the standard Bayes' theorem (\ref{eq:Bayesonsubsets}) provided $p(A)$ and $q(B)$ are non-zero. One upshot of our diagrammatic perspective is that it avoids the issue of measure zero in the statement of the theorem and relegates it to the uniqueness of the associated Bayesian inverse. A much simpler derivation of~(\ref{eq:Bayesonsubsets}) will be given from the $C^*$-algebraic perspective later in Example~\ref{ex:bayesclassicalalgebra}.
\er

\bprf
[Proof of Bayes' theorem]
The existence of a Bayesian inverse is guaranteed by the formula
\be
g_{yx}:=\begin{cases}f_{yx}p_{x}/q_{y}&\mbox{ if $q_{y}>0$}\\1/|X|&\mbox{ if $q_{y}=0$}\end{cases}.
\ee
Almost everywhere uniqueness of a Bayesian inverse is immediate from this formula, but it can also be seen from the string diagram perspective.%
%footnote
\footnote{The reader should appreciate how simple this proof is with Cho--Jacobs' carefully chosen definition of a.e.\ equivalence. It is as if the notion of a.e.\ equivalence was \emph{built} for Bayesian inference! The same proof will work in the infinite-dimensional setting as well as the quantum setting. Compare, for instance, a typically standard proof from measure theory as in \cite[Proposition~3.2]{VaTa07}. Also compare the proof of uniqueness here with the proof of uniqueness for non-commutative disintegrations~\cite[Theorem~5.1]{PaRu19}, which is closely related to the notion of Bayesian inverses~\cite{PaBayes}.
}
%end footnote
 To see this, let $g':Y\stoch X$ be another Bayesian inverse of $f$. Then
\be
\vcenter{\hbox{%
\begin{tikzpicture}[font=\small]
\node[state] (q) at (0,0) {$q$};
\node[copier] (copier) at (0,0.3) {};
\node[arrow box] (g) at (-0.5,0.95) {$g'$};
\coordinate (X) at (-0.5,1.5);
\coordinate (Y) at (0.5,1.5);
\draw (q) to (copier);
\draw (copier) to[out=165,in=-90] (g);
\draw (copier) to[out=30,in=-90] (Y);
\draw (g) to (X);
\path[scriptstyle]
node at (-0.7,1.45) {$X$}
node at (0.7,1.45) {$Y$};
\end{tikzpicture}}}
\qquad
=
\qquad
\vcenter{\hbox{%
\begin{tikzpicture}[font=\small]
\node[state] (p) at (0,0) {$p$};
\node[copier] (copier) at (0,0.3) {};
\node[arrow box] (f) at (0.5,0.95) {$f$};
\coordinate (X) at (-0.5,1.5);
\coordinate (Y) at (0.5,1.5);
\draw (p) to (copier);
\draw (copier) to[out=150,in=-90] (X);
\draw (copier) to[out=15,in=-90] (f);
\draw (f) to (Y);
\path[scriptstyle]
node at (-0.7,1.45) {$X$}
node at (0.7,1.45) {$Y$};
\end{tikzpicture}}}
\qquad
=
\qquad
\vcenter{\hbox{%
\begin{tikzpicture}[font=\small]
\node[state] (q) at (0,0) {$q$};
\node[copier] (copier) at (0,0.3) {};
\node[arrow box] (g) at (-0.5,0.95) {$g$};
\coordinate (X) at (-0.5,1.5);
\coordinate (Y) at (0.5,1.5);
\draw (q) to (copier);
\draw (copier) to[out=165,in=-90] (g);
\draw (copier) to[out=30,in=-90] (Y);
\draw (g) to (X);
\path[scriptstyle]
node at (-0.7,1.45) {$X$}
node at (0.7,1.45) {$Y$};
\end{tikzpicture}}}
\quad.
\ee
The first equality holds because $g'$ is a Bayesian inverse of $f$ and the second holds because $g$ is a Bayesian inverse of $f$. 
\eprf

\bx
\label{ex:pchops}
A simple example will illustrate why in practice one distinguishes between the sets $X$ and $Y$. Suppose I go to the grocery store and that my probability of going in a given week depends on whether or not there is a good sale. In this case, we can let $X:=\{\text{sale},\text{no sale}\}$ and $Y:=\{\text{I go},\text{I don't go}\}$. Suppose, further, that the statistics for whether there is a good sale in a given week is known (perhaps based on data and averaged out). This is the prior. Suppose $p_{\text{sale}}=0.3$ (and hence $p_{\text{no sale}}=0.7$) and also suppose if there is a sale, I am $90\%$ likely to go, whereas, if there is no sale, I am $60\%$ likely go to. These statistics are represented by a stochastic map $f:X\stoch Y$ and define the likelihood. They can be used to deduce the probability of me going to the store this week 
\be
q_{\text{I go}}
=f_{\text{I go},\text{sale}}p_{\text{sale}}+f_{\text{I go},\text{no sale}}p_{\text{no sale}}
=(0.9)(0.3)+(0.6)(0.7)
=0.69, 
\ee
which defines the marginal likelihood $\{\bullet\}\xstoch{q}Y$. 
Now, if you see me at the grocery store in a given week (the evidence), is it more likely that there is a sale this week? Based on this observation, you can infer that the probability of there being a sale has increased. Your Bayesian updated hypothesis based on seeing me there is now 
\be
g_{\text{sale},\text{I go}}=\frac{f_{\text{I go},\text{sale}}p_{\text{sale}}}{q_{\text{I go}}}
=\frac{(0.9)(0.3)}{0.69}\approx0.391.
\ee
Similarly, if you know that I did not go to the store this week, it is less likely that there is a sale, and your updated hypothesis is
\be
g_{\text{sale},\text{I don't go}}=\frac{f_{\text{I don't go},\text{sale}}p_{\text{sale}}}{q_{\text{I don't go}}}
=\frac{(0.1)(0.3)}{0.31}\approx0.097.
\ee
\ex

\br
\label{rmk:whyBayesclassical}
Bayes' theorem has many categorical formulations. We will comment on several such formulations, ones that we are aware of that seem closely related to ours. 
Culbertson and Sturtz were the first to use category theory to model Bayesian statistical inference. They proved a version of Bayes' theorem that constructs conditionals from joint probability measures~\cite[Theorem~3.2]{CuSt14}. Note that their equation~(19) is precisely what we call the Bayes condition.
Indeed, given the pair $(\{\bullet\}\xstoch{p}X,X\xstoch{f}Y)$, one can define the joint measure via the composite $\{\bullet\}\xstoch{p}X\xrightarrow{\D_{X}}X\times X\xstoch{\id_{X}\times f}X\times Y$. To go from a joint measure $\{\bullet\}\xstoch{s}X\times Y$ to a conditional $Y\xstoch{g}X$, one first obtains a disintegration $Y\xstoch{h}X\times Y$ of the pair $(X\times Y\xrightarrow{\pi_{Y}}Y,\{\bullet\}\xstoch{s}X\times Y)$ and takes the composite $Y\xstoch{h}X\times Y\xrightarrow{\pi_{X}}X$, which is the Bayesian inverse of $(f,p)$. 
A.e.\ uniqueness of Bayesian inverses was also addressed~\cite[Theorem~4.1]{CuSt14}, though this part was formulated in a more standard measure-theoretic manner. 
An elegant string-diagrammatic interpretation of Bayesian inverses (the first we are aware of) was proposed by Fong at the end of his masters thesis~\cite{Fo12}. However, Fong did not provide a full statement of Bayes' theorem, which required a fully string-diagrammatic formulation of a.e.\ uniqueness that was not available at the time. 
Clerc, Danos, Dahlqvist, and Ilias 
recast Bayes' theorem in a category of a.e.\ equivalence classes of measure-preserving Markov kernels~\cite[Theorem~2]{CDDG17}. 
Furthermore, by working with Banach cones and Markov operators, they were able to avoid the assumption that their measure spaces were standard Borel in the infinite-dimensional setting~\cite[Theorem~7]{CDDG17}. 

A completely string diagrammatic formulation of a type of Bayes' theorem (one relating joint distributions to conditionals), including its uniqueness properties, was proposed recently by Cho and Jacobs~\cite{ChJa18}. They are the first to have provided a completely diagrammatic formulations of a.e.\ equivalence (cf.~\cite[Section~3]{ChJa18}). 
Independently, we developed an algebraic definition of a.e.\ equivalence motivated by the GNS construction, which was suitable for our purposes of a non-commutative disintegration theory~\cite{PaRu19}, a precursor to Bayesian inversion. The first author was able to show the equivalence of Cho--Jacobs' definition to the operator algebraic one, substantiating both definitions for a strong candidate of a.e.\ equivalence in the non-commutative setting~\cite{PaBayes}. 

The diagrammatic formulation of Cho and Jacobs' perspective on Bayes' theorem was  on the relationship between joint marginals and conditionals (though the relationship between conditionals already appeared in Jacobs' earlier work on conjugate priors and Bayesian updating~\cite{Ja20}). This is indeed important in practice, particularly when one wants to analyze causal inference based on statistical correlation, as was done in more recent work of Jacobs, Kissinger, and Zanasi~\cite{JaKiZa}. 
Our perspective is to avoid describing Bayesian inversion as a way of going between joint distributions and conditionals and completely bypass the joint distributions, but merely work with the Bayes diagram as a \emph{condition} that must be satisfied for a morphism to be called a Bayesian inference. Classically, the two approaches are equivalent: given a conditional, one can immediately construct the joint distribution and then construct the other conditional. Quantum mechanically, however, the distinction is pivotal. This will be discussed more in Remark~\ref{rmk:whyBayesquantum}. Briefly, one will be able to construct Bayesian inference much more easily than constructing the joint distribution, which is very likely to have negative probabilities. The reason for this is due to the copy map, which is not a positive map in the non-commutative setting. Nevertheless, the key idea in \emph{formulating} the Bayes' condition is the copy map. We took this copy map seriously and learned that this lead to some surprising results in the theory of positivity, operator algebras, quantum theory, and their connection to category theory~\cite{PaBayes}. We will also see much of this in the present work. 
\er

In the remaining part of this section, we review some properties of Bayesian inverses and disintegrations. First, if a Bayesian inverse exists, it is necessarily probability-preserving. This means that if one observes new evidence that agrees with the evidence obtained as the marginal likelihood from the prior and likelihood, then the Bayesian update agrees with the prior. 

\bn
\label{prop:Bayesimpliesstatepreserving}
Given finite sets $X$ and $Y$, a probability measure $\{\bullet\}\xstoch{p}X$, and a stochastic map $X\xstoch{f}Y$ with Bayesian inverse $Y\xstoch{g}X$, then  
\be
\label{eq:Bayesstatepreserving}
\xy0;/r.25pc/:
(0,7.5)*+{\{\bullet\}}="o";
(-10,-7.5)*+{X}="X";
(10,-7.5)*+{Y}="Y";
{\ar@{~>}"o";"X"_{p}};
{\ar@{~>}"o";"Y"^{f\circ p}};
{\ar@{~>}"Y";"X"^{g}};
{\ar@{=}(4,-4);(-2,-1)};
\endxy
\quad,\qquad\text{i.e.}\qquad
\vcenter{\hbox{%
\begin{tikzpicture}[font=\small]
\node[state] (p) at (0,0) {$p$};
\node (X) at (0,1.0) {};
\draw (p) to (X);
\end{tikzpicture}}}
\qquad
=
\qquad
\vcenter{\hbox{%
\begin{tikzpicture}[font=\small]
\node[state] (p) at (0,0) {$p$};
\node[arrow box] (f) at (0,0.55) {$f$};
\node[arrow box] (g) at (0,1.35) {$g$};
\draw (p) to (f);
\draw (f) to (g);
\draw (g) to (0,1.85);
\end{tikzpicture}}}
\quad.
\ee
\en

\bprf
[Proof of Proposition \ref{prop:Bayesimpliesstatepreserving}]
By using Bayes' diagram together with the properties of the copy and discard maps, we obtain commutativity of the diagram 
\be
\xy0;/r.25pc/:
(0,15)*+{\{\bullet\}}="1";
(-25,15)*+{Y}="Y";
(25,15)*+{X}="X";
(-25,0)*+{Y\times Y}="YY";
(25,0)*+{X\times X}="XX";
(0,0)*+{X\times Y}="XY";
(0,-15)*+{X}="Xb";
{\ar@{~>}"1";"X"^{p}};
{\ar@{~>}"1";"Y"_{f\circ p}};
{\ar"Y";"YY"_{\Delta_{Y}}};
{\ar"X";"XX"^{\Delta_{X}}};
{\ar@{~>}"YY";"XY"_{g\times\id_{Y}}};
{\ar@{~>}"XX";"XY"^{\id_{X}\times f}};
%(0,7.5)*{=\joinrel=\joinrel=};
{\ar"XY";"Xb"|-{\pi_{X}}};
{\ar"XX";"Xb"|-{\pi_{X}}};
%{\ar@{~>}"YY";"Xb"_{g}};
{\ar@{~}@/_1.0pc/"Y";(-38,0)};
{\ar@{~}@/_0.3pc/(-38,0);(-32,-10)};
{\ar@{~>}@/_0.8pc/(-32,-10);"Xb"_(0.1){g}};
{\ar@{-}@/^1.0pc/"X";(38,0)};
{\ar@{-}@/^0.3pc/(38,0);(32,-10)};
{\ar@/^0.8pc/(32,-10);"Xb"^(0.1){\id_{X}}};
\endxy
\quad,
\ee
which proves the proposition. 
It may be helpful to compare this with a string diagrammatic proof
\be
\vcenter{\hbox{%
\begin{tikzpicture}[font=\small]
\node[state] (p) at (0,0) {$p$};
\node (X) at (0,1.0) {};
\draw (p) to (X);
\end{tikzpicture}}}
\qquad
=
\qquad
\vcenter{\hbox{%
\begin{tikzpicture}[font=\small]
\node[state] (p) at (0,0) {$p$};
\node[copier] (copier) at (0,0.3) {};
\coordinate (X) at (-0.5,1.5);
\coordinate (c) at (0.5,0.91);
\node[discarder] (Y) at (c) {};
\draw (p) to (copier);
\draw (copier) to[out=150,in=-90] (X);
\draw (copier) to[out=15,in=-90] (c);
\end{tikzpicture}}}
\qquad
=
\qquad
\vcenter{\hbox{%
\begin{tikzpicture}[font=\small]
\node[state] (p) at (0,0) {$p$};
\node[copier] (copier) at (0,0.3) {};
\node[arrow box] (f) at (0.5,0.95) {$f$};
\coordinate (X) at (-0.5,1.55);
\node[discarder] (Y) at (0.5,1.4) {};
\draw (p) to (copier);
\draw (copier) to[out=150,in=-90] (X);
\draw (copier) to[out=15,in=-90] (f);
\draw (f) to (Y);
\end{tikzpicture}}}
\qquad
=
\qquad
\vcenter{\hbox{%
\begin{tikzpicture}[font=\small]
\node[state] (q) at (0,0) {$q$};
\node[copier] (copier) at (0,0.3) {};
\coordinate (c) at (0.5,0.91);
\node[discarder] (Y) at (c) {};
\node[arrow box] (g) at (-0.5,0.95) {$g$};
\coordinate (X) at (-0.5,1.55);
\node[discarder] (Y) at (c) {};
\draw (q) to (copier);
\draw (copier) to[out=165,in=-90] (g);
\draw (g) to (X);
\draw (copier) to[out=15,in=-90] (c);
\end{tikzpicture}}}
\qquad
=
\qquad
\vcenter{\hbox{%
\begin{tikzpicture}[font=\small]
\node[state] (q) at (0,0) {$q$};
\node[arrow box] (g) at (0,0.75) {$g$};
\draw (q) to (g);
\draw (g) to (0,1.55);
\end{tikzpicture}}}
\quad,
\ee
where $q:=f\circ p$. 
\eprf

\bt
Let $p,f,$ and $q$ be as in Theorem~\ref{thm:classicalBayestheorem}. 
\begin{enumerate}[i.]
\itemsep0pt
\item
If $f$ is an isomorphism, then $g:=f^{-1}$ is a Bayesian inverse of $(f,p)$. 
\item
If $f$ is deterministic, then $g$ is a Bayesian inverse of $(f,p)$ if and only if $g$ is a disintegration of $(f,p)$. 
\item
A Bayesian inverse of a composite is the composite of Bayesian inverses. 
\item
A Bayesian inverse of a Bayesian inverse is a.e.\ equivalent to the original map. 
\end{enumerate}
\et

\bprf
All of these claims are proved in~\cite{PaBayes}. 
\eprf

%%%%%%%%%%%%%%%%%%%%%%%%%%%%%%%%%%%%%
\section{Quantum a.e.\ equivalence, measure zero, and Bayesian inference} 
%\subsection{Measure zero in non-commutative probability}
\label{sec:measurezero}
%%%%%%%%%%%%%%%%%%%%%%%%%%%%%%%%%%%%%

Before we state our quantum version of Bayes' theorem, we describe how our operator-algebraic notion of a.e.\ equivalence for finite-dimensional $C^*$-algebras introduced in \cite{PaRu19} agrees with the diagrammatic definition of a.e.\ equivalence introduced by Cho and Jacobs \cite{ChJa18}  in Proposition~\ref{thm:ncaeequivalence}. The work of Cho and Jacobs focused exclusively on commutative probability, so it is quite surprising that their notion of a.e.\ equivalence agrees with ours, which was motivated by the theory of operator algebras and the Gelfand--Naimark--Segal construction. 
This notion is crucial for formulating Bayesian statistics and quantum probability in general. 

\bd
Given a $C^*$-algebra $\mA,$ 
a \define{positive element} of $\mA$ 
is an element $a\in\mA$ for which there exists an $x\in\mA$
such that $a=x^*x.$
The set of positive elements in $\mA$ 
is denoted by $\mA^{+}.$
An element $a\in\mA$ is \define{self-adjoint} iff $a^*=a$. 
Positivity defines a partial order on self-adjoint elements and one writes $a\ge a'$ iff $a-a'\in\mA^{+}$. 
Given another $C^*$-algebra $\mB,$ a \define{positive map}
$\vf:\mB\stoch\mA$ is a linear map such that 
$\vf(\mB^{+})\subseteq\mA^{+}.$ 
A positive unital map $\omega:\mA\stoch\C$ is called a \define{state} (on $\mA$). 
A state $\omega$ is \define{faithful} iff $\omega(a^*a)=0$, with $a\in\mA$, implies $a=0$. 
A linear map $\vf:\mB\stoch\mA$ is \define{$*$-preserving} iff $\vf(b^*)=\vf(b)^*$ for all $b\in\mB$. 
For the $C^*$-algebra of $m\times m$ matrices $\mathcal{M}_{m}(\C)$, which will be referred to as a \define{matrix algebra}, the involution is complex conjugation and will be denoted by $\dag$ instead of $*$. 
Given $n\in\N,$ 
a linear map $\vf:\mB\stoch\mA$ is \define{$n$-positive} iff 
$\id_{\mathcal{M}_{n}(\C)}\otimes\vf:\mathcal{M}_{n}(\C)\otimes\mB\stoch\mathcal{M}_{n}(\C)\otimes\mA$ 
is positive. The map 
$\vf$ is \define{completely positive} iff $\vf$ is $n$-positive for all $n\in\N.$
A completely positive (unital) map will be abbreviated as
a CP (CPU) map. 
Let $E^{(m)}_{ij}\in\mM_{m}(\C)$ be the standard $ij$-th matrix unit (with $1$ in the $ij$-th entry and $0$ otherwise). If it is clear from context, the shorthand $E_{ij}$ may be used instead of $E_{ij}^{(m)}$. 
If $\psi:\mathcal{M}_{n}(\C)\stoch\mathcal{M}_{m}(\C)$ is a linear map, then the matrix \be
\mathrm{Choi}(\psi):=\sum_{i,j}E_{ij}^{(n)}\otimes \psi\left(E_{ij}^{(n)}\right)\in\mathcal{M}_{n}(\C)\otimes\mathcal{M}_{m}(\C)
\ee
is called the \define{Choi matrix} of $\psi$. 
The \define{Hilbert--Schmidt inner product} on $\mathcal{M}_{m}(\C)$ is given by 
\be
\mathcal{M}_{m}(\C)\ni A,B\mapsto\<A,B\>:=\tr(A^{\dag}B).
\ee
If $\mathcal{M}_{n}(\C)\xstoch{F}\mathcal{M}_{m}(\C)$ is a linear map, its \define{Hilbert--Schmidt dual} will be denoted by $F^{*}$ and is uniquely characterized by the condition
\be
\label{eq:HSdualcondition}
\tr\big(F^*(A)B\big)=\tr\big(AF(B)\big)\qquad\forall\;A\in\mathcal{M}_{m}(\C),\;\;B\in\mathcal{M}_{n}(\C). 
\ee
If $S\subseteq\mA$ is a subset of a $C^*$-algebra $\mA$,  
the \define{commutant} of $S$ inside $\mA$ is the unital algebra
\be
S':=\{a\in\mA\;:\;as=sa\;\forall\;s\in S\}.
\ee
Since the commutant depends on the embedding algebra, 
$S'$ will often be written as $S'\subseteq\mA.$
\ed

We will expect the reader is \emph{vaguely} familiar with some facts about CP maps such as ``a map is CP if and only if its Choi matrix is positive (semidefinite)'' and ``all positive maps are $*$-preserving''~\cite[Section~1.1]{St13}. The seminal paper of Choi is an excellent and concise reference~\cite{Ch75}. 
Also note that we use the term positive for what some may call positive semidefinite. 

\br
The definition of complete positivity is well-motivated in the physics literature (tensoring with a trivial system should not change the positivity of a map)~\cite{Kr83}. Mathematically, there are good categorical reasons to use completely positive maps as well. 
In the category of finite-dimensional $C^*$-algebras, the tensor product of positive maps is not necessarily a positive map. In fact, the largest subcategory of the category of positive maps between finite-dimensional $C^*$-algebras that is closed under the tensor product is the category of finite-dimensional $C^*$-algebras and completely positive maps. This follows from Proposition~\ref{prop:whyCP} below.%
%footnote
\footnote{We learned Proposition~\ref{prop:whyCP} in Lecture~3 of Reinhard Werner's course on quantum information theory~\cite{Werner17}.}
%end footnote
\er

\bn
\label{prop:whyCP}
Fix finite-dimensional $C^*$-algebras $\mA$ and $\mB$ and let $\vf:\mB\stoch\mA$ be a positive map. Then the following conditions are equivalent. 
\begin{enumerate}[i.]
\itemsep0pt
\item
$\vf$ is completely positive. 
\item
$\vf\otimes\id_{\mathcal{M}_{n}(\C)}$ is positive for all $n\in\N$.
\item
$\vf\otimes\psi$ is positive for all positive $\psi:\mD\stoch\mC$ with $\mC$ and $\mD$ arbitrary finite-dimensional $C^*$-algebras. 
\end{enumerate}
\en

The proof is short enough that we include it. 

\bprf
{\color{white}{you found me!}}

\noindent
(i$\Leftrightarrow$ii)
The swap map%
%footnote
\footnote{Recall, the swap map is the unique linear map $\mA\otimes\mB\stoch\mB\otimes\mA$ uniquely determined by the assignment $A\otimes B\mapsto B\otimes A$.} 
%end footnote
is positive because it is a $^*$-isomorphism. The composite of positive maps is positive. 

\noindent
(iii$\Rightarrow$i)
Set $\psi=\id_{\mathcal{M}_{n}(\C)}.$ Use the equivalence between i. and ii. 

\noindent
(i$\Rightarrow$iii) Since $\vf\otimes\psi=(\vf\otimes\id_{\mC})\circ(\id_{\mB}\otimes\psi)$ by the interchange law, it suffices to show that each factor is positive. Since every finite-dimensional $C^*$-algebra is ($^*$-isomorphic to) a direct sum of matrix algebras, it suffices to assume $\mC=\bigoplus_{x\in X}\mathcal{M}_{m_{x}}(\C)$ for some finite set $X$ and positive integers $\{m_{x}\}$. By the distributivity between direct sum and the tensor product, $\vf\otimes\id_{\mC}=\vf\otimes\bigoplus_{x\in X}\id_{\mathcal{M}_{m_{x}}(\C)}\cong\bigoplus_{x\in X}\vf\otimes\id_{\mathcal{M}_{m_{x}}(\C)}$. This is positive if and only if each $\vf\otimes\id_{\mathcal{M}_{m_{x}}(\C)}$ is, which is true by assumption ii. A similar argument holds for the second factor $\id_{\mB}\otimes\psi$.  
\eprf

\bc
Let $\mathcal{M}$ be the monoidal category of finite-dimensional $C^*$-algebras and linear maps (with the standard tensor product) and let $\mathcal{P}$ be the subcategory of finite-dimensional $C^*$-algebras and positive maps. The largest subcategory of $\mathcal{P}$ that is closed under the tensor product is the category of finite-dimensional $C^*$-algebras and completely positive maps. 
\ec

Now that we spent enough time discussing positivity, let us consider some useful examples of maps that are not positive. 

\begin{notation}
Given any finite-dimensional $C^*$-algebra $\mA$, let $\mu_{\mA}:\mA\otimes\mA\stoch\mA$ be the multiplication map uniquely determined by 
\be
\xy0;/r.25pc/:
(0,-5)*+{\mA}="X";
(0,5)*+{\mA\otimes\mA}="XX";
{\ar@{~>}"XX";"X"_{\mu_{\mA}}};
\endxy
\quad
\xy0;/r.25pc/:
(0,-5)*+{aa'}="X";
(0,5)*+{a\otimes a'}="XX";
{\ar@{|->}"XX";"X"};
\endxy
\quad
\vcenter{\hbox{
\begin{tikzpicture}%[scale=1.5]
\newcommand\dist{0.6};
\node at (0,0) {$\bullet$};
\draw (0,0) to [out=180,in=270] (-\dist,\dist);
\draw (0,0) to [out=0,in=270] (\dist,\dist);
\draw (0,0) to [out=270,in=90] (0,-\dist);
\end{tikzpicture}
}}
\ee
This map is linear and unital, but it is not a $^*$-homomorphism unless $\mA$ is commutative. In fact, $\mu_{A}$ is not even positive in general. This is because the product of two positive matrices need not be positive.
\end{notation}

\bd
\label{defn:ncaeequivalence}
Let $\mA$ and $\mB$ be finite-dimensional $C^*$-algebras (or more generally, von Neumann algebras), let $\mA\xstoch{\w}\C$ be a state on $\mA$, and let $F,G:\mB\stoch\mA$ be linear maps. Then $F$ is said to be \define{$\w$-a.e. equivalent to} $G$ iff
\be
\label{eq:ncaeequivalencediagram}
\xy;/r.25pc/:
(-30,0)*+{\C}="0";
(-15,7.5)*+{\mA}="Xt";
(-15,-7.5)*+{\mA}="Xb";
(7.5,7.5)*+{\mA\otimes \mA}="XXt";
(7.5,-7.5)*+{\mA\otimes\mA}="XXb";
(30,0)*+{\mA\otimes\mB}="XY";
{\ar@{~>}"Xt";"0"_{\w}};
{\ar@{~>}"Xb";"0"^{\w}};
{\ar@{~>}"XXt";"Xt"_(0.55){\mu_{\mA}}};
{\ar@{~>}"XXb";"Xb"^(0.55){\mu_{\mA}}};
{\ar@{~>}"XY";"XXt"_{\id_{\mA}\otimes F}};
{\ar@{~>}"XY";"XXb"^{\id_{\mA}\otimes G}};
{\ar@{=}(-4,3);(-4,-3)};
\endxy
,\qquad\text{i.e.}\qquad
\vcenter{\hbox{
\begin{tikzpicture}[font=\small]
\node[state] (p) at (0,0) {$\w$};
\node[copier] (copier) at (0,0.3) {};
\node[arrow box] (f) at (0.5,0.95) {$F$};
\coordinate (X) at (-0.5,1.5);
\coordinate (Y) at (0.5,1.5);
\draw (p) to (copier);
\draw (copier) to[out=150,in=-90] (X);
\draw (copier) to[out=15,in=-90] (f);
\draw (f) to (Y);
\path[scriptstyle]
node at (-0.7,1.45) {$\mA$}
node at (0.7,1.45) {$\mB$};
\end{tikzpicture}}}
\quad=\quad
\vcenter{\hbox{
\begin{tikzpicture}[font=\small]
\node[state] (p) at (0,0) {$\w$};
\node[copier] (copier) at (0,0.3) {};
\node[arrow box] (g) at (0.5,0.95) {$G$};
\coordinate (X) at (-0.5,1.5);
\coordinate (Y) at (0.5,1.5);
\draw (p) to (copier);
\draw (copier) to[out=150,in=-90] (X);
\draw (copier) to[out=15,in=-90] (g);
\draw (g) to (Y);
\path[scriptstyle]
node at (-0.7,1.45) {$\mA$}
node at (0.7,1.45) {$\mB$};
\end{tikzpicture}}}
.
\ee
In this case, the notation $F\underset{\raisebox{.3ex}[0pt][0pt]{\scriptsize$\w$}}{=}G$ will be used. As an equation, $F\underset{\raisebox{.3ex}[0pt][0pt]{\scriptsize$\w$}}{=}G$ is equivalent to $\w(AF(B))=\w(AG(B))$ for all $A\in\mA$ and $B\in\mB$. 
\ed

\begin{notation}
In what follows, let $P_{\w}$ denote the \define{support} of a state $\w:\mA\stoch\C$ with $\mA$ a finite-dimensional $C^*$-algebra (or von Neumann algebra). The support is the smallest projection in $\mA$ such that 
\be
\label{eq:supportstateidentities}
\w(A)=\w(P_{\w}A)=\w(AP_{\w})=\w(P_{\w}AP_{\w})\qquad\forall\;A\in\mA. 
\ee
In particular, if we write $P_{\w}^{\perp}:=1_{\mA}-P_{\w}$ for the orthogonal complement, then $\w(P_{\w}^{\perp}A)=0$ and $\w(AP_{\w}^{\perp})=0$ for all $A\in\mA$.
\end{notation}

\bn
\label{thm:ncaeequivalence}
Let $\mA$ and $\mB$ be finite-dimensional $C^*$-algebras (or von Neumann algebras), let $\mA\xstoch{\w}\C$ be a state on $\mA$, and let $F,G:\mB\stoch\mA$ be linear maps. Then the following conditions are equivalent. 
\begin{enumerate}[i.]
\itemsep0pt
\item
$F\underset{\raisebox{.3ex}[0pt][0pt]{\scriptsize$\w$}}{=}G$. 
\item
$F(B)P_{\w}=G(B)P_{\w}$ for all $B\in\mB$. 
\item
$\w\Big(\big(F(B)-G(B)\big)^*\big(F(B)-G(B)\big)\Big)=0$ for all $B\in\mB$, i.e.\ $F(B)-G(B)$ is in the null space of $\w$ for all $B\in\mB$. 
\end{enumerate}
\en

\bprf
See~\cite[Proposition~5.15]{PaBayes} for a proof.
\eprf

\br
\label{rmk:aeCAlg}
The equivalence between i.\ and iii.\ in Proposition~\ref{thm:ncaeequivalence} holds more generally for \emph{all} $C^*$-algebras~\cite{PaBayes}. However, ii.\ is no longer equivalent to i.\ nor iii.\ because general $C^*$-algebras do not have enough projections. 
\er

\br
We had two reasons for avoiding the diagrammatic definition~(\ref{eq:ncaeequivalencediagram}) in~\cite{PaRu19}. The first is that the multiplication map $\mu_{\mA}:\mA\otimes\mA\stoch\mA$ is not a positive map and therefore not a quantum channel. In fact, the no-broadcasting theorem states that a CPU map $\D_{\mA}:\mA\otimes\mA\stoch\mA$ satisfying $\D_{\mA}(1_{\mA}\otimes A)=A=\D_{\mA}(A\otimes1_{\mA})$ for all $A\in\mA$ exists if and only if $\mA$ is commutative (cf.~\cite[Theorem~6]{Ma10} and~\cite[Theorem~4.20]{PaBayes}). The second reason we avoided the Cho--Jacobs definition of a.e.\ equivalence is that it was not clear to us at the time whether the notion was equivalent to our definition in terms of null spaces.
In the present work, we find that working with $\mu_{\mA}$ is much more crucial due to its explicit appearance in our \emph{statement} of Bayes' theorem. Using $\mu_{\mA}$ means we must leave the Kleisli category consisting of CPU maps in the non-commutative setting~\cite{We17}. 
However, our morphisms of interest will generally remain in the subcategory of CPU maps and we will use the structure of the larger quantum Markov categories in which they live~\cite{PaBayes}. 
\er

\br
As discussed in Remark~\ref{eq:twonotionsofaeequivalence}, Cho and Jacobs prove that 
\be
\vcenter{\hbox{
\begin{tikzpicture}[font=\small]
\node[state] (p) at (0,0) {$p$};
\node[copier] (copier) at (0,0.3) {};
\node[arrow box] (f) at (0.5,0.95) {$f$};
\coordinate (X) at (-0.5,1.5);
\coordinate (Y) at (0.5,1.5);
\draw (p) to (copier);
\draw (copier) to[out=150,in=-90] (X);
\draw (copier) to[out=15,in=-90] (f);
\draw (f) to (Y);
\path[scriptstyle]
node at (-0.7,1.45) {$X$}
node at (0.7,1.45) {$Y$};
\end{tikzpicture}}}
\quad=\quad
\vcenter{\hbox{
\begin{tikzpicture}[font=\small]
\node[state] (p) at (0,0) {$p$};
\node[copier] (copier) at (0,0.3) {};
\node[arrow box] (g) at (0.5,0.95) {$g$};
\coordinate (X) at (-0.5,1.5);
\coordinate (Y) at (0.5,1.5);
\draw (p) to (copier);
\draw (copier) to[out=150,in=-90] (X);
\draw (copier) to[out=15,in=-90] (g);
\draw (g) to (Y);
\path[scriptstyle]
node at (-0.7,1.45) {$X$}
node at (0.7,1.45) {$Y$};
\end{tikzpicture}}}
\qquad
\iff
\qquad
\vcenter{\hbox{
\begin{tikzpicture}[font=\small]
\node[state] (p) at (0,0) {$p$};
\node[copier] (copier) at (0,0.3) {};
\node[arrow box] (f) at (-0.5,0.95) {$f$};
\coordinate (X) at (0.5,1.5);
\coordinate (Y) at (-0.5,1.5);
\draw (p) to (copier);
\draw (copier) to[out=30,in=-90] (X);
\draw (copier) to[out=165,in=-90] (f);
\draw (f) to (Y);
\path[scriptstyle]
node at (-0.7,1.45) {$Y$}
node at (0.7,1.45) {$X$};
\end{tikzpicture}}}
\quad=\quad
\vcenter{\hbox{
\begin{tikzpicture}[font=\small]
\node[state] (p) at (0,0) {$p$};
\node[copier] (copier) at (0,0.3) {};
\node[arrow box] (g) at (-0.5,0.95) {$g$};
\coordinate (X) at (0.5,1.5);
\coordinate (Y) at (-0.5,1.5);
\draw (p) to (copier);
\draw (copier) to[out=30,in=-90] (X);
\draw (copier) to[out=165,in=-90] (g);
\draw (g) to (Y);
\path[scriptstyle]
node at (-0.7,1.45) {$Y$}
node at (0.7,1.45) {$X$};
\end{tikzpicture}}}
\ee
in any classical Markov category. 
However, their proof involves an identity that does not hold in the quantum setting of arbitrary linear maps $f,g,$ and $p$. Nevertheless, the analogous `if and only if' statement holds in the quantum setting assuming $f,g,$ and $p$ are $*$-preserving due to the interplay between the multiplication map $\mu_{\mA}:\mA\otimes\mA\stoch\mA$, the involution $*$ on $\mA$, and the swap map from the symmetric monoidal structure on finite-dimensional $C^*$-algebras. This is proved in~\cite{PaBayes}. In fact, there is another condition when these are equivalent, which will be useful for us later. 
\er

\blem
\label{lem:faithfulstateimpliesunique}
Let $F,G:\mA\stoch\mB$ be two linear maps  of $C^*$-algebras and let $\xi:\mB\stoch\C$ be a faithful state. Then 
\be
\vcenter{\hbox{
\begin{tikzpicture}[font=\small]
\node[state] (p) at (0,0) {$\xi$};
\node[copier] (copier) at (0,0.3) {};
\node[arrow box] (f) at (0.5,0.95) {$F$};
\coordinate (X) at (-0.5,1.5);
\coordinate (Y) at (0.5,1.5);
\draw (p) to (copier);
\draw (copier) to[out=150,in=-90] (X);
\draw (copier) to[out=15,in=-90] (f);
\draw (f) to (Y);
\path[scriptstyle]
node at (-0.3,1.45) {$\mB$}
node at (0.7,1.45) {$\mA$};
\end{tikzpicture}}}
\quad=\quad
\vcenter{\hbox{
\begin{tikzpicture}[font=\small]
\node[state] (p) at (0,0) {$\xi$};
\node[copier] (copier) at (0,0.3) {};
\node[arrow box] (g) at (0.5,0.95) {$G$};
\coordinate (X) at (-0.5,1.5);
\coordinate (Y) at (0.5,1.5);
\draw (p) to (copier);
\draw (copier) to[out=150,in=-90] (X);
\draw (copier) to[out=15,in=-90] (g);
\draw (g) to (Y);
\path[scriptstyle]
node at (-0.3,1.45) {$\mB$}
node at (0.7,1.45) {$\mA$};
\end{tikzpicture}}}
\quad
\iff
\quad
\vcenter{\hbox{
\begin{tikzpicture}[font=\small]
\node[state] (p) at (0,0) {$\xi$};
\node[copier] (copier) at (0,0.3) {};
\node[arrow box] (f) at (-0.5,0.95) {$F$};
\coordinate (X) at (0.5,1.5);
\coordinate (Y) at (-0.5,1.5);
\draw (p) to (copier);
\draw (copier) to[out=30,in=-90] (X);
\draw (copier) to[out=165,in=-90] (f);
\draw (f) to (Y);
\path[scriptstyle]
node at (-0.7,1.45) {$\mA$}
node at (0.3,1.45) {$\mB$};
\end{tikzpicture}}}
\quad=\quad
\vcenter{\hbox{
\begin{tikzpicture}[font=\small]
\node[state] (p) at (0,0) {$\xi$};
\node[copier] (copier) at (0,0.3) {};
\node[arrow box] (g) at (-0.5,0.95) {$G$};
\coordinate (X) at (0.5,1.5);
\coordinate (Y) at (-0.5,1.5);
\draw (p) to (copier);
\draw (copier) to[out=30,in=-90] (X);
\draw (copier) to[out=165,in=-90] (g);
\draw (g) to (Y);
\path[scriptstyle]
node at (-0.7,1.45) {$\mA$}
node at (0.3,1.45) {$\mB$};
\end{tikzpicture}}}
\quad
\iff
\quad
F=G.
\ee
\elem

\bprf
It is clear that the condition $F=G$ implies both diagrammatic conditions. The first condition implies the last by Proposition~\ref{thm:ncaeequivalence}, Remark~\ref{rmk:aeCAlg}, and the definition of a faithful state. The second condition, which reads
\be
\xi\big(F(A)B\big)=\xi\big(G(A)B\big)\qquad\forall\;A\in\mA,\;B\in\mB,
\ee 
implies the last for the following reason. Fixing $A$ and setting $B:=F(A)^*-G(A)^*$ and using this equality gives%
%\footnote
\footnote{Note that we did not assume $F$ or $G$ are $*$-preserving in this calculation.}
%end footnote
\be
0=\xi\Big(\big(F(A)-G(A)\big)\big(F(A)^*-G(A)^*\big)\Big)
=\xi\Big(\big(F(A)^*-G(A)^*\big)^*\big(F(A)^*-G(A)^*\big)\Big).
\ee
By faithfulness of $\xi$, $F(A)^*-G(A)^*=0$, i.e. $F(A)=G(A)$. Since $A$ was arbitrary, $F=G$. This proves all conditions are equivalent. 
\eprf

\bd
Let $(\mA,\w)$ and $(\mB,\xi)$ be finite-dimensional $C^*$-algebras equipped with states.
Let $F:\mB\stoch\mA$ be a CPU state-preserving map, i.e.\ 
$\w\circ F=\xi$. 
A \define{disintegration} of $(F,\omega,\xi)$
is a CPU map
$G:\mA\stoch\mB$ such that 
\be
\xy0;/r.25pc/:
(0,-7.5)*+{\C}="C";
(-12.5,7.5)*+{\mA}="H";
(12.5,7.5)*+{\mB}="K";
{\ar@{~>}"H";"C"_{\w}};
{\ar@{~>}"K";"C"^{\xi}};
{\ar@{~>}"H";"K"^{G}};
{\ar@{=}(-3,0);(5,4.5)};
\endxy
\qquad\text{and}\qquad
\xy0;/r.25pc/:
(0,-7.5)*+{\mA}="C";
(-12.5,7.5)*+{\mB}="H";
(12.5,7.5)*+{\mB}="K";
{\ar@{~>}"H";"C"_{F}};
{\ar@{~>}"C";"K"_{G}};
{\ar"H";"K"^{\id_{\mB}}};
{\ar@{=}(-3,0);(5,4.5)_{\xi}};
\endxy
.
\ee
\ed

\bd
\label{defn:quantumBayesianinverse}
Let $\mB\xstoch{F}\mA$ be a CPU map between finite-dimensional $C^*$-algebras, let $\mA\xstoch{\w}\C$ be a state, and set $\xi:=\w\circ F$. A \define{Bayesian inverse} of $(F,\w)$ is a CPU map $\mA\xstoch{G}\mB$ such that 
\be
\label{eq:quantumBayesdiagram}
\xy0;/r.25pc/:
(0,-7.5)*+{\C}="1";
(-25,-7.5)*+{\mB}="B";
(25,-7.5)*+{\mA}="A";
(-25,7.5)*+{\mB\otimes\mB}="BB";
(25,7.5)*+{\mA\otimes\mA}="AA";
(0,7.5)*+{\mA\otimes\mB}="AB";
{\ar@{~>}"A";"1"^{\w}};
{\ar@{~>}"B";"1"_{\xi}};
{\ar@{~>}"BB";"B"_{\mu_{\mB}}};
{\ar@{~>}"AA";"A"^{\mu_{\mA}}};
{\ar@{~>}"AB";"BB"_{G\otimes\id_{\mB}}};
{\ar@{~>}"AB";"AA"^{\id_{\mA}\otimes F}};
(0,0)*{=\joinrel=\joinrel=};
\endxy
,
\ee
i.e.\
\be
\label{eq:quantumBayesequation}
\xi\big(G(A)B\big)=\w\big(AF(B)\big)
\ee
for all $A\in\mA$ and $B\in\mB$. 
The diagram in~(\ref{eq:quantumBayesdiagram}) is referred to as \define{Bayes' diagram} or the (quantum) \define{Bayes condition}.
More generally, a \emph{linear} map $G$ satisfying the Bayes condition will be called a \define{Bayes map}. 
In the language of (quantum) Bayesian statistics, the state $\w$ is called the \define{prior state}, the CPU map $F$ is called the \define{likelihood}, the state $\xi$ is called the \define{marginal likelihood}, and the CPU map $G$ is called the \define{weighted likelihood}/\define{posterior}. 
It is helpful to summarize this diagrammatically as 
\be
\xy0;/r.35pc/:
(0,-7.5)*+{\C}="C";
(10,7.5)*+{\mA}="A";
(-10,7.5)*+{\mB}="B";
{\ar@{~>}"A";"C"^{\w=\text{prior}}};
{\ar@{~>}"B";"C"_{\text{marginal likelihood}=\xi}};
{\ar@/^1.0pc/@{~>}"A";"B"_{G}^{\text{posterior}}};
{\ar@/^1.0pc/@{~>}"B";"A"_{F}^{\text{likelihood}}};
\endxy
\ee
If one obtains \emph{new} \define{evidence} in the form of a state $\mB\xstoch{\xi'}\C$, then the (quantum) \define{Bayesian update/Jeffrey conditioning} is the state on $\mA$ obtained from the composite $\mA\xstoch{G}\mB\xstoch{\xi'}\C$.
\ed

\br
We do not claim that our definition of a quantum Bayesian inverse is \emph{the} quantum analogue of a classical Bayesian inverse, which implicitly implies there is no other possibility. We stress that it is \emph{a} quantum generalization. More generally, one should be cautious in making statements of the form ``$X$ is \emph{the} quantum generalization of $Y$.'' Nevertheless, we will occasionally say a linear map $G$ satisfies \emph{the} quantum Bayes condition, which will specifically mean that~(\ref{eq:quantumBayesequation}) is satisfied. 
\er

\bn
Let $F,\w,$ and $\xi$ be as in Definition~\ref{defn:quantumBayesianinverse}. 
\begin{enumerate}[i.]
\itemsep0pt
\item
If $G$ is a Bayes map for $(F,\omega)$, then $\w=\xi\circ G$. 
\item
If $G$ is a $*$-preserving unital Bayes map for $(F,\omega)$, then it is necessarily $\xi$-a.e.\ unique. 
\item
If $F$ is a $^*$-isomorphism, then $G=F^{-1}$ is a Bayesian inverse of $(F,\w)$.
\item
If $F$ is a $^*$-homomorphism and has a disintegration $G$, then $G$ is a Bayesian inverse of $(F,\w)$. 
\item
If $F$ is a $^*$-homomorphism and has a Bayesian inverse $G$ of $(F,\w)$, then $G$ is a disintegration of $(F,\w)$.
\item
The composite of $*$-preserving Bayes maps is a $*$-preserving Bayes maps of the composite. 
\item
A $*$-preserving Bayes maps of a $*$-preserving Bayes maps is a.e.\ equivalent to the original map. 
\end{enumerate} 
\en

\bprf
The first item can be proven immediately using unitality of $F$ from~(\ref{eq:quantumBayesequation}) or string diagrammatically from causality as in Proposition~\ref{prop:Bayesimpliesstatepreserving}. 
The second item follows immediately from the diagrammatic definition of a.e.\ equivalence. The rest of the claims (along with others not listed here) are proved in~\cite{PaBayes}. 
\eprf

The uniqueness property will be important when we try to find Bayes maps that are positive, and not just linear. 

\bx
\label{ex:bayesclassicalalgebra}
Consider the case where $\mA=\C^{X}$ and $\mB=\C^{Y}$ with $X$ and $Y$ finite sets. Then CPU maps $\mB\xstoch{F}\mA$ and $\mA\xstoch{G}\mB$ correspond to stochastic maps $X\xstoch{f}Y$ and $Y\xstoch{g}X$, respectively~\cite{Pa17}. Furthermore, states $\mA\xstoch{\w}\C$ and $\mB\xstoch{\xi}\C$ correspond to probability measures $p$ on $X$ and $q$ on $Y$, respectively. For each $x\in X$, let $e_{x}\in\C^{X}$ denote the function 
\be
X\ni x'\mapsto e_{x}(x'):=\de_{xx'}
\ee
and similarly for $e_{y}\in\C^{Y}$. 
If we now represent subsets (events) $A\subseteq X$ and $B\subseteq Y$ by their corresponding indicator functions $\chi_{A}\in\C^{X}$ and $\chi_{B}\in\C^{Y}$, respectively, we see that the Bayes condition gives
\be
\xi\big(G(\chi_{A})\chi_{B}\big)=\w\big(\chi_{A}F(\chi_{B})\big).
\ee 
It is helpful to evaluate these expressions more explicitly. We obtain  
\be
G\left(\chi_{A}\right)=\sum_{x\in A\subseteq X}G(e_{x})=\sum_{x\in A\subseteq X}\sum_{y\in Y}g_{xy}e_{y},
\ee 
while 
\be
F\left(\chi_{B}\right)=\sum_{y\in B\subseteq Y}F(e_{y})=\sum_{y\in B\subseteq Y}\sum_{x\in X}f_{yx}e_{x}. 
\ee
Hence, 
\be
\xi\big(G(\chi_{A})\chi_{B}\big)
=\sum_{x\in A\subseteq X}\sum_{y\in Y}\sum_{y'\in B\subseteq Y}g_{xy}\xi\left(e_{y}e_{y'}\right)
=\sum_{x\in A\subseteq X}\sum_{y\in B\subseteq Y}g_{xy}\xi(e_{y})
%=\sum_{x\in A\subseteq X}\sum_{y\in B\subseteq Y}g_{xy}q_{y}
=\sum_{y\in B\subseteq Y}g_{y}(A)q_{y}
\ee
while
\be
\w\big(\chi_{A}F(\chi_{B})\big)
=\sum_{x'\in A\subseteq X}\sum_{y\in B\subseteq Y}\sum_{x\in X}f_{yx}\w\left(e_{x'}e_{x}\right)
=\sum_{x\in A\subseteq X}\sum_{y\in B\subseteq Y}f_{yx}\w(e_{x})
=\sum_{x\in A\subseteq X}f_{x}(B)p_{x}
\ee
The equality of these two expressions is precisely Bayes' Theorem mentioned earlier in Remark~\ref{rmk:whyBayes}. 
\ex

We will provide more examples of when Bayesian inverses exist, but first we 
provide some lemmas that will be useful later. 

\blem
\label{lem:bayesianinversesgeneralexamples}
If $F:\mB\stoch\mA$ is a CPU map between $C^*$-algebras, and if $(F,\w)$ admits a Bayesian inverse $G$, then given any unitaries $U\in\mA$ and $V\in\mB$, the CPU map $\mathrm{Ad}_{V^{\dag}}\circ G\circ\mathrm{Ad}_{U^{\dag}}$ is a Bayesian inverse of $(\mathrm{Ad}_{U}\circ F\circ\mathrm{Ad}_{V},\w\circ\mathrm{Ad}_{U^{\dag}})$. Conversely, if $\mathrm{Ad}_{V^{\dag}}\circ G\circ\mathrm{Ad}_{U^{\dag}}$ is a Bayesian inverse of $(\mathrm{Ad}_{U}\circ F\circ\mathrm{Ad}_{V},\w\circ\mathrm{Ad}_{U^{\dag}})$, then $G$ is a Bayesian inverse of $(F,\w$). 
\elem

\br
In the notation of Lemma~\ref{lem:bayesianinversesgeneralexamples}, if we set $\xi:=\w\circ F$, we can depict these data diagrammatically as 
\be
\xy0;/r.25pc/:
(-10,6)*+{\mB}="B1";
(10,6)*+{\mA}="A1";
(-30,6)*+{\mB}="B2";
(30,6)*+{\mA}="A2";
(0,-6)*+{\C}="C";
{\ar@{~>}@/_0.7pc/"B1";"A1"^{F}};
{\ar@/_0.7pc/"B2";"B1"^{\mathrm{Ad}_{V}}};
{\ar@/_0.7pc/"A1";"A2"^{\mathrm{Ad}_{U}}};
{\ar@{~>}"B1";"C"_{\xi}};
{\ar@{~>}"A1";"C"^{\w}};
{\ar@{~>}@/^1.0pc/"A2";"C"^{\w\circ\mathrm{Ad}_{U^{\dag}}}};
{\ar@{~>}@/_1.0pc/"B2";"C"_{\xi\circ\mathrm{Ad}_{V}}};
{\ar@/_0.7pc/"A2";"A1"_{\mathrm{Ad}_{U^{\dag}}}};
{\ar@{~>}@/_0.7pc/"A1";"B1"_{G}};
{\ar@/_0.7pc/"B1";"B2"_{\mathrm{Ad}_{V^{\dag}}}};
\endxy
\ee
to help the reader visualize all the maps involved. In particular, when $\mB=\mathcal{M}_{n}(\C)$ and $\mA=\mathcal{M}_{m}(\C)$, since the states $\w=:\tr(\rho\;\cdot\;)$ and $\xi=:\tr(\s\;\cdot\;)$ can be described by density matrices, this means that we can find unitary matrices $U$ and $V$ that diagonalize these density matrices and find the Bayesian inverse of the associated maps after diagonalization. 
When $\mA$ and $\mB$ are direct sums of matrix algebras, the unitaries act componentwise on the factors. 
\er

\blem
\label{lem:comultiplication}
Set $\mA:=\mathcal{M}_{m}(\C)$. Then 
\be
\mu_{\mA}^*(E_{ik})=\sum_{j=1}^{m}E_{ij}\otimes E_{jk}, 
\ee
where the $E_{ij}$ are the matrix units and $\mu_{\mA}^*$ denotes the dual of the multiplication map $\mu_{A}:\mA\otimes\mA\stoch\mA$ with respect to the Hilbert--Schmidt inner product. 
\elem

\bprf
We set $\a_{ghnopq}$ to be the coefficients appearing in 
\be
\mu_{\mA}^*(E_{gh})=:\sum_{n,o,p,q}\a_{ghnopq}E_{no}\otimes E_{pq}.
\ee
We also let $\<\;\cdot\;,\;\cdot\;\>$ denote the Hilbert--Schmidt inner product
(both on the tensor product and on the original algebra). 
Then we have
\be
\xy0;/r.25pc/:
(-25,30)*+{\<\mu_{\mA}(E_{ij}\otimes E_{kl}),E_{gh}\>}="1";
(-45,15)*+{\de_{jk}\<E_{il},E_{gh}\>}="2";
(-50,-5)*+{\de_{jk}\tr(E_{il}^{\dag}E_{gh})}="3";
(-35,-20)*+{\de_{jk}\de_{ig}\de_{lh}}="4";
(0,-30)*+{\a_{ghijkl}}="5";
(35,-20)*+{\sum\limits_{n,o,p,q}\a_{ghnopq}\de_{in}\de_{jo}\de_{kp}\de_{lq}}="6";
(50,-5)*+{\sum\limits_{n,o,p,q}\a_{ghnopq}\tr(E_{ij}^{\dag}E_{no})\tr(E_{kl}^{\dag}E_{pq})}="7";
(45,15)*+{\sum\limits_{n,o,p,q}\<E_{ij}\otimes E_{kl},\a_{ghnopq}E_{no}\otimes E_{pq}\>}="8";
(25,30)*+{\<E_{ij}\otimes E_{kl},\mu_{\mA}^*(E_{gh})\>}="9";
{\ar@{=}@/_1.0pc/"9";"1"};
{\ar@{=}@/_0.7pc/"1";"2"};
{\ar@{=}@/_0.5pc/"2";"3"};
{\ar@{=}@/_1.0pc/"3";"4"};
{\ar@{=}@/_1.0pc/"5";"6"};
{\ar@{=}@/_1.0pc/"6";"7"};
{\ar@{=}@/_0.5pc/"7";"8"};
{\ar@{=}@/_0.7pc/"8";"9"};
\endxy
\ee
Therefore, 
\be
\mu_{\mA}^*(E_{gh})=\sum_{i,j,k,l}\a_{ghijkl}E_{ij}\otimes E_{kl}=\sum_{i,j,k,l}\de_{jk}\de_{ig}\de_{lh}E_{ij}\otimes E_{kl}
=\sum_{j}E_{gj}\otimes E_{jh}
\ee
as needed. 
\eprf

\br
The map $\mu_{\mA}^*$ is the co-multiplication map of the Frobenius structure on $\mA$ as described in Section~3.5 of Coecke and Spekkens~\cite{CoSp12}. 
Since $\mu_{\mA}$ is in general not completely positive, neither is $\mu_{\mA}^*$ (recall, a map on finite-dimensional $C^*$-algebras is CP if and only if its dual with respect to the Hilbert--Schmidt inner product is CP). 
\er

\blem
\label{lem:nullspaces}
Let $F:\mB\stoch\mA$ be a linear map, let $\w:\mA\stoch\C$ be a state, and set $\xi:=\w\circ F$. If $F$ is positive unital, then
\be
F\big(P_{\xi}^{\perp}\mathcal{B}P_{\xi}^{\perp}\big)\subseteq P_{\w}^{\perp}\mathcal{A}P_{\w}^{\perp}.
\ee
If $F$ is 2-positive, then 
\be
\label{eq:nullspacetonullspace}
F(\mathcal{N}_{\xi})\subseteq\mathcal{N}_{\w},
\ee
\elem

\bprf
Lemma~3.121 in~\cite{PaRu19} is the first statement and Proposition~3.106 from~\cite{PaRu19} is almost the second statement. Technically, Proposition~3.106 from~\cite{PaRu19} assumed $F$ was 2-positive unital, but 2-positivity is enough because in this case, the Kadison--Cauchy--Schwarz inequality reads $F(B)^*F(B)\le\lVert F(1_{\mB})\rVert F(B^*B)$ for all $B\in\mB$.
\eprf

\br
Condition~(\ref{eq:nullspacetonullspace}) is not generally true if $F$ is only positive unital. For example, take $F$ to be the transpose map on $2\times2$ matrices and $\rho=\left[\begin{smallmatrix}1&0\\0&0\end{smallmatrix}\right]$ with $\omega=\tr(\rho\;\cdot\;)=\xi$. 
\er

%%%%%%%%%%%%%%%%%%%%%%%%%%%%%%%%%%%%%
\section{Examples of quantum Bayesian inference} 
\label{sec:examples}
%%%%%%%%%%%%%%%%%%%%%%%%%%%%%%%%%%%%%

Before proving general existence and uniqueness theorems for quantum Bayesian inference, we examine several special cases and their physical relevance. These examples will also supply additional intuition for the more general results that will come later.

%%%%%%%%%%%%%%%%%%%%%%%%%%%%%%%%%%%%%
\subsection{The bit flip channel} 
%%%%%%%%%%%%%%%%%%%%%%%%%%%%%%%%%%%%%

%The present example shows that CPU maps need not have (CPU) Bayesian inverses. 
Set 
$
\mA:=\mathcal{M}_{2}(\C), %\quad
\mB:=\mathcal{M}_{2}(\C), %\quad
\rho=\left[\begin{smallmatrix}p_{1}&0\\0&p_{2}\end{smallmatrix}\right],$
%\quad\text{ and }\quad
and 
$F:=\l_{1}\mathrm{id}_{\mathcal{M}_{2}(\C)}+\l_{2}\mathrm{Ad}_{\left[\begin{smallmatrix}0&1\\1&0\end{smallmatrix}\right]}
,
$
where $0\le p_{1},p_{2}\le1,$ $0<\l_{1},\l_{2}<1$, and $p_{1}+p_{2}=\l_{1}+\l_{2}=1$ (the reason for choosing $\l_{1},\l_{2}\in(0,1)$ is because if one of them is $0$ then $F$ is $\mathrm{Ad}_{U}$ with $U$ unitary, which is invertible, and hence has a Bayesian inverse, namely $\mathrm{Ad}_{U^\dag}$). 
The map $F$ here is called the \define{bit flip channel}~\cite[Section~8.3.3]{NiCh11}. 
Then
\be
\s:=F^*(\rho)=\begin{bmatrix}\l_{1}p_{1}+\l_{2}p_{2}&0\\0&\l_{1}p_{2}+\l_{2}p_{1}\end{bmatrix}=:\begin{bmatrix}q_{1}&0\\0&q_{2}\end{bmatrix}. 
\ee
Hence, $F$ pulls back the state $\w:=\tr(\rho\;\cdot\;)$ to $\xi:=\tr(\sigma\;\cdot\;)$. In what follows, we will discover when a Bayesian inverse of $(F,\omega)$ exists. 
First note that $F^*=F$ and 
\be
F(E_{ij})=\begin{cases}
\l_{1}E_{ij}+\l_{2}E_{ji}&\mbox{ if $i\ne j$}\\
\l_{1}E_{11}+\l_{2}E_{22}&\mbox{ if $i=j=1$}\\
\l_{1}E_{22}+\l_{2}E_{11}&\mbox{ if $i=j=2$}\\
\end{cases}
.
\ee
The unital linear functional $\z:=\w\circ\mu_{\mA}\circ(\id_{\mA}\otimes F)$, the right side of the rectangle in~(\ref{eq:quantumBayesdiagram}), is uniquely determined by the matrix 
\be
\begin{split}
\t&:=\z^*(1)=(\id_{\mA}\otimes F^*)\Big(\mu_{\mA}^*\big(\underbrace{\w^*(1)}_{\rho}\big)\Big)
=(\id_{\mA}\otimes F^*)\big(\mu_{\mA}^*(p_{1}E_{11}+p_{2}E_{22})\big)\\
&\overset{\text{Lem~\ref{lem:comultiplication}}}{=\joinrel=\joinrel=\joinrel=\joinrel=}(\id_{\mA}\otimes F^*)(p_{1}E_{11}\otimes E_{11}+p_{1}E_{12}\otimes E_{21}+p_{2}E_{21}\otimes E_{12}+p_{2}E_{22}\otimes E_{22})\\
&=p_{1}E_{11}\otimes%(\l_{1}E_{11}+\l_{2}E_{22})
\begin{bmatrix}\l_{1}&0\\0&\l_{2}\end{bmatrix}
+p_{1}E_{12}\otimes%(\l_{1}E_{21}+\l_{2}E_{12})
\begin{bmatrix}0&\l_{2}\\\l_{1}&0\end{bmatrix}
+p_{2}E_{21}\otimes\begin{bmatrix}0&\l_{1}\\\l_{2}&0\end{bmatrix}
+p_{2}E_{22}\otimes\begin{bmatrix}\l_{2}&0\\0&\l_{1}\end{bmatrix},
\end{split}
\ee
which simplifies to%
%footnote
\footnote{Note that $\t$ is not a density matrix unless $p_{1}=p_{2}$. %I have checked this
Nevertheless, we do not in general expect $\t$ to be a density matrix (see Remark~\ref{rmk:whyBayesquantum} for further discussion). Hence, we do not impose that $\t$ need to be a density matrix in what follows. 
}
%end footnote
\be
\label{eq:jointdensitymatrixbitflip}
\t=\begin{bmatrix}
p_{1}\l_{1}&0&0&p_{1}\l_{2}\\
0&p_{1}\l_{2}&p_{1}\l_{1}&0\\
0&p_{2}\l_{1}&p_{2}\l_{2}&0\\
p_{2}\l_{2}&0&0&p_{2}\l_{1}\\
\end{bmatrix}
.
\ee
A Bayesian inverse $G$ of $F$ must satisfy
\be
\t=(G^*\otimes\id_{\mB})\big(\mu_{\mB}^*(\s)\big)
,
\ee
because this reproduces the left part of the rectangle in~(\ref{eq:quantumBayesdiagram}) upon taking the dual. 
The general form of a CP map $G$ is 
\be
G=\sum_{j}\mathrm{Ad}_{G_{j}},\qquad\text{ where }\qquad
G_{j}=\begin{bmatrix}g_{j;11}&g_{j;12}\\g_{j;21}&g_{j;22}\end{bmatrix}
.
\ee
Therefore, 
\be
\begin{split}
(G^*\otimes\id_{\mB})\big(\mu_{\mB}^*(\s)\big)
&=(G^*\otimes\id_{\mB})\big(q_{1}E_{11}\otimes E_{11}+q_{1}E_{12}\otimes E_{21}+q_{2}E_{21}\otimes E_{12}+q_{2}E_{22}\otimes E_{22}\big)\\
&=\sum_{j}
\begin{bmatrix}[1.3]
q_{1}|g_{j;11}|^2&q_{2}\ov{g_{j;21}}g_{j;11}&q_{1}\ov{g_{j;11}}g_{j;12}&q_{2}\ov{g_{j;21}}g_{j;12}\\
q_{1}\ov{g_{j;11}}g_{j;21}&q_{2}|g_{j;21}|^2&q_{1}\ov{g_{j;11}}g_{j;22}&q_{2}\ov{g_{j;21}}g_{j;22}\\
q_{1}\ov{g_{j;12}}g_{j;11}&q_{2}\ov{g_{j;22}}g_{j;11}&q_{1}|g_{j;12}|^2&q_{2}\ov{g_{j;22}}g_{j;12}\\
q_{1}\ov{g_{j;12}}g_{j;21}&q_{2}\ov{g_{j;22}}g_{j;21}&q_{1}\ov{g_{j;12}}g_{j;22}&q_{2}|g_{j;22}|^2\\
\end{bmatrix}
.
\end{split}
\ee
Equating this with $\t$ in~(\ref{eq:jointdensitymatrixbitflip}) gives several relations between the $p$'s and $\l$'s. 
In particular, the non-vanishing off-diagonal terms give 
\be
\label{eq:twobytwoconditionsa}
\begin{split}
p_{1}\l_{2}&=q_{2}\sum_{j}\ov{g_{j;21}}g_{j;12},\quad
p_{1}\l_{1}=q_{1}\sum_{j}\ov{g_{j;11}}g_{j;22},\quad\\
p_{2}\l_{1}&=q_{2}\sum_{j}\ov{g_{j;22}}g_{j;11},\quad
p_{2}\l_{2}=q_{1}\sum_{j}\ov{g_{j;12}}g_{j;21}.
\end{split}
\ee
Multiplying the left equations by $q_{1}$ and the right equations by $q_{2}$ gives
\be
p_{1}\l_{2}q_{1}=q_{2}q_{1}\sum_{j}\ov{g_{j;21}}g_{j;12}=q_{2}q_{1}\sum_{j}\ov{g_{j;12}}g_{j;21}=q_{2}p_{2}\l_{2}.
\ee
since the equations (\ref{eq:twobytwoconditionsa}) are invariant under complex conjugation. Since $\l_{2}\ne0,$ this implies $p_{1}q_{1}=p_{2}q_{2}$. Expanding this out using the definition of the $q$'s gives $\l_{1}p_{1}^2+\l_{2}p_{1}p_{2}=\l_{1}p_{2}^2+\l_{2}p_{1}p_{2}$, which entails $p_{1}=p_{2}=\frac{1}{2}$. Therefore, a necessary condition for a Bayesian inverse to exist is that $p_{1}=p_{2}=\frac{1}{2}$, which entails $q_{1}=q_{2}=\frac{1}{2}$. In this case, $G=F$ is such a Bayesian inverse as can be checked.

\br
It is a curious fact that Bayesian inference does not exist in the simple situation covered in this example of the bit flip map. This is particularly surprising since a classical analogue%
%footnote
\footnote{This is not an analogue in the rigorous sense of category theory but only in the sense of what some may consider a classical analogue based on the fact that we have used diagonal density matrices. Such analogies have been shown to be misleading in other cases as well and have led to the discovery of concepts such as the quantum discord~\cite{HaZu01}.}
%end footnote
 of this situation may be considered to be the following. If $\mathcal{M}_{2}(\C)$ is replaced with $\C^{2},$ the states $\w$ and $\xi$ are replaced by the (dual) probability vectors $\begin{bmatrix}p_{1}&p_{2}\end{bmatrix}$ and $\begin{bmatrix}q_{1}&q_{2}\end{bmatrix}$, and $F$ is the (doubly) stochastic matrix given by $F=\left[\begin{smallmatrix}\l_{1}&\l_{2}\\\l_{2}&\l_{1}\end{smallmatrix}\right]$, then the relationships $q_{1}=\l_{1}p_{1}+\l_{2}p_{2}$ and $q_{2}=\l_{2}p_{1}+\l_{1}p_{2}$ still hold. Furthermore, a Bayesian inference exists regardless of the values of $p_{1}$ and $p_{2}$. Namely, 
\be
G=\begin{bmatrix}[1.3]
p_{1}\l_{1}/q_{1}&p_{2}\l_{2}/q_{1}\\
p_{1}\l_{2}/q_{2}&p_{2}\l_{1}/q_{2}
\end{bmatrix}
.
\ee
Note that $q_{1},q_{2}\in(0,1)$ because $\l_{1},\l_{2}\in(0,1)$. 
\er

%%%%%%%%%%%%%%%%%%%%%%%%%%%%%%%%%%%%%
\subsection{Positive operator-valued measures} 
%%%%%%%%%%%%%%%%%%%%%%%%%%%%%%%%%%%%%

\bd
A \define{positive operator-valued measure} (POVM) on the Hilbert space $\C^{m}$ is a collection of positive operators $\{ F_{y}\}_{y\in Y}$ in $\mathcal{M}_{m}(\C)$, where $Y$ is a finite set and $\sum\limits_{y\in Y}F_{y}=\mathds{1}_{m}$. Equivalently, these data are described by a (necessarily completely) positive unital map $F:\C^{Y}\stoch\mathcal{M}_{m}(\C)$, since one can identify $F_{y}$ with $F(e_{y}),$ where $e_{y}$ is the function on $Y$ whose value at $y$ is $1$ and is zero everywhere else.
If $\mA$ is a finite-dimensional $C^*$-algebra, an \define{operator system} in $\mA$ is a (complex) vector subspace $\mathcal{O}\subseteq\mA$ that contains $1_{\mA}$ and is closed under the involution of $\mA$. 
If $\mA\xstoch{\vf}\mA$ is a linear map, the \define{fixed point set} of $\vf$ is the vector subspace of $\mA$ given by 
\be
\mathrm{Fix}(\vf):=\big\{A\in\mA\;:\;\vf(A)=A\big\}. 
\ee
If $\mA=\mathcal{M}_{m}(\C)$ is a matrix algebra and $\vf=\sum_{\a}\mathrm{Ad}_{V_{\a}}$ is a Kraus decomposition of $\vf$, the \define{interaction algebra} associated with the Kraus operators $\{V_{\a}\}$ is given by
$\mathrm{Alg}(\{V_{\a}\}),$
the algebra generated by the Kraus operators, i.e.\ all polynomials in the $\{V_{\a}\}$. 
\ed

We recall some important facts due to Kribs~\cite{Kr03} (see also~\cite[Remark~2.1 and Theorem~3.4]{HKL03}). 

\bt
\label{thm:Kribs}
Let $\mathcal{M}_{m}(\C)\xstoch{\vf}\mathcal{M}_{m}(\C)$ be a completely positive unital and trace-preserving map with a Kraus decomposition $\vf=\sum_{\a}\mathrm{Ad}_{V_{\a}}$. Then 
\begin{enumerate}[i.]
\itemsep0pt
\item
$\mathrm{Alg}(\{V_{\a}\})$ is a (unital) $C^*$-subalgebra of $\mathcal{M}_{m}(\C)$ and depends only on the map $\vf$ and not the specific choice of Kraus operators (as such, it will henceforth be denoted by $\mathrm{Alg}(\vf)$).
\item
The fixed point set of $\vf$ is a (unital) $C^*$-subalgebra of $\mathcal{M}_{m}(\C)$ and equals the commutant of the interaction algebra, i.e.\
\be
\mathrm{Fix}(\vf)=\mathrm{Alg}(\vf)'\subseteq\mathcal{M}_{m}(\C). 
\ee 
\end{enumerate}
\et

The equivalent formulation of a POVM as a CPU map allows us to ask when a POVM has a Bayesian inverse. For this, we have a thorough and detailed result given by Proposition~\ref{thm:BayesianinversePOVM} below. 

\blem
\label{lem:matrixunitsandtrace}
Let $A_{ij}$ denote the $ij$-th entry of $A\in\mM_{m}(\C)$. Then $A_{ij}=\tr(AE_{ji})$. 
Furthermore, if $F:\mathcal{M}_{n}(\C)\stoch\mathcal{M}_{m}(\C)$ is a linear map, then $F$ is $*$-preserving if and only if $F^*$, the Hilbert--Schmit dual of $F$, is $*$-preserving. 
\elem

\bprf
Using Dirac bra-ket notation gives
\be
\tr(AE_{ji})=\sum_{k}\<k|AE_{ji}|k\>
=\sum_{k}\<k|A|j\>\<i|k\>
=\sum_{k}\<i|k\>\<k|A|j\>
=\<i|A|j\>
=A_{ij},
\ee
where the completeness relation $\sum_{k=1}^{m}|k\>\<k|=\mathds{1}_{m}$ has been used in the fourth step.
The second claim follows from the first. Namely, if $F$ is $*$-preserving, then
\be
\begin{split}
\big(F^*(A^\dag)\big)_{ij}
&=\tr\big(F^*(A^\dag)E_{ji}\big)
=\tr\big(A^\dag F(E_{ji})\big)
=\overline{\tr\big(F(E_{ji})^\dag A\big)}\\
&=\overline{\tr\big(F(E_{ij})A\big)}
=\overline{\tr\big(E_{ij}F^*(A)\big)}
=\tr\big(F^*(A)^{\dag}E_{ji}\big)
=\big(F^*(A)^{\dag}\big)_{ij}
\end{split}
\ee
by cyclicity of trace, the $*$-preserving property of trace, and the $*$-preserving property of $F$. A similar calculation proves the converse. 
\eprf

\bn
\label{thm:BayesianinversePOVM}
Let $\mathcal{M}_{m}(\C)\xstoch{\w=\tr(\rho\;\cdot\:)}\C$ be a state on $\mathcal{M}_{m}(\C)$, let $Y$ be a finite set, let $\C^{Y}\xstoch{\xi}\C$ be a state on $\C^{Y},$ and let $\C^{Y}\xstoch{F}\mathcal{M}_{m}(\C)$ be a state-preserving POVM. Set $F_{y}:= F(e_{y})$ and $q_{y}:=\tr(\rho F_{y})$ for each $y\in Y$. Then the following are equivalent. 
\begin{enumerate}[i.]
\itemsep0pt
\item 
\label{item:POVM1}
A Bayesian inverse of $(F,\w)$ exists.
\item
\label{item:POVM2}
The commutation rule
\be
\label{eq:BayesianinversePOVMcondition}
[\rho,F_{y}]=0\quad\forall\;y\in Y\setminus N_{q}
%\qquad\text{and}\qquad
%F_{y}\rho=0\quad\forall\;y\in N_{q}. 
\ee
holds.
\item
\label{item:POVM3}
The density matrix $\rho$ is in the commutant of the algebra generated by the operator system 
\be
\mathrm{span}\{F_{y}\;:\;y\in Y\},\qquad\text{ i.e.\ }\qquad 
\rho\in F\left(\C^{Y}\right)'\subseteq\mathcal{M}_{m}(\C).
\ee 
\item
\label{item:POVM4}
The density matrix $\rho$ satisfies 
\be
\sum_{y\in Y}\sqrt{F_{y}}\rho\sqrt{F_{y}}=\rho,
\ee
i.e.\ $\rho$ is in the fixed point set for the CP map $\sum_{y\in Y}\mathrm{Ad}_{\sqrt{F_{y}}}$. 
\end{enumerate}
When one (and hence all) of these conditions is satisfied, a Bayesian inverse is $\xi$-a.e.\ uniquely determined and a representative is given by the formula
\be
\label{eq:BayesianinversePOVMformula}
\begin{split}
\mathcal{M}_{m}(\C)&\xstoch{G}\C^{Y}\\
A&\mapsto G(A):=\sum_{y\in Y\setminus N_{q}}\frac{\tr(\rho A F_{y})}{q_{y}}e_{y}+\sum_{y\in N_{q}}\frac{1}{m}\tr(A)e_{y}. 
\end{split}
\ee
\en

\bprf
Set $\mA:=\mathcal{M}_{m}(\C)$ and $\mB:=\C^{Y}$. 
Note that $\xi(e_{y})=q_{y}$ for all $y\in Y$. 
Hence, for $y\in N_{q}$, which means $e_{y}\in P_{\xi}^{\perp}\mathcal{B}P_{\xi}^{\perp}$, Lemma~\ref{lem:nullspaces}
implies $F_{y}\in P_{\w}^{\perp}\mathcal{A}P_{\w}^{\perp}$. Hence, $F_{y}\rho=0$ for $y\in N_{q}$ and therefore $\tr(\rho AF_{y})=0= q_{y}G(A)_{y}$ so that the Bayes condition holds automatically for all $y\in N_{q}$ for any linear map $\mA\xstoch{G}\mB$. Here, $G(A)_{y}:=(\mathrm{ev}_{y}\circ G)(A)$ denotes the $y$-th component of $G(A)$.

\noindent
(\ref{item:POVM1}$\Rightarrow$\ref{item:POVM2})
Suppose a Bayesian inverse $G$ exists. %Upon setting $B:=F(e_{y})\equiv F_{y}$,  
The Bayes condition entails $\tr(\rho A F_{y})=\xi\big(G(A)e_{y}\big)=q_{y} G(A)_{y}$ for all $y\in Y$ and for all $A\in\mA$. 
Thus, 
\be
G(A)_{y}=\frac{\tr(\rho A F_{y})}{q_{y}}\qquad\forall\;y\in Y\setminus N_{q},\;\;\forall\;A\in\mA. 
\ee
Since $G$ is CP, its $y$ component $\mathrm{ev}_{y}\circ G$ is also CP. In particular, $\mathrm{ev}_{y}\circ G$ is $*$-preserving, which means $\overline{G(A^{\dagger})_{y}}=G(A)_{y}$ for all $A\in\mA$. This implies 
\be
\tr(\rho A F_{y})=\tr(F_{y} A \rho)
\qquad\forall\;y\in Y\setminus N_{q},\;\;\forall\;A\in\mA
\ee
because the trace is $*$-preserving and because $F_{y}$ and $\rho$ are positive (and hence self-adjoint). By cyclicity of the trace, this is equivalent to 
\be
\tr\big([F_{y},\rho] A\big)=0\qquad\forall\;y\in Y\setminus N_{q},\;\;\forall\;A\in\mA. 
\ee
By plugging in $A=E_{ij}$ for all the values of $i,j\in\{1,\dots,m\}$, this shows all components of the matrix $[F_{y},\rho]$ must vanish, i.e.\
\be
[F_{y},\rho]=0\qquad\forall\;y\in Y\setminus N_{q}.
\ee

\noindent
(\ref{item:POVM1}$\Leftarrow$\ref{item:POVM2})
Suppose that condition~(\ref{eq:BayesianinversePOVMcondition}) holds. In what follows, we will prove that $G$ as defined in~(\ref{eq:BayesianinversePOVMformula}) is a Bayesian inverse of $(F,\w)$. 
The Bayes condition is satisfied by~(\ref{eq:BayesianinversePOVMformula}) because $q_{y}G(A)_{y}=\tr(\rho A F_{y})$ for all $y\in Y\setminus N_{q}$, and the Bayes condition holds automatically on $y\in N_{q}$ by the first paragraph in the proof. 
Furthermore, $G$ is easily seen to be unital. What is left to prove is that $G$ is in fact CP. Since the trace is CP, it suffices to focus on the formula when $y\in Y\setminus N_{q}$. Note that since both $\rho$ and $F_{y}$ are positive and commute by assumption~\ref{item:POVM2}, their product $\rho F_{y}=F_{y}\rho$ is positive. Hence, 
\be
\tr(\rho_{y} AF_{y})=\tr\left(\mathrm{Ad}_{\sqrt{\rho F_{y}}}(A)\right)\qquad\forall\;y\in Y\setminus N_{q},\;\;\forall\;A\in\mA.
\ee
This shows $G$ is CP. 

\noindent
(\ref{item:POVM2}$\Leftrightarrow$\ref{item:POVM3})
This follows immediately from the functional calculus, the fact that the $F_{y}$ are positive operators, and since $[F_{y},\rho]=0$ holds automatically for all $y\in N_{q}$. 

\noindent
(\ref{item:POVM3}$\Rightarrow$\ref{item:POVM4})
Let $\vf:=\sum_{y\in Y}\mathrm{Ad}_{\sqrt{F_{y}}}$. Then $\vf$ is completely positive unital and trace-preserving. Indeed, unitality follows from the fact that $F$ is a POVM. Furthermore, trace-preservation of $\vf$ follows from the fact that $\vf^*=\vf$ and $\vf$ is unital if and only if $\vf^*$ is trace-preserving. By Kribs' theorem (Theorem~\ref{thm:Kribs}), $\mathrm{Fix}(\vf)=\mathrm{Alg}(\vf)'\subseteq\mA$. 
Now, assume $\rho$ satisfies condition~\ref{item:POVM3}. Then, by the functional calculus, $\rho\in\mathrm{Alg}(\vf)'\subseteq\mA$. Hence, by Kribs' theorem, $\rho\in\mathrm{Fix}(\vf)$. In fact, combining earlier observations, we have shown
\be
\sum_{y\in Y}\sqrt{F_{y}}\rho\sqrt{F_{y}}=\rho=\sum_{y\in Y\setminus N_{q}}\sqrt{F_{y}}\rho\sqrt{F_{y}}.
\ee

\noindent
(\ref{item:POVM4}$\Rightarrow$\ref{item:POVM3})
Assume $\rho\in\mathrm{Fix}(\vf)$ so that $\rho$ satisfies~\ref{item:POVM4}. Then by Kribs' theorem again, $\rho\in\mathrm{Alg}(\vf)'\subseteq\mA$. Since the commutant $\mathrm{Alg}(\vf)'\subseteq\mA$ equals the commutant $F\big(\C^{Y}\big)'\subseteq\mA$ (by the functional calculus and properties of commutants), $\rho$ satisfies~\ref{item:POVM3}. 
\eprf

\br
Proposition~\ref{thm:BayesianinversePOVM} generalizes our result for projection-valued measures (PVMs)~\cite[Theorem~6.28]{PaRu19}. Indeed, if $F$ defines a PVM (i.e.\ a unital $^*$-homomorphism, where each $F_{y}$ is an orthogonal projection), then $[F_{y},\rho]=0$ for all $y\in Y$ holds if and only if $\rho=\sum_{y\in Y}F_{y}\rho F_{y},$ i.e.\ $\rho$ equals its L\"uders projection with respect to the PVM. 
\er

\br
Unitality, self-adjointness, and the Bayes condition were sufficient to imply that $G$ (when evaluated on $Y\setminus N_{q}$) is CP in the proof of Proposition~\ref{thm:BayesianinversePOVM}. 
This remark will be relevant later when we discuss our more general quantum Bayes' theorem. 
\er

%%%%%%%%%%%%%%%%%%%%%%%%%%%%%%%%%%%%%
\subsection{Ensemble of states} 
%%%%%%%%%%%%%%%%%%%%%%%%%%%%%%%%%%%%%

The following setup is in some sense dual to the previous example. A \define{collection/ensemble of states} (sometimes called an \define{ensemble preparation}) on a quantum system is a CPU map $\mathcal{M}_{n}(\C)\xstoch{F}\C^{X}$. 
%\AJP{The idea for doing this example came to me while reading Leifer--Spekkens (2013).} 
Given a state $\C^{X}\xstoch{\w}\C$ and a state $\mathcal{M}_{n}(\C)\xstoch{\xi=\tr(\s\;\cdot\;)}\C$ such that $\w\circ F=\xi$, one could ask when a Bayesian inverse of $(F,\omega)$ exists. This question could not be analyzed in the disintegration setting because there are no $^*$-homomorphisms from $\mathcal{M}_{n}(\C)$ into $\C^{X}$ when $n>1$. The reason $F$ is called a collection of states is because for each $x\in X$, the functional $\mathrm{ev}_{x}\circ F$ defines a state on $\mathcal{M}_{n}(\C)$. Let $\s_{x}$ be the density matrix associated to this state. Since $F$ is state-preserving, this means that $\s=\sum_{x\in X}p_{x}\s_{x}$, where $p_{x}:=\w(e_{x})$. This follows from 
\be
\tr(\s B)=\w\big(F(B)\big)=\w\left(\sum_{x\in X}F_{x}(B)e_{x}\right)=\sum_{x\in X}p_{x}\tr(\s_{x}B)
\qquad\forall\;B\in\mathcal{M}_{n}(\C).
\ee
Thus, we have a convex combination of states forming $\xi$. A Bayesian inverse in this situation is a CPU map $\C^{X}\xstoch{G}\mathcal{M}_{n}(\C)$, which itself is determined by the images of the basis vectors $\{e_{x}\}$ in $\C^{X}$.

\blem
\label{lem:inverseuptosupport}
Let $\sigma\in\mathcal{M}_{n}(\C)$ be a density matrix with associated support denoted by $P$. There exists a unique positive (semidefinite) matrix $\hat{\sigma}$ whose support equals $P$ and such that $\sigma\hat{\sigma}=P=\hat{\sigma}\sigma$. When $P=\mathds{1}_{n}$, then $\hat{\sigma}=\sigma^{-1}$. 
\elem

\bprf[Proof sketch]
This is a simple linear algebra exercise: diagonalize $\s$, invert the non-zero entries, and then undiagonalize. 
\eprf

\begin{notation}
The matrix $\hat{\s}$ in Lemma~\ref{lem:inverseuptosupport} is called the \define{pseudoinverse} of $\s$.
\end{notation}

\bn
Using the notation of the previous paragraphs, a Bayesian inverse $G$ for the pair $(\mathcal{M}_{n}(\C)\xstoch{F}\C^{X},\C^{X}\xstoch{\w}\C)$ exists if and only if 
\be
\label{eq:ensemblecommutes}
p_{x}[\s,\s_{x}]=0\qquad\forall\;x\in X.
\ee
When this commuting condition holds, 
\be
\label{eq:Bayesianinverseforensemble}
G(e_{x})=p_{x}\left(\hat{\sigma}\s_{x}+P_{\xi}^{\perp}\right)\qquad\forall\;x\in X.
\ee
determines a Bayesian inverse of $(F,\w)$. 
\en

\bprf
In what follows, we set $G_{x}:=G(e_{x})$. 

\noindent
($\Rightarrow$)
Assume a Bayesian inverse $G$ exists. 
Since $G$ is positive and unital, this means $G_{x}\ge0$ for all $x\in X$ and $\sum_{x\in X}G_{x}=\mathds{1}_{n},$ respectively. 
Plugging in $B:=E_{ji}\in\mathcal{M}_{n}(\C)$ and $A:=e_{x}\in\C^{X}$ into the Bayes condition~(\ref{eq:quantumBayesequation}) gives 
\be
(\s G_{x})_{ij}=\tr(\s G_{x}E_{ji})
\overset{\text{(\ref{eq:quantumBayesequation})}}{=\joinrel=\joinrel=}\w\big(e_{x}F(E_{ji})\big)
=p_{x}\mathrm{ev}_{x}\big(F(E_{ji})\big)
=p_{x}\tr(\s_{x}E_{ji})
=p_{x}(\s_{x})_{ij}
,
\ee
i.e. 
\be
\label{eq:sigmaGxpx}
\s G_{x}=p_{x}\s_{x}\qquad\forall\;x\in X.
\ee
Since the right-hand-side of this is self-adjoint, the left-hand-side must be as well. This entails the condition 
\be
\label{eq:Gxsigma}
[G_{x},\sigma]=0\qquad\forall\;x\in X.
\ee 
Thus, 
\be
p_{x}\s\s_{x}\overset{\text{(\ref{eq:sigmaGxpx})}}{=\joinrel=\joinrel=}
\s\s G_{x}
\overset{\text{(\ref{eq:Gxsigma})}}{=\joinrel=\joinrel=}\s G_{x}\s
\overset{\text{(\ref{eq:sigmaGxpx})}}{=\joinrel=\joinrel=}p_{x}\s_{x}\s,
\ee
which proves~(\ref{eq:ensemblecommutes}).

\noindent
($\Leftarrow$)
Conversely, suppose this commuting condition holds. 
Then $p_{x}[P_{\xi},\s_{x}]=0$ by the functional calculus for $\s$. Hence, 
\be
p_{x}\s_{x}\hat{\s}%=p_{x}\s_{x}P_{\xi}\hat{\s}
=p_{x}P_{\xi}\s_{x}\hat{\s}=\hat{\s}(p_{x}\s\s_{x})\hat{\s}=\hat{\s}(p_{x}\s_{x}\s)\hat{\s}=p_{x}\hat{\s}\s_{x}P_{\xi}%=p_{x}\hat{\s}P_{\xi}\s_{x}
=p_{x}\hat{\s}\s_{x}.
\ee
Thus, the matrix $G_{x}$ is positive because 
the product of two commuting positive operators is positive and since projections are positive.
The map $G$ is unital because 
\be
G(1_{X})=\sum_{x\in X}G_{x}
=\hat{\sigma}\sum_{x\in X}p_{x}\s_{x}+\sum_{x\in X}p_{x}P_{\xi}^{\perp}
=\hat{\s}\s+P_{\xi}^{\perp}
=P_{\xi}+P_{\xi}^{\perp}
=\mathds{1}_{n}. 
\ee
To see that $G$ satisfies the Bayes condition, 
first note that $P_{\mathrm{ev}_{x}\circ F}\le P_{\xi}$ because $[p_{x}\s_{x},\s]=0$ and $p_{x}\s_{x}\le\s$ imply $[P_{\mathrm{ev}_{x}\circ F},P_{\xi}]=0$. Hence, 
\be
\tr\big(\s G(e_{x})B)%=\tr(\s G_{x}B)
=p_{x}\tr(P_{\xi}\s_{x}B)
=p_{x}\tr(\s_{x}B)
=\w\big(e_{x}F(B)\big).
\ee
Since $x\in X$ and $B\in\mathcal{M}_{n}(\C)$ were arbitrary, this proves the claim. 
\eprf

%%%%%%%%%%%%%%%%%%%%%%%%%%%%%%%%%%%%%
\subsection{After wave collapse} 
%%%%%%%%%%%%%%%%%%%%%%%%%%%%%%%%%%%%%

Set $\mA:=\mathcal{M}_{m}(\C)$, let $H\in\mA$ be self-adjoint with spectrum $\s(H),$ and let $\{P_{\l}\}_{\l\in\s(H)}$ be the orthogonal projections onto the corresponding eigenspaces of $H$. Set $F:=\sum_{\l\in\s(H)}\mathrm{Ad}_{P_{\l}}$. Let $\rho$ be a density matrix on $\mA$ with corresponding state $\w:=\tr(\rho\;\cdot\;)$. Let $\s=\sum_{\l\in\s(H)}P_{\l}\rho P_{\l}$ be the density matrix associated to $\w\circ F$. Set $\xi:=\tr(\s\;\cdot\;)$. In this section, we will analyze the conditions for $(F,\w)$ to have a Bayesian inverse. This setup describes what happens to a system $\mA$ after the observable $H$ has been measured and the result of the measurement has been forgotten or is hidden (cf.\ Lecture~11 of~\cite{Werner17}). One can also think of $\s$ as describing the statistical ensemble of density matrices that the original density matrix $\rho$ ``collapses'' into upon measurement of the observable $H$. The corresponding non-negative numbers $\tr(\rho P_{\l})$ are the probabilities that the eigenvalue $\l$ is obtained after the measurement. In a single run of such an experiment, the density matrix $\rho$ reduces to $P_{\l}\rho P_{\l}$. 

\bn
\label{thm:wavecollapse}
Using the notation in the previous paragraph, $(F,\w)$ has a Bayesian inverse $G$ if and only if 
$\rho=\s$. When a Bayesian inverse exists, it is unique and is given by $G=F^*=F=\sum_{\l}\mathrm{Ad}_{P_{\l}}$.
\en

The proof of this proposition will be postponed until after our main theorem (Theorem~\ref{thm:Bayesianinversionmatrixalgebracase}), more precisely after Lemma~\ref{lem:pseudoinverseofChoi} and Definition~\ref{defn:orthogonalkraus}.

%%%%%%%%%%%%%%%%%%%%%%%%%%%%%%%%%%%%%
\section{A quantum Bayes' theorem for matrix algebras}
\label{sec:quantumBayes}
%%%%%%%%%%%%%%%%%%%%%%%%%%%%%%%%%%%%%

In this section, we will find conditions for the existence and uniqueness of Bayesian inference on matrix algebras. 

%%%%%%%%%%%%%%%%%%%%%%%%%%%%%%%%%%%%%
\subsection{Bayes maps and the supported corner}
\label{sec:Bayesmapssupc}
%%%%%%%%%%%%%%%%%%%%%%%%%%%%%%%%%%%%%

\bn
\label{prop:qBayesianinverseformula}
Let $\mB:=\mathcal{M}_{n}(\C)\xstoch{F}\mathcal{M}_{m}(\C)=:\mA$ be a CPU map, let $\mA\xstoch{\w=\tr(\rho\;\cdot\;)}\C$ be a state represented by a density matrix $\rho$, and set $\xi:=\w\circ F=\tr(\s\;\cdot\;)$, where $\s$ is a density matrix representing $\xi$. Let $P_{\xi}$ denote the support of $\xi$. Then
a Bayes map $\mA\xstoch{G}\mB$ must necessarily satisfy 
\be
\label{eq:qBayesianinversepartialsupport}
P_{\xi}G(A)=\hat{\sigma}F^*(\rho A)\qquad\forall\;A\in\mA, 
\ee
or equivalently
\be
\label{eq:qBayesianinversepartialsupportdual}
G^*(BP_{\xi})=F(B\hat{\sigma})\rho\qquad\forall\;B\in\mB.
\ee
Conversely, any linear map $G$ satisfying (\ref{eq:qBayesianinversepartialsupport}) is a Bayes map.  
In particular, if $\s$ is invertible, then 
\be
\label{eq:qBayesianinversefullsupport}
G(A)=\sigma^{-1}F^*(\rho A)\qquad\forall\;A\in\mA
\ee
is the unique linear (necessarily unital) Bayes map.
\en

\bprf
Suppose $G$ is a Bayes map for $(F,\omega)$. Plugging in $B:=E_{ji}$ into~(\ref{eq:quantumBayesequation}) gives 
\be
\big(\s G(A)\big)_{ij}
=\tr\big(\s G(A)E_{ji}\big)
\overset{\text{(\ref{eq:quantumBayesequation})}}{=\joinrel=\joinrel=}
\tr\big(\rho A F(E_{ji})\big)
\overset{\text{(\ref{eq:HSdualcondition})}}{=\joinrel=\joinrel=}
\tr\big(F^*(\rho A) E_{ji}\big)
=\big(F^*(\rho A)\big)_{ij}
\ee
for all $A\in\mA$. Since this equation is true for all $i$ and $j$, it holds as matrix equations, i.e.\ 
$\s G(A)=F^*(\rho A)$ for all $A\in\mathcal{M}_{m}(\C)$. 
Multiplying both sides by $\hat{\s}$ on the left gives~(\ref{eq:qBayesianinversepartialsupport}). 

Let us now prove that (\ref{eq:qBayesianinversepartialsupportdual}) is equivalent to (\ref{eq:qBayesianinversepartialsupport}). 
This argument is similar to earlier ones involving the trace (see the proof of Lemma~\ref{lem:comultiplication} for example). 
Assuming~(\ref{eq:qBayesianinversepartialsupport}), we have%
%footnote
\footnote{After~(\ref{eq:bayesanddual}), we will stop explicitly writing when~(\ref{eq:HSdualcondition}) and cyclicity of trace are used.}
%end footnote
\be
\label{eq:bayesanddual}
\xy0;/r.40pc/:
(10.84040286651,12.7938524157)*+{\big(\hat{\s}F^*(\rho A)\big)_{ij}}="1";%at 7pi/18
(21.3205080757,4)*+{\tr\big(\hat{\s}F^*(\rho A)E_{ji}\big)}="2";
(21.6961550602,-4.47296355334)*+{\tr\big(\rho AF(E_{ji}\hat{\s})\big)}="3";
(10.8557521937,-12.3208888624)*+{\tr\big(AF(E_{ij}\hat{\s})\rho\big)}="4";%at -5pi/18
(-10.8557521937,-12.3208888624)*+{\tr\big(AG^*(E_{ji}P_{\xi})\big)\;\;}="6";
(-21.6961550602,-4.47296355334)*+{\tr\big(G(A)E_{ji}P_{\xi}\big)}="7";
(-21.3205080757,4)*+{\tr\big(P_{\xi} G(A)E_{ji}\big)}="8";
(-10.84040286651,12.7938524157)*+{\big(P_{\xi} G(A)\big)_{ij}}="9";
{\ar@{=}@/^0.55pc/"9";"1"^{\text{(\ref{eq:qBayesianinversepartialsupport})}}};
{\ar@{=}@/^0.55pc/"1";"2"^(0.6){\text{Lem~\ref{lem:matrixunitsandtrace}}}};
{\ar@{=}@/^0.45pc/"2";"3"^{\text{(\ref{eq:HSdualcondition})}}_{\text{cyclicity of trace}}};
{\ar@{=}@/^0.55pc/"3";"4"_{\text{cyclicity of trace}\;}};
{\ar@{=}@/^0.55pc/"6";"7"^(0.70){\text{(\ref{eq:HSdualcondition})}}};
{\ar@{=}@/^0.45pc/"7";"8"_{\text{cyclicity of trace}}};
{\ar@{=}@/^0.55pc/"8";"9"^(0.4){\text{Lem~\ref{lem:matrixunitsandtrace}}}};
\endxy
\ee
Since this is true for all $A\in\mA$, we obtain $G^*(E_{ji}P_{\xi})=F(E_{ji}\hat{\s})\rho$. 
Since this is true for all $i,j$ this proves~(\ref{eq:qBayesianinversepartialsupportdual}), the bottom line in~(\ref{eq:bayesanddual}). Reading the calculation upwards proves the converse.  

Now suppose that $G$ is a linear map satisfying~(\ref{eq:qBayesianinversepartialsupport}). Then
\be
\label{eq:Bayesholdsfromotherassumptions}
\xy0;/r.33pc/:
(6.84040286651,18.7938524157)*+{\;\;\;\;\;\;\;\tr\big(\rho AF(B)\big)}="1";%at 7pi/18
(17.3205080757,10)*+{\tr\big(\rho AF(B)P_{\w}\big)}="2";%at 3pi/18
(19.6961550602,-1.47296355334)*+{\tr\big(\rho AF(BP_{\xi}+BP_{\xi}^{\perp})P_{\w}\big)}="3";%at -pi/18+higher
(12.8557521937,-12.3208888624)*+{\tr\big(\rho AF(BP_{\xi})P_{\w}\big)}="4";%at -5pi/18
(0,-19)*+{\tr\big(\rho AF(BP_{\xi})\big)}="5";%at -pi/2
(-12.8557521937,-12.3208888624)*+{\tr\big(F^*(\rho A)BP_{\xi}\big)\;\;}="6";
(-19.6961550602,-1.47296355334)*+{\tr\big(\s\hat{\s} F^*(\rho A)B\big)}="7";
(-17.3205080757,10)*+{\tr\big(\s P_{\xi}G(A)B\big)}="8";
(-6.84040286651,18.7938524157)*+{\tr\big(\s G(A)B\big)\;\;\;\;\;\;\;\;}="9";
{\ar@{=}@/^0.55pc/"1";"2"};
{\ar@{=}@/^0.55pc/"2";"3"^{\mathds{1}_{n}=P_{\xi}+P_{\xi}^{\perp}}};
{\ar@{=}@/^0.55pc/"3";"4"^(0.4){\text{(\ref{eq:nullspacetonullspace})}}};
{\ar@{=}@/^0.55pc/"4";"5"};
{\ar@{=}@/^0.55pc/"5";"6"};
{\ar@{=}@/^0.55pc/"6";"7"^(0.6){\text{Lemma~\ref{lem:inverseuptosupport}}}};
{\ar@{=}@/^0.55pc/"7";"8"^{\text{(\ref{eq:qBayesianinversepartialsupport})}}};
{\ar@{=}@/^0.55pc/"8";"9"};
\endxy
,
\ee
which is the Bayes condition. 

Finally, the formula~(\ref{eq:qBayesianinversefullsupport}) and uniqueness of $G$ when $\s$ is invertible follows from~(\ref{eq:qBayesianinversepartialsupport}) and Lemma~\ref{lem:faithfulstateimpliesunique}. 
\eprf 

Proposition~\ref{prop:qBayesianinverseformula} tells us that when $\s$ is invertible, a unique unital Bayes map exists, but we have made no such claim that this map is positive. Without requiring positivity, we already have a solution to the existence of  unital Bayes maps.

\bc
Let $\mA,\mB,F,\w,\rho,\xi,\s,$ and $P_{\xi}$ be as in Proposition~\ref{prop:qBayesianinverseformula}.
Then the unital linear map $\mA\xstoch{G}\mB$ uniquely determined by the projections
\be
P_{\xi}G(A):=\hat{\sigma}F^*(\rho A)
,\quad
P_{\xi}^{\perp}G(A)P_{\xi}:=P_{\xi}^{\perp}F^*(A\rho)\hat{\sigma}
,\quad\text{ and }\quad
P_{\xi}^{\perp}G(A)P_{\xi}^{\perp}:=\frac{1}{m}\tr(A)P_{\xi}^{\perp}
\ee
for all $A\in\mA$ 
is a unital Bayes map for $(F,\w)$. 
\ec

\bprf
First note that the projections specify $G$ everywhere since 
$G(A)=P_{\xi}G(A)+P_{\xi}^{\perp}G(A)P_{\xi}+P_{\xi}^{\perp}G(A)P_{\xi}^{\perp}$. 
The map $G$ satisfies condition~(\ref{eq:qBayesianinversepartialsupport}) and therefore satisfies the Bayes condition by Proposition~\ref{prop:qBayesianinverseformula}. The calculation 
\be
G(\mathds{1}_{m})=\hat{\sigma}F^*(\rho)+P_{\xi}^{\perp}F^*(\rho)\hat{\sigma}+\frac{1}{m}\tr(\mathds{1}_{m})P_{\xi}^{\perp}
=\hat{\sigma}\sigma+P_{\xi}^{\perp}\s\hat{\sigma}+P_{\xi}^{\perp}
=P_{\xi}+P_{\xi}^{\perp}=\mathds{1}_{n}
\ee
proves that $G$ is unital.
\eprf

Therefore, causal quantum Bayes maps exist, but they might not be implementable as quantum operations. In what follows, we will find conditions on $F$ and $\rho$ that guarantee when a (CPU) Bayesian inverse exists. 

\bd
\label{defn:supportedcorner}
Let $\mB\xstoch{\xi}\C$ be a state with associated support $P_{\xi}$. The subalgebra $P_{\xi}\mB P_{\xi}$ will be called the \define{supported corner} of $\mB$ with respect to the state $\xi$. 
\ed

When the density matrix $\s$ is invertible, a necessary condition for a Bayes map to be completely positive is that it preserve the involution (this is the implication~\ref{item:sa1}$\Rightarrow$\ref{item:sa6} in Proposition~\ref{prop:selfadjointimpliesCP} below). This result is closely related to a theorem of Nakamura, Takesaki, and Umegaki, which states that a positive conditional expectation is automatically completely positive~\cite{NTU60}.
In our case, we are working with morphisms that are a bit weaker than conditional expectations (the latter of which can be viewed as assuming $F$ to be an injective $^*$-homomorphism) and we will only assume that (the supported corner of) our Bayes maps are $*$-preserving rather than positive.%
\footnote{We would like to thank Luca Giorgetti and Alessio Ranallo for discussions clarifying some of these points. Further details elaborating these, and other, points will appear elsewhere.}

\bn
\label{prop:selfadjointimpliesCP}
Let $F, G,\omega, \rho,\xi$, and $\sigma$ be as in Proposition~\ref{prop:qBayesianinverseformula} 
with $G$ a Bayes map for $(F,\omega)$. Then the following conditions are equivalent.
\begin{enumerate}[i.]
\itemsep0pt
\item
\label{item:sa1}
$\mathrm{Ad}_{P_{\xi}}\circ G$ is $*$-preserving.
\item
\label{item:sa2}
$P_{\xi} F^*(\rho A)\sigma=\sigma F^*(A\rho)P_{\xi}$ for all $A\in\mathcal{M}_{m}(\C)$. 
\item
\label{item:sa3}
$F(\s B)\rho=\rho F(B\s)$ for all $B\in P_{\xi}\mathcal{M}_{n}(\C)P_{\xi}$. 
\item
\label{item:sa4}
Let $U\in\mathcal{M}_{m}(\C)$ be a unitary matrix diagonalizing $\rho$, i.e.\ $U\rho U^{\dag}=\boldsymbol{\rho}$, with $\boldsymbol{\rho}$ diagonal. Let ${}_{U}F:=\mathrm{Ad}_{U}\circ F$. 
Then, the Choi matrix of 
$F^*_{U^{\dag}}:=F^*\circ\mathrm{Ad}_{U^{\dag}}$
satisfies
\be
(\boldsymbol{\rho}\otimes\hat{\sigma})\mathrm{Choi}\big(F^*_{U^{\dag}}\big)(\mathds{1}_{m}\otimes P_{\xi})
=(\mathds{1}_{m}\otimes P_{\xi})\mathrm{Choi}\big(F^*_{U^{\dag}}\big)(\boldsymbol{\rho}\otimes\hat{\sigma})
\ee
(hence, $\mathrm{Choi}\big(F^*_{U^{\dag}}\big)$ commutes with $\boldsymbol{\rho}\otimes\hat{\s}=\boldsymbol{\rho}\otimes\s^{-1}$ when $P_{\xi}=\mathds{1}_{n}$).
\item
\label{item:sa5}
Let $U\in\mathcal{M}_{m}(\C)$ be a unitary matrix diagonalizing $\rho$ as in item~\ref{item:sa4} and let $V\in\mathcal{M}_{n}(\C)$ be a unitary matrix simultaneously diagonalizing $\s,\hat{\s},$ and $P_{\xi}$, i.e.\ $V^{\dag}\s V=\boldsymbol{\s}$, $V^{\dag}\hat{\s} V=\hat{\boldsymbol{\s}}$, and $V^{\dag}P_{\xi}V=\boldsymbol{P_\xi}$, with bold symbols representing diagonal matrices and states.%
%footnote
\footnote{This can be done because $\s,\hat{\s},$ and $P_{\xi}$ are self-adjoint and commute with each other and are therefore simultaneously diagonalizable via the same unitary.}
%end footnote
 Let ${}_{U}F_{V}:=\mathrm{Ad}_{U}\circ F\circ\mathrm{Ad}_{V}$ and ${}_{V^{\dag}}G_{U^{\dag}}:=\mathrm{Ad}_{V^{\dag}}\circ G\circ\mathrm{Ad}_{U^{\dag}}$. Then, 
the Choi matrix of ${}_{U}F_{V}\circ\mathrm{Ad}_{\boldsymbol{P_\xi}}$ 
satisfies 
\be
\big[\mathrm{Choi}\big({}_{U}F_{V}\circ\mathrm{Ad}_{\boldsymbol{P_\xi}}\big),\hat{\boldsymbol{\sigma}}\otimes\boldsymbol{\rho}\big]=0. 
\ee
\item
\label{item:sa6}
$\mathrm{Ad}_{P_{\xi}}\circ G$ is completely positive. %, i.e.\ $G$ restricted to the supported corner is CP.
\end{enumerate}
\en

\bprf
{\color{white}{you found me!}}

\noindent
(\ref{item:sa1}$\Rightarrow$\ref{item:sa2}) Suppose that $\mathrm{Ad}_{P_{\xi}}\circ G$ is $*$-preserving. Then
\be
\hat{\sigma}F^*(\rho A^{\dag})P_{\xi}
\overset{\text{(\ref{eq:qBayesianinversepartialsupport})}}{=\joinrel=\joinrel=}P_{\xi}G\big(A^{\dag}\big)P_{\xi}
\overset{\text{\ref{item:sa1}}}{=\joinrel=}
\big(P_{\xi}G(A)P_{\xi}\big)^{\dag}
\overset{\text{(\ref{eq:qBayesianinversepartialsupport})}}{=\joinrel=\joinrel=}\big(\hat{\sigma}F^*(\rho A)P_{\xi}\big)^{\dag}
=P_{\xi}F^*(A^{\dag}\rho)\hat{\s},
\ee
where we have also used that $F^*$ is $*$-preserving because $F^*$ is CP. Multiplying the left and right by $\sigma$ and using the properties of $\sigma,P_{\xi},$ and $\hat{\sigma}$ gives the claim.

\noindent
(\ref{item:sa2}$\Rightarrow$\ref{item:sa3})
Since $P_{\xi} F^*(\rho A)\sigma=\sigma F^*(A\rho)P_{\xi}$ is a matrix equation, the corresponding $ij$ components are also equal. By using the properties of the trace, 
\be
\xy0;/r.40pc/:
(10,12)*+{\big(\sigma F^*(A\rho)P_{\xi}\big)_{ij}}="1";
(19,4)*+{\tr\big(\s F^*(A\rho)P_{\xi}E_{ji}\big)}="2";
(19,-4)*+{\tr\big(F^*(A\rho)P_{\xi}E_{ji}P_{\xi}\sigma\big)}="3";
(13,-12)*+{\tr\big(\rho F(P_{\xi}E_{ji}P_{\xi}\sigma)A\big)}="4";
(-13,-12)*+{\tr\big(F(\sigma P_{\xi}E_{ji}P_{\xi})\rho A\big)}="6";
(-19,-4)*+{\tr\big(\sigma P_{\xi}E_{ji}P_{\xi}F^*(\rho A)\big)}="7";
(-19,4)*+{\tr\big(P_{\xi} F^*(\rho A)\sigma E_{ji}\big)}="8";
(-10,12)*+{\big(P_{\xi} F^*(\rho A)\sigma\big)_{ij}}="9";
{\ar@{=}@/^0.55pc/"1";"2"^(0.65){\text{Lem~\ref{lem:matrixunitsandtrace}}}};
{\ar@{=}@/^0.35pc/"2";"3"^{\s=P_{\xi}\s}};
{\ar@{=}@/^0.35pc/"3";"4"};
{\ar@{=}@/^0.35pc/"6";"7"};
{\ar@{=}@/^0.35pc/"7";"8"^{\s=\s P_{\xi}}};
{\ar@{=}@/^0.55pc/"8";"9"^(0.35){\text{Lem~\ref{lem:matrixunitsandtrace}}}};
{\ar@{=}@/_0.25pc/"1";"9"_{\text{\ref{item:sa2}}}};
\endxy
.
\ee
The bottom of this gives
\be
\tr\Big(\big(F(\sigma P_{\xi}E_{ji}P_{\xi})\rho-\rho F(P_{\xi}E_{ji}P_{\xi}\sigma)\big) A\Big)=0\qquad\forall\;A\in\mathcal{M}_{m}(\C).
\ee
For this to be true, the term to the left of $A$ must vanish. In other words, 
\be
F(\sigma P_{\xi}E_{ji}P_{\xi})\rho=\rho F(P_{\xi}E_{ji}P_{\xi}\sigma)
\ee
for all $i,j$. Hence, \ref{item:sa3} holds. 

\noindent
(\ref{item:sa3}$\Rightarrow$\ref{item:sa1})
Since $\mathrm{Ad}_{P_{\xi}}\circ G$ is $*$-preserving if and only if $G^*\circ\mathrm{Ad}_{P_{\xi}}$ is $*$-preserving by Lemma~\ref{lem:matrixunitsandtrace}, it suffices to prove the latter. In this case, 
\be
\begin{split}
G^*(P_{\xi}B^{\dag}P_{\xi})
&\overset{\text{(\ref{eq:qBayesianinversepartialsupportdual})}}{=\joinrel=\joinrel=}F(P_{\xi}B^{\dag}\hat{\s})\rho
=\big(\rho F(\hat{\s}BP_{\xi})\big)^{\dag}
\overset{\text{Lem~\ref{lem:inverseuptosupport}}}{=\joinrel=\joinrel=\joinrel=\joinrel=}\big(\rho F(\hat{\s}B\hat{\s}\s)\big)^{\dag}\\
&\overset{\text{\ref{item:sa3}}}{=\joinrel=}
\big(F(\s\hat{\s}B\hat{\s})\rho\big)^{\dag}
\overset{\text{(\ref{eq:qBayesianinversepartialsupportdual})}}{=\joinrel=\joinrel=}G^*(\s\hat{\s}BP_{\xi})^{\dag}
=G^*(P_{\xi}BP_{\xi})^{\dag}. 
\end{split}
\ee

\noindent
(\ref{item:sa1}$\Leftrightarrow$\ref{item:sa4})
Recall that $G$ is a Bayesian inverse of $(F,\w)$ if and only if $G_{U^{\dag}}:=G\circ\mathrm{Ad}_{U^\dag}$ is a Bayesian inverse of $({}_{U}F,\boldsymbol{\w}:=\w\circ\mathrm{Ad}_{U^\dag})$ (cf.\ Lemma~\ref{lem:bayesianinversesgeneralexamples}). Let $\{p_{i}\}$ denote the eigenvalues of $\rho$. 
Recall that a linear map $\vf$ between matrix algebras is $*$-preserving if and only if its Choi matrix is self-adjoint, $\Choi(\vf)^{\dag}=\Choi(\vf)$. 
In this case, $\mathrm{Ad}_{P_{\xi}}\circ G$ is $*$-preserving if and only if $\mathrm{Ad}_{P_{\xi}}\circ G_{U^{\dag}}$ is $*$-preserving if and only if $\Choi(\mathrm{Ad}_{P_{\xi}}\circ G_{U^{\dag}})$ is self-adjoint, and the latter is given by 
\be
\begin{split}
\mathrm{Choi}\big(\mathrm{Ad}_{P_{\xi}}\circ G_{U^{\dag}}\big)
&=\sum_{i,j}E_{ij}^{(m)}\otimes P_{\xi}G_{U^{\dag}}\big(E_{ij}^{(m)}\big)P_{\xi}
\overset{\text{(\ref{eq:qBayesianinversepartialsupport})}}{=\joinrel=\joinrel=}\sum_{i,j}E_{ij}^{(m)}\otimes \hat{\sigma}F^*_{U^{\dag}}\big(\boldsymbol{\rho} E_{ij}^{(m)}\big)P_{\xi}\\
&=\sum_{i,j}p_{i}E_{ij}^{(m)}\otimes \hat{\sigma}F^*_{U^{\dag}}\big(E_{ij}^{(m)}\big)P_{\xi}
=\sum_{i,j}\boldsymbol{\rho} E_{ij}^{(m)}\otimes \hat{\sigma}F^*_{U^{\dag}}\big(E_{ij}^{(m)}\big)P_{\xi}\\
&=(\boldsymbol{\rho}\otimes\hat{\sigma})\mathrm{Choi}\big(F^*_{U^{\dag}}\big)(\mathds{1}_{m}\otimes P_{\xi}).
\end{split}
\ee
Taking the adjoint of this and equating it to itself proves the claim.

\noindent
(\ref{item:sa1}$\Leftrightarrow$\ref{item:sa5})
Set 
\be
\hat{q}_{j}:=\begin{cases}q_{j}^{-1}&\mbox{ when $q_{j}>0$}\\
0&\mbox{ when $q_{j}=0$}
\end{cases}
\;\;,
\ee
where $\{q_{j}\}$ consists of the eigenvalues of $\s$. 
The Choi matrix of  ${}_{U}G^*_{V}\circ\mathrm{Ad}_{\boldsymbol{P_\xi}}$ is equal to 
\be
\begin{split}
\mathrm{Choi}&\big({}_{U}G^*_{V}\circ\mathrm{Ad}_{\boldsymbol{P_\xi}}\big)
=\sum_{i,j}E_{ij}^{(n)}\otimes UG^*\big(V\boldsymbol{P_\xi}E_{ij}^{(n)}\boldsymbol{P_\xi}V^{\dag}\big)U^{\dag}\\
&=\sum_{i,j}E_{ij}^{(n)}\otimes UG^*\big(P_{\xi}VE_{ij}^{(n)}V^{\dag}P_{\xi}\big)U^{\dag}
\overset{\text{(\ref{eq:qBayesianinversepartialsupportdual})}}{=\joinrel=\joinrel=}\sum_{i,j}E_{ij}^{(n)}\otimes UF\big(P_{\xi}VE_{ij}^{(n)}V^{\dag}\hat{\sigma}\big)\rho U^{\dag}\\
&=\sum_{i,j}E_{ij}^{(n)}\otimes UF\big(V\boldsymbol{P_\xi}\!\!\underbrace{E_{ij}^{(n)}\hat{\boldsymbol{\sigma}}}_{E_{ij}^{(n)}\hat{q}_{j}\boldsymbol{P_\xi}}\!\!V^{\dag}\big)U^{\dag}U\rho U^{\dag}
=\sum_{i,j}E_{ij}^{(n)}\hat{q}_{j}\otimes UF\big(V\boldsymbol{P_\xi}E_{ij}^{(n)}\boldsymbol{P_\xi}\big)U^{\dag}\boldsymbol{\rho}\\
&=\sum_{i,j}E_{ij}^{(n)}\hat{\s}\otimes {}_{U}F_{V}\!\left(\mathrm{Ad}_{\boldsymbol{P_\xi}}\big(E_{ij}^{(n)}\big)\right)\boldsymbol{\rho}
=\mathrm{Choi}\big({}_{U}F_{V}\circ\mathrm{Ad}_{\boldsymbol{P_\xi}}\big)(\hat{\boldsymbol{\sigma}}\otimes\boldsymbol{\rho}).
\end{split}
\ee
This is self-adjoint if and only if $G^*\circ\mathrm{Ad}_{P_{\xi}}$ is $*$-preserving, which holds if and only if $\mathrm{Ad}_{P_{\xi}}\circ G$ is $*$-preserving. 

\noindent
(\ref{item:sa1}$\Rightarrow$\ref{item:sa6})
In this part of the proof, we will assume that $\rho$ and $\s$ have been diagonalized with all non-vanishing eigenvalues appearing on the top-left block. The reason we can do this is the following. As in Lemma~\ref{lem:bayesianinversesgeneralexamples}, let $U$ be a unitary such that $U^{\dag}\boldsymbol{\rho}U=\rho$ with $\boldsymbol{\rho}$ diagonal, and let $V$ be a unitary such that $V^{\dag}\s V=\boldsymbol{\s}, V^{\dag}P_{\xi}V=\boldsymbol{P_\xi},$ and $V^{\dag}\hat{\s}V=\hat{\boldsymbol{\s}}$, where the bold matrices are diagonal. By Lemma~\ref{lem:bayesianinversesgeneralexamples}, $G$ is a Bayesian inverse of $(F,\w)$ if and only if ${}_{V^{\dag}}G_{U^{\dag}}$ is a Bayesian inverse of $({}_{U}F_{V},\boldsymbol{\w}:=\omega\circ\mathrm{Ad}_{U^{\dag}})$. Since $U$ and $V$ are unitary, $\mathrm{Ad}_{P_{\xi}}\circ G$ is CP (or $*$-preserving) if and only if $\mathrm{Ad}_{V^{\dag}}\circ\mathrm{Ad}_{P_{\xi}}\circ G\circ\mathrm{Ad}_{U^{\dag}}$ is CP (or $*$-preserving). Since the latter map equals
\be
\mathrm{Ad}_{V^{\dag}}\circ\mathrm{Ad}_{P_{\xi}}\circ G\circ\mathrm{Ad}_{U^{\dag}}
=\mathrm{Ad}_{\boldsymbol{P_\xi}}\circ\mathrm{Ad}_{V^{\dag}}\circ G\circ\mathrm{Ad}_{U^{\dag}}
=\mathrm{Ad}_{\boldsymbol{P_\xi}}\circ {}_{V^{\dag}}G_{U^{\dag}},
\ee
the map $\mathrm{Ad}_{P_{\xi}}\circ G$ is CP (or $*$-preserving) if and only if $\mathrm{Ad}_{\boldsymbol{P_\xi}}\circ {}_{V^{\dag}}G_{U^{\dag}}$ is. This has established that we can assume $\rho$ and $\s$ are diagonal. As such, 
let $r$ be the rank of $\s$. 
Write $F=\sum_{\a}\mathrm{Ad}_{V_{\a}}$, where $V_{\a}$ is an $m\times n$ matrix whose $ij$-th entry will be written as $V^{\a}_{ij}$ for the purposes of this proof (and later proofs). 
Then the Choi matrix of $\mathrm{Ad}_{P_{\xi}}\circ G$ is given by 
\be
\label{eq:ChoimatrixofBayesmatrixcasecompact}
\begin{split}
\mathrm{Choi}&\big(\mathrm{Ad}_{P_{\xi}}\circ G\big)
=\sum_{i,j}E_{ij}^{(m)}\otimes P_{\xi}G\big(E_{ij}^{(m)}\big)P_{\xi}
\overset{\text{(\ref{eq:qBayesianinversepartialsupport})}}{=\joinrel=\joinrel=}
\sum_{i,j}E_{ij}^{(m)}\otimes \hat{\sigma}F^*\big(\rho E_{ij}^{(m)}\big)P_{\xi}\\
&=\sum_{i,j}E_{ij}^{(m)}\otimes \hat{\sigma}F^*\big(p_{i} E_{ij}^{(m)}\big)P_{\xi}
=\sum_{i,j}E_{ij}^{(m)}\otimes \left(p_{i}\hat{\sigma}\sum_{\a}V_{\a}^{\dag}E_{ij}^{(m)}V_{\a}P_{\xi}\right)\\
&=\sum_{\a}\sum_{i,j}E_{ij}^{(m)}\otimes\left(
\begin{bmatrix}p_{i}/q_{1}&\cdots&0&0\\\vdots&\ddots&&\vdots\\
0&\cdots&p_{i}/q_{r}&0\\
0&\cdots&0&0_{n-r}\end{bmatrix}
\begin{bmatrix}
\overline{V^{\a}_{i1}}V^{\a}_{j1}&\cdots&\overline{V^{\a}_{i1}}V^{\a}_{jn}\\
\vdots&&\vdots\\
\overline{V^{\a}_{in}}V^{\a}_{j1}&\cdots&\overline{V^{\a}_{in}}V^{\a}_{jn}
\end{bmatrix}\begin{bmatrix}\mathds{1}_{r}&0\\0&0_{n-r}\end{bmatrix}\right)\\
&=
\sum_{\a}\sum_{i,j}E_{ij}^{(m)}\otimes
\begin{bmatrix}
\frac{p_{i}}{q_{1}}\overline{V^{\a}_{i1}}V^{\a}_{j1}&\cdots&\frac{p_{i}}{q_{1}}\overline{V^{\a}_{i1}}V^{\a}_{jr}
&0\\
\vdots&&\vdots&\vdots\\
\frac{p_{i}}{q_{r}}\overline{V^{\a}_{ir}}V^{\a}_{j1}&\cdots&\frac{p_{i}}{q_{r}}\overline{V^{\a}_{ir}}V^{\a}_{jr}&0\\
0&\cdots&0&0_{n-r}\\
\end{bmatrix}
,
\end{split}
\ee
while taking the adjoint of the Choi matrix gives
\be
\label{eq:adjointChoi}
\mathrm{Choi}\big(\mathrm{Ad}_{P_{\xi}}\circ G\big)^{\dag}
=\sum_{\a}\sum_{i,j}E_{ij}^{(m)}\otimes
\begin{bmatrix}
\frac{p_{j}}{q_{1}}\overline{V^{\a}_{i1}}V^{\a}_{j1}&\cdots&\frac{p_{j}}{q_{r}}\overline{V^{\a}_{i1}}V^{\a}_{jr}
&0\\
\vdots&&\vdots&\vdots\\
\frac{p_{j}}{q_{1}}\overline{V^{\a}_{ir}}V^{\a}_{j1}&\cdots&\frac{p_{j}}{q_{r}}\overline{V^{\a}_{ir}}V^{\a}_{jr}&0\\
0&\cdots&0&0_{n-r}\\
\end{bmatrix}
\ee
due to the summation over all $i,j$. 
Self-adjointness of the Choi matrix 
equates~(\ref{eq:ChoimatrixofBayesmatrixcasecompact}) with~(\ref{eq:adjointChoi}) and gives
\be
\label{eq:selfadjointnessChoicondition}
\frac{p_{i}}{q_{k}}\sum_{\a}\overline{V^{\a}_{ik}}V^{\a}_{jl}=\frac{p_{j}}{q_{l}}\sum_{\a}\overline{V^{\a}_{ik}}V^{\a}_{jl}
\qquad\forall\;k,l\in\{1,\dots,r\},\;\;i,j\in\{1,\dots,m\}. 
\ee
This condition is so strong that it actually implies positivity of this matrix. Indeed, if we set 
\be
\setcounter{MaxMatrixCols}{20}
\mathfrak{V}_{\a}:=
\begin{bmatrix}
\sqrt{\frac{p_{1}}{q_{1}}}V^{\a}_{11}&\cdots&\!\!\sqrt{\frac{p_{1}}{q_{r}}}V^{\a}_{1r}
&0&\cdots&0
&{\small\bullet\;\bullet\;\bullet}&
\sqrt{\frac{p_{m}}{q_{1}}}V^{\a}_{m1}&\cdots&\!\!\sqrt{\frac{p_{m}}{q_{r}}}V^{\a}_{mr}
&0&\cdots&0
\end{bmatrix}
,
\ee
then 
\be
\label{eq:Choiissumpositive}
\sum_{\a}\mathfrak{V}_{\a}^{\dag}\mathfrak{V}_{\a}=
\sum_{\a}\sum_{i,j}E_{ij}^{(m)}\otimes\begin{bmatrix}\sqrt{\frac{p_{i}p_{j}}{q_{1}^2}}\overline{V^{\a}_{i1}}V^{\a}_{j1}&\cdots&\sqrt{\frac{p_{i}p_{j}}{q_{1}q_{r}}}\overline{V^{\a}_{i1}}V^{\a}_{jr}
&0\\
\vdots&&\vdots&\vdots\\
\sqrt{\frac{p_{i}p_{j}}{q_{r}q_{1}}}\overline{V^{\a}_{ir}}V^{\a}_{j1}&\cdots&\sqrt{\frac{p_{i}p_{j}}{q_{r}^2}}\overline{V^{\a}_{ir}}V^{\a}_{jr}&0\\
0&\cdots&0&0_{n-r}\\\end{bmatrix}
=\mathrm{Choi}\big(\mathrm{Ad}_{P_{\xi}}\circ G\big).
\ee
To see the last equality in~(\ref{eq:Choiissumpositive}), first notice that  condition~(\ref{eq:selfadjointnessChoicondition}) can be rewritten as 
\be
p_{i}q_{l}\b_{ijkl}=p_{j}q_{k}\b_{ijkl},\quad\text{where}\quad\b_{ijkl}:=\sum_{\a}\overline{V^{\a}_{ik}}V^{\a}_{jl}. 
\ee
If $\b_{ijkl}=0$, then $\sqrt{p_{i}q_{l}}\b_{ijkl}=\sqrt{p_{j}q_{k}}\b_{ijkl}$. 
If $\b_{ijkl}\ne0$, then $p_{i}q_{l}=p_{j}q_{k}$ so again $\sqrt{p_{i}q_{l}}\b_{ijkl}=\sqrt{p_{j}q_{k}}\b_{ijkl}$. Thus, we have
\be
\label{eq:betaijkl}
\sqrt{\frac{p_{i}}{q_{k}}}\b_{ijkl}=\sqrt{\frac{p_{j}}{q_{l}}}\b_{ijkl}\qquad\forall\;k,l\in\{1,\dots,r\},\;i,j\in\{1,\dots,m\}.
\ee 
Hence,
\be
\sqrt{\frac{p_{i}p_{j}}{q_{k}q_{l}}}\b_{ijkl}
\overset{\text{(\ref{eq:betaijkl})}}{=\joinrel=\joinrel=}\frac{p_{i}}{q_{k}}\b_{ijkl}, 
\ee
which proves~(\ref{eq:Choiissumpositive}). 
This shows that $\mathrm{Ad}_{P_{\xi}}\circ G$ is completely positive by Choi's theorem. 

\noindent
(\ref{item:sa6}$\Rightarrow$\ref{item:sa1}) This direction is immediate since all positive matrices are self-adjoint. 
\eprf

\bc
\label{cor:KrausdecompBayesmatrixcase}
Let $F, G,\omega, \rho,\xi$, and $\sigma$ be as in Proposition~\ref{prop:qBayesianinverseformula} with $G$ a Bayes map for $(F,\omega)$. Let $F=\sum_{\a}\mathrm{Ad}_{V_{\a}}$ be a Kraus decomposition of $F$. Suppose further that any of the equivalent conditions in Proposition~\ref{prop:selfadjointimpliesCP} hold. Then $\mathrm{Ad}_{P_{\xi}}\circ G$ has a Kraus decomposition given by 
\be
\mathrm{Ad}_{P_{\xi}}\circ G
=\mathrm{Ad}_{\sqrt{\hat{\s}}}\circ F^*\circ\mathrm{Ad}_{\sqrt{\rho}}
=\sum_{\a}\mathrm{Ad}_{\sqrt{\hat{\s}}V_{\a}^{\dag}\sqrt{\rho}}. 
\ee
\ec

\bprf
A Choi matrix for $\mathrm{Ad}_{P_{\xi}}\circ G$ was constructed in~(\ref{eq:ChoimatrixofBayesmatrixcasecompact}) in the proof of Proposition~\ref{prop:selfadjointimpliesCP}. It was shown that $\mathrm{Choi}(\mathrm{Ad}_{P_{\xi}}\circ G)=\sum_{\a}\mathfrak{V}_{\a}^{\dag}\mathfrak{V}_{\a}$. Hence, if we demand $\mathrm{Ad}_{P_{\xi}}\circ G$ has a Kraus decomposition of the form $\mathrm{Ad}_{P_{\xi}}\circ G=\sum_{\b}\mathrm{Ad}_{W_{\b}}$, then it must be the case that 
\be
(\mathrm{Ad}_{P_{\xi}}\circ G)(E_{ij})
=\sum_{\b}W_{\b}E_{ij}W_{\b}^{\dag}
=\sum_{\b}W_{\b}e_{i}e_{j}^{\dag}W_{\b}^{\dag}
=\sum_{\b}(W_{\b}e_{i})(W_{\b}e_{j})^{\dag}
\ee
But this is precisely the $ij$-th block of the Choi matrix. Hence, we can set 
\be
W_{\a}:=
\begin{bmatrix}
\sqrt{\frac{{p}_{1}}{q_{1}}}\overline{V_{11}^{\a}}&\cdots&\sqrt{\frac{{p}_{m}}{q_{1}}}\overline{V_{m1}^{\a}}\\
\vdots&&\vdots\\
\sqrt{\frac{{p}_{1}}{q_{r}}}\overline{V_{1r}^{\a}}&\cdots&\sqrt{\frac{{p}_{m}}{q_{r}}}\overline{V_{mr}^{\a}}\\
0&\cdots&0\\
\vdots&&\vdots\\
0&\cdots&0\\
\end{bmatrix}
=\sqrt{\hat{\s}}V_{\a}^{\dag}\sqrt{\rho}. 
\ee
Note that the expression on the right still holds even if we did not choose a basis in which $\s$ and $\rho$ have been diagonalized. To see this, we temporarily use the notation from the proof of~(\ref{item:sa1}$\Rightarrow$\ref{item:sa6}) in Proposition~\ref{prop:selfadjointimpliesCP}. In doing so, we have
\be
\begin{split}
\mathrm{Ad}_{P_{\xi}}\circ G
&=\mathrm{Ad}_{V}\circ\mathrm{Ad}_{\boldsymbol{P_\xi}}\circ {}_{V^{\dag}}G_{U^{\dag}}\circ\mathrm{Ad}_{U}
=\mathrm{Ad}_{V}\circ\mathrm{Ad}_{\sqrt{\hat{\boldsymbol{\s}}}}\circ {}_{V^{\dag}}F^*_{U^{\dag}}\circ\mathrm{Ad}_{\sqrt{\boldsymbol{\rho}}}\circ\mathrm{Ad}_{U}\\
&=\mathrm{Ad}_{\sqrt{\hat{\s}}}\circ\mathrm{Ad}_{V}\circ {}_{V^{\dag}}F^*_{U^{\dag}}\circ\mathrm{Ad}_{U}\circ\mathrm{Ad}_{\sqrt{\rho}}
=\sum_{\a}\mathrm{Ad}_{\sqrt{\hat{\s}}V_{\a}^{\dag}\sqrt{\rho}}
\end{split}
\ee
by the way in which the bold versions of these mathematical objects have been defined. 
\eprf

\br
Let $\rho\in\mathcal{M}_{m}(\C)$ be a density matrix and 
$\mathcal{M}_{n}(\C)\xstoch{F}\mathcal{M}_{m}(\C)$ a CP map with Kraus operator decomposition $F=\sum_{\a}\mathrm{Ad}_{V_{\a}}$. Set $\s:=F^*(\rho)$. The map 
$L:=\sum_{\a}\mathrm{Ad}_{\sqrt{\hat{\s}}V_{\a}^{\dag}\sqrt{\rho}}$ is what Leifer calls the Bayesian inverse of $F$~\cite{Le06,Le07}, which he obtains by a variant of the Choi--Jamiolkowski isomorphism. This map was later described diagrammatically 
by Coecke and Spekkens~\cite{CoSp12} and Jacobs~\cite{Ja18EPTCS}
and was further enhanced in the work of Leifer and Spekkens~\cite{LeSp13}. There are several important comments to be made with regards to Leifer's map versus ours. 
First, the notion of a.e.\ equivalence was not discussed in any of these works and therefore the uniqueness of Bayesian inverses was not addressed. In particular, how should the Bayesian inverse be defined off the support (what we call the supported corner)? 
Second, the map $H$ is not in general unital unless $\s$ is invertible. Furthermore, even if $\s$ is invertible, the Bayesian inverse of the Bayesian inverse is \emph{not} in general a.e.\ equivalent to the original map $F$ unless $\rho$ is \emph{also} invertible. However, this restricts the analysis of CPU Bayesian inversion to states that have full support and therefore ignores, in particular, pure states. Third, our arrival at Corollary~\ref{cor:KrausdecompBayesmatrixcase} comes at a cost. While Leifer makes no assumptions about the relationship between the density matrices and the map $F$ to construct his Bayesian inverse $L$, we have enforced additional (in general non-trivial) constraints as in Proposition~\ref{prop:selfadjointimpliesCP} to accomplish this task. This is due to our reliance on the categorical expression of the classical Bayes' theorem presented in Theorem~\ref{thm:classicalBayestheorem}, which was used to define Bayesian inversion in the non-commutative setting in Definition~\ref{defn:quantumBayesianinverse}. We make no claims as to which perspective is correct or wrong, and we suspect both have their appropriate domains of applicability. 
One interesting curiosity is the arrival at the form of the Bayesian inverse in Corollary~\ref{cor:KrausdecompBayesmatrixcase} through a more structural perspective. 
Another benefit of our categorical approach is that it offers an explicit construction of the Bayesian inverse off of the support. Therefore, our definition is capable of handling density matrices regardless of their support. This will be explained in Theorem~\ref{thm:Bayesianinversionmatrixalgebracase} after we go through a few more examples and preparation. 
\er

\bn
\label{prop:Krausrankone}
A Bayesian inverse always exists for any pair of CPU maps of the form 
\be
\mathcal{M}_{n}(\C)\xstoch{F=\mathrm{Ad}_{V}}\mathcal{M}_{m}(\C)\xstoch{\omega=\tr(\rho\;\cdot\;)}\C,
\ee
where $V^{\dag}:\C^{n}\to\C^{m}$ is an isometry. A representative for such Bayesian inverse is given by 
\be
\label{eq:bayesianinverseisometry}
\mathcal{M}_{m}(\C)\ni A\xmapsto{G} V^{\dag}AV+\frac{1}{m}\tr(A)(V^{\dag}V)^{\perp}. 
\ee
\en

Before proving this, some explanation is needed. Note that since $V^{\dag}$ is an isometry, the $n\times n$ matrix $V^{\dag}V$ is a projection. An interesting point about this result is that a Bayesian inverse of a Kraus rank one CPU map need not be Kraus rank one. We first prove a little lemma that will be used elsewhere as well. 

\blem
\label{lem:supportLEpisometry}
Let $V:\C^{n}\to\C^{m}$ be a coisometry (i.e.\ $VV^{\dag}=\mathds{1}_{m}$) and let $\mathcal{M}_{m}(\C)\xstoch{\omega=\tr(\rho\;\cdot\;)}\C$ be a state. Set $\s:=V^{\dag}\rho V$ and write the corresponding state on $\mathcal{M}_{n}(\C)$ as $\xi$, so that $\xi=\w\circ F$. 
Then $V^{\dag}V$ is a projection satisfying  
\be
P_{\xi}\le V^{\dag}V
\quad
\text{ and }
\quad 
(V^{\dag}V)^{\perp}\le P_{\xi}^{\perp}. 
\ee
In particular, 
\be
P_{\xi}=P_{\xi}V^{\dag}V=V^{\dag}VP_{\xi}
\quad
\text{ and }
\quad 
(V^{\dag}V)^{\perp}=P_{\xi}^{\perp}(V^{\dag}V)^{\perp}=(V^{\dag}V)^{\perp}P_{\xi}^{\perp}.
\ee 
\elem

\bprf
[Proof of Lemma~\ref{lem:supportLEpisometry}]
The fact that $V^{\dag}V$ is a projection follows from $V^{\dag}$ being an isometry. 
Since 
\be
V^{\dag}V\s=V^{\dag}VV^{\dag}\rho V
=V^{\dag}\rho V
=\s
=\s^{\dag}
=\big(V^{\dag}V\s\big)^{\dag}
=\s^{\dag}\big(V^{\dag}V\big)^{\dag}
=\s V^{\dag}V
,
\ee
$V^{\dag}V$ is a projection satisfying $\s V^{\dag}V=\s=V^{\dag}V\s$. Because $P_{\xi}$ is the \emph{smallest} projection satisfying this condition, $P_{\xi}\le V^{\dag}V$. The other claims follow from the properties of projections and orthogonal complements. 
\eprf

\bprf
[Proof of Proposition~\ref{prop:Krausrankone}]
The proof involves several small steps which will be done one at a time. These include showing that $G$ is unital, completely positive, and finally that it satisfies the Bayes condition. The first two are simple to prove. Indeed, 
\be
G(\mathds{1}_{m})=V^{\dag}V+\frac{1}{m}\tr(\mathds{1}_{m})(V^{\dag}V)^{\perp}
=V^{\dag}V+(V^{\dag}V)^{\perp}=\mathds{1}_{n}
\ee
proves unitality. Complete positivity can be proved in several ways. The first term is clearly completely positive while the latter term is the composite
\be
\label{eq:traceprojection}
\mathcal{M}_{m}(\C)\xstoch{\frac{1}{m}\tr(\;\cdot\;)}\C\xrightarrow{(V^{\dag}V)^{\perp}}\mathcal{M}_{m}(\C)
\ee
of CP maps, which is CP. The last map in~(\ref{eq:traceprojection}) sends $\l\in\C$ to $\l(V^{\dag}V)^{\perp}$ which is a (not necessarily unital) $^*$-homomorphism and hence CP. 

Finally, the Bayes condition holds because
\be
\begin{split}
\tr\big(\s G(A) B\big)&=\tr\left(\s\left(V^{\dag}AV+\frac{1}{m}\tr(A)(V^{\dag}V)^{\perp}\right)B\right)\\
&=\tr\left(V^{\dag}\rho VV^{\dag}AVB\right)
+\frac{1}{m}\tr(A)\tr\left(\s (V^{\dag}V)^{\perp} B\right)\\
&\overset{\text{Lem~\ref{lem:supportLEpisometry}}}{=\joinrel=\joinrel=\joinrel=\joinrel=}
\tr(\rho A VBV^{\dag})+\frac{1}{m}\tr(A)\tr\left(\s P_{\xi}^{\perp}(V^{\dag}V)^{\perp} B\right)\\
&=\tr\big(\rho AF(B)\big)
\end{split}
\ee
for all $A\in\mathcal{M}_{m}(\C)$ and $B\in\mathcal{M}_{n}(\C)$. 
\eprf

\br
It may seem that Proposition~\ref{prop:Krausrankone} seems to contradict Proposition~\ref{prop:selfadjointimpliesCP} since we were able to find a Bayesian inverse for $(\mathrm{Ad}_{V},\w)$ without any additional commutativity requirements. However, it turns out that these commutativity requirements are satisfied. Indeed, 
\be
F(\s B)\rho=V\s B V^{\dag}\rho
=V\s P_{\xi}V^{\dag}V B V^{\dag}\rho V V^{\dag}
=V\s V^{\dag}V B\s V^{\dag}
=\rho F(B\s)
\ee
by Lemma~\ref{lem:supportLEpisometry}. Also, the formula for $P_{\xi}G(A)$ in~(\ref{eq:qBayesianinversepartialsupport}) seems quite different from the formula~(\ref{eq:bayesianinverseisometry}). Nevertheless, they agree because
\be
P_{\xi}G(A)=P_{\xi}V^{\dag}AV+\frac{1}{m}\tr(A)P_{\xi}(V^{\dag}V)^{\perp}
=\hat{\s}\s V^{\dag}AV
=\hat{\s}V^{\dag}\rho VV^{\dag}AV
=\hat{\s}V^{\dag}\rho AV,
\ee
where Lemma~\ref{lem:supportLEpisometry} was used again to eliminate the second term in the second expression. 
\er

%%%%%%%%%%%%%%%%%%%%%%%%%%%%%%%%%%%%%
\subsection{Matrix completion and Bayesian inversion}
\label{sec:matrixcomp}
%%%%%%%%%%%%%%%%%%%%%%%%%%%%%%%%%%%%%

We will now begin to provide further characterizations for when Bayesian inverses exist and how to construct Bayesian inverses when the density matrices have non-vanishing nullspaces. For this, we recall a useful theorem from linear algebra. 

\blem
\label{lem:positivitySchur}
Let 
\be
\mathfrak{M}=\begin{bmatrix}\mathfrak{A}&\mathfrak{B}\\\mathfrak{B}^{\dag}&\mathfrak{C}\end{bmatrix}
\ee
be a self-adjoint matrix, where $\mathfrak{A}$ is an $m\times m$ matrix, $\mathfrak{C}$ is an $n\times n$ matrix, and $\mathfrak{B}$ is therefore $m\times n$. Then $\mathfrak{M}\ge0$ if and only if the following three conditions hold: 
\begin{enumerate}[i.]
\itemsep0pt
\item
$\mathfrak{A}\ge0$,
\item
$\ker(\mathfrak{A})\subseteq\ker(\mathfrak{B}^{\dag})$, and 
\item
$\mathfrak{C}-\mathfrak{B}^{\dag}\hat{\mathfrak{A}}\mathfrak{B}\ge0$, where $\hat{\mathfrak{A}}$ is the pseudoinverse of $\mathfrak{A}$. 
\end{enumerate}
Furthermore, when $\mathfrak{M}\ge0$, then 
\be
\begin{bmatrix}\mathfrak{A}&\mathfrak{B}\\\mathfrak{B}^{\dag}&\mathfrak{C}\end{bmatrix}
=\begin{bmatrix}
\mathfrak{A}^{1/2}&0\\
\mathfrak{B}^{\dag}\hat{\mathfrak{A}}^{1/2}&(\mathfrak{C}-\mathfrak{B}^{\dag}\hat{\mathfrak{A}}\mathfrak{B})^{1/2}
\end{bmatrix}
\begin{bmatrix}
\mathfrak{A}^{1/2}&0\\
\mathfrak{B}^{\dag}\hat{\mathfrak{A}}^{1/2}&(\mathfrak{C}-\mathfrak{B}^{\dag}\hat{\mathfrak{A}}\mathfrak{B})^{1/2}
\end{bmatrix}^{\dag}
.
\ee
\elem

\bprf
See Theorem~4.3 in~\cite{Ga19} for an exceptionally clear review of the concepts and for the proof (see also Theorem~5.2 in~\cite{Wo12}).%
%footnote
\footnote{Technically, Gallier assumes the matrices are real and his second assumption reads $(\mathds{1}_{m}-\mathfrak{A}\hat{\mathfrak{A}})\mathfrak{B}=0$. The reality condition poses no issue if one uses adjoints instead of transpose. The condition $(\mathds{1}_{m}-\mathfrak{A}\hat{\mathfrak{A}})\mathfrak{B}=0$ is equivalent to $\ker(\mathfrak{A})\subseteq\ker(\mathfrak{B}^{\dag})$ by the following argument. Assume $(\mathds{1}_{m}-\mathfrak{A}\hat{\mathfrak{A}})\mathfrak{B}=0$ holds. Taking the adjoint gives $\mathfrak{B}^{\dag}=\mathfrak{B}^{\dag}\hat{\mathfrak{A}}\mathfrak{A}$. Thus, if $v\in\ker(\mathfrak{A})$, then $\mathfrak{B}^{\dag}v=\mathfrak{B}^{\dag}\hat{\mathfrak{A}}\mathfrak{A}v=0$. Conversely, suppose $\ker(\mathfrak{A})\subseteq\ker(\mathfrak{B}^{\dag})$. Note that by the properties of the pseudoinverse, $\mathfrak{A}(\mathds{1}_{m}-\hat{\mathfrak{A}}\mathfrak{A})=\mathfrak{A}-\mathfrak{A}\hat{\mathfrak{A}}\mathfrak{A}=\mathfrak{A}-\mathfrak{A}=0$, which shows that $\mathrm{image}(\mathds{1}_{m}-\hat{\mathfrak{A}}\mathfrak{A})\subseteq\ker(\mathfrak{A})\subseteq\ker(\mathfrak{B}^{\dag})$. Hence, $\mathfrak{B}^{\dag}(\mathds{1}_{m}-\hat{\mathfrak{A}}\mathfrak{A})=0$, which gives the desired result by taking adjoints. For us, it will be more computationally convenient to check a kernel condition than to compute the pseudoinverse of a matrix. 
}
%end footnote
The last equality follows from matrix multiplication and the fact that $P_{\mathfrak{A}}\mathfrak{B}=\mathfrak{B},$ where $P_{\mathfrak{A}}$ is the support of $\mathfrak{A}$. This is because 
\be
\mathrm{image}(\mathfrak{B})=\ker(\mathfrak{B}^{\dag})^{\perp}\subseteq\ker(\mathfrak{A})^{\perp}=\mathrm{image}(\mathfrak{A}^{\dag})=\mathrm{image}(\mathfrak{A})
\ee
by the second condition and the relationship between kernels and images of operators and their adjoints. 
\eprf

\br
\label{rmk:positivitySchur}
Lemma~\ref{lem:positivitySchur} will occasionally be used slightly differently than as stated. 
We will be presented with a matrix $\mathfrak{M}$ and a projection $P$ and will decompose $\mathfrak{M}$ via
\be
\mathfrak{M}=\underbrace{P\mathfrak{M}P}_{\mathfrak{A}}+\underbrace{P\mathfrak{M}P^{\perp}}_{\mathfrak{B}}+\underbrace{P^{\perp}\mathfrak{M}P}_{\mathfrak{D}}+\underbrace{P^{\perp}\mathfrak{M}P^{\perp}}_{\mathfrak{C}}. 
\ee
This decomposition can be viewed as expressing $\mathfrak{M}$ via a decomposition as in Lemma~\ref{lem:positivitySchur} by the adjoint action with some unitary and removing the appropriate zero entries. More precisely, 
let $\{v_{1},\dots,v_{s}\}$ and $\{v_{s+1},\dots,v_{m+n}\}$ be orthonormal bases of $\mathrm{image}(P)$ and $\mathrm{image}(P^{\perp})$, respectively. Let $U$ be the  unitary uniquely determined by $Uv_{k}=e_{k}$ (the standard basis) for all $k$. Then there exist matrices $\uline{\mathfrak{A}}$, $\uline{\mathfrak{B}}$, $\uline{\mathfrak{D}}$, and $\uline{\mathfrak{C}}$ such that 
\be
U\mathfrak{A}U^{\dag}=\begin{bmatrix}\uline{\mathfrak{A}}&0\\0&0\end{bmatrix},\quad
U\mathfrak{B}U^{\dag}=\begin{bmatrix}0&\uline{\mathfrak{B}}\\0&0\end{bmatrix},\quad
U\mathfrak{D}U^{\dag}=\begin{bmatrix}0&0\\\uline{\mathfrak{D}}&0\end{bmatrix},\quad\text{and}\quad
U\mathfrak{C}U^{\dag}=\begin{bmatrix}0&0\\0&\uline{\mathfrak{C}}\end{bmatrix}
\ee
(Appendix~\ref{appendix:permutation} provides a special case that will be used in the proof of Theorem~\ref{thm:Bayesianinversionmatrixalgebracase}).
Then, $\mathfrak{M}$ is self-adjoint if and only if ${\mathfrak{A}}^{\dag}={\mathfrak{A}}$, ${\mathfrak{D}}={\mathfrak{B}}^{\dag}$, and ${\mathfrak{C}}^{\dag}={\mathfrak{C}}$. Furthermore, $\mathfrak{M}\ge0$ if and only if 
\begin{enumerate}[i.]
\itemsep0pt
\item
$\mathfrak{A}\ge0$,
\item
$\ker(\mathfrak{A})\subseteq\ker(\mathfrak{B}^{\dag})$, and 
\item
$\mathfrak{C}-\mathfrak{B}^{\dag}\hat{\mathfrak{A}}\mathfrak{B}\ge0$, where $\hat{\mathfrak{A}}$ is the pseudoinverse of $\mathfrak{A}$.
\end{enumerate}
\er

\begin{notation}
Let $\tr_{\mathcal{M}_{m}(\C)}:\mathcal{M}_{m}(\C)\otimes\mathcal{M}_{n}(\C)\stoch\mathcal{M}_{n}(\C)$ denote the \define{partial trace}, which traces out the first factor and is uniquely determined by the formula
\be
\tr_{\mathcal{M}_{m}(\C)}(A\otimes B):=\tr(A)B\qquad\forall\;A\in\mathcal{M}_{m}(\C),\;B\in\mathcal{M}_{n}(\C).
\ee
Equivalently, $\tr_{\mathcal{M}_{m}(\C)}$ is the Hilbert--Schmidt dual of the $*$-homomorphism
$C\mapsto\mathds{1}_{m}\otimes C$. 
\end{notation}

\blem
\label{eq:extendingC}
Let $A\in\mathcal{M}_{m}(\C)\otimes\mathcal{M}_{n}(\C)$ with $A\ge0$ and $\tr_{\mathcal{M}_{m}(\C)}(A)\le\mathds{1}_{n}$. Then there exists a $B\in\mathcal{M}_{m}(\C)\otimes\mathcal{M}_{n}(\C)$ with $B\ge A$ and $\tr_{\mathcal{M}_{m}(\C)}(B)=\mathds{1}_{n}$. 
\elem

\bprf
Take 
\be
B:=A+\frac{1}{m}\Big(\mathds{1}_{m}\otimes\big(\mathds{1}_{n}-\tr_{\mathcal{M}_{m}(\C)}(A)\big)\Big).
\ee
Since $\mathds{1}_{n}-\tr_{\mathcal{M}_{m}(\C)}(A)\big)\ge0$ by assumption, $B\ge A$. Secondly, $\tr_{\mathcal{M}_{m}(\C)}(B)=\mathds{1}_{n}$. 
\eprf

\bt
\label{thm:Bayesianinversionmatrixalgebracase}[A Bayes' theorem for matrix algebras]
Let $F, \rho,\omega,\sigma,\xi$, and $P_{\xi}$ be as in Proposition~\ref{prop:qBayesianinverseformula}. 
Set
\be
\mathfrak{A}:=\sum_{i,j=1}^{m}E_{ij}^{(m)}\otimes\hat{\s}F^*\big(\rho E_{ij}^{(m)}\big)P_{\xi}
\quad\text{ and }\quad
\mathfrak{B}:=\sum_{i,j=1}^{m}E_{ij}^{(m)}\otimes\hat{\s}F^*\big(\rho E_{ij}^{(m)}\big)P_{\xi}^{\perp},
\ee
which are matrices in $\mathcal{M}_{m}(\C)\otimes\mathcal{M}_{n}(\C)$. 
Then $(F,\w)$ has a (CPU) Bayesian inverse if and only if
\be
\label{eq:theexistencecondition}
\mathfrak{A}^{\dag}=\mathfrak{A}
\qquad\text{ and }\qquad
\tr_{\mathcal{M}_{m}(\C)}\left(\mathfrak{B}^{\dag}\hat{\mathfrak{A}}\mathfrak{B}\right)\le P_{\xi}^{\perp}.
\ee 
If, in addition, $\tr_{\mathcal{M}_{m}(\C)}\left(\mathfrak{B}^{\dag}\hat{\mathfrak{A}}\mathfrak{B}\right)=P_{\xi}^{\perp}$, then the Bayesian inverse is unique (a-priori, it is only $\xi$-a.e.\ unique). 
\et

\bprf
{\color{white}{you found me!}}

\noindent
($\Rightarrow$)
Suppose that $(F,\w)$ has a Bayesian inverse $G$. Then by Proposition~\ref{prop:qBayesianinverseformula}
\be
\label{eq:ChoiholdsAB}
(\mathds{1}_{m}\otimes P_{\xi})\big(\Choi(G)\big)(\mathds{1}_{m}\otimes P_{\xi})=\mathfrak{A}
\quad\text{ and }\quad
(\mathds{1}_{m}\otimes P_{\xi})\big(\Choi(G)\big)(\mathds{1}_{m}\otimes P_{\xi}^{\perp})=\mathfrak{B}.
\ee
Since $G$ is CP, $\mathfrak{A}\ge0$, and therefore $\mathfrak{A}^{\dag}=\mathfrak{A}$. 
Now, set
\be
\mathfrak{C}:=(\mathds{1}_{m}\otimes P_{\xi}^{\perp})\big(\Choi(G)\big)(\mathds{1}_{m}\otimes P_{\xi}^{\perp}).
\ee
Since $G$ is CP,
\be
\Choi(G)=
\mathfrak{A}+\mathfrak{B}+\mathfrak{B}^{\dag}+\mathfrak{C}\ge0.
\ee
By Lemma~\ref{lem:positivitySchur}, $\mathfrak{C}\ge\mathfrak{B}^{\dag}\hat{\mathfrak{A}}\mathfrak{B}$. This combined with unitality of $G$ gives
\be
P_{\xi}^{\perp}=\tr_{\mathcal{M}_{m}(\C)}(\mathfrak{C})\ge\tr_{\mathcal{M}_{m}(\C)}(\mathfrak{B}^{\dag}\hat{\mathfrak{A}}\mathfrak{B}).
\ee

\noindent
($\Leftarrow$)
Conversely, suppose that~(\ref{eq:theexistencecondition}) holds. In what follows, we will construct a Bayesian inverse $G$. 
Since a linear map $G$ is determined by its Choi matrix, we will define $G$ by constructing its Choi matrix and we will use Lemma~\ref{lem:positivitySchur} to prove this Choi matrix is positive.  
Set 
\be
\label{eq:ChoiTLTR}
(\mathds{1}_{m}\otimes P_{\xi})\big(\Choi(G)\big)(\mathds{1}_{m}\otimes P_{\xi}):=\mathfrak{A}
,\quad
(\mathds{1}_{m}\otimes P_{\xi})\big(\Choi(G)\big)(\mathds{1}_{m}\otimes P_{\xi}^{\perp}):=\mathfrak{B},
\ee
\be
(\mathds{1}_{m}\otimes P_{\xi}^{\perp})\big(\Choi(G)\big)(\mathds{1}_{m}\otimes P_{\xi}):=\mathfrak{B}^{\dag}, 
\ee
and 
\be
\label{eq:definingCforChoi}
\mathfrak{C}\equiv(\mathds{1}_{m}\otimes P_{\xi}^{\perp})\big(\Choi(G)\big)(\mathds{1}_{m}\otimes P_{\xi}^{\perp}):=\mathfrak{B}^{\dag}\hat{\mathfrak{A}}\mathfrak{B}+\frac{1}{m}\Big(\mathds{1}_{m}\otimes\big(P_{\xi}^{\perp}-\tr_{\mathcal{M}_{m}(\C)}(\mathfrak{B}^{\dag}\hat{\mathfrak{A}}\mathfrak{B})\big)\Big).
\ee
Since the Choi matrix has now been specified, this defines a linear map $G$.
By~(\ref{eq:ChoiTLTR}) and Proposition~\ref{prop:qBayesianinverseformula}, $G$ is a Bayes map. 
Since $\mathfrak{A}^{\dag}=\mathfrak{A}$, Proposition~\ref{prop:selfadjointimpliesCP} implies $\mathfrak{A}\ge0$ because
\be
\mathrm{Ad}_{P_{\xi}}\circ G=(\mathds{1}_{m}\otimes P_{\xi})\big(\Choi(G)\big)(\mathds{1}_{m}\otimes P_{\xi})=\mathfrak{A}.
\ee
This proves the first condition in Lemma~\ref{lem:positivitySchur} (cf.\ Remark~\ref{rmk:positivitySchur}). To prove the second condition, namely $\ker(\mathfrak{A})\subseteq\ker(\mathfrak{B}^{\dag})$, choose a basis in which $\rho$ and $\sigma$ have been diagonalized as in the proof of (\ref{item:sa1}$\Rightarrow$\ref{item:sa6}) in Proposition~\ref{prop:selfadjointimpliesCP}. In addition, write $F=\sum_{\a}\mathrm{Ad}_{V_{\a}}$ and use the notation from the proof of (\ref{item:sa1}$\Rightarrow$\ref{item:sa6}) in Proposition~\ref{prop:selfadjointimpliesCP}. Now, by applying a suitable permutation matrix $U$ (cf.\ Remark~\ref{rmk:positivitySchur} and Appendix~\ref{appendix:permutation}), $\mathfrak{A}$ and $\mathfrak{B}$ can be written as
\be
U\mathfrak{A}U^{\dag}=\begin{bmatrix}\uline{\mathfrak{A}}&0\\0&0\end{bmatrix}
\qquad\text{ and }\qquad
U\mathfrak{B}U^{\dag}=\begin{bmatrix}0&\uline{\mathfrak{B}}\\0&0\end{bmatrix},
\ee
where
(recall that $\mathfrak{A}$ was computed earlier in (\ref{eq:ChoimatrixofBayesmatrixcasecompact}), but we use its form in (\ref{eq:adjointChoi}))
\be
\uline{\mathfrak{A}}
=
\sum_{\a}
\begin{bmatrix}
\begin{matrix}
\frac{p_{1}}{q_{1}}\overline{V^{\a}_{11}}V^{\a}_{11}&\cdots&\frac{p_{1}}{q_{r}}\overline{V^{\a}_{11}}V^{\a}_{1r}\\
\vdots&&\vdots\\
\frac{p_{1}}{q_{1}}\overline{V^{\a}_{1r}}V^{\a}_{11}&\cdots&\frac{p_{1}}{q_{r}}\overline{V^{\a}_{1r}}V^{\a}_{1r}\\
\end{matrix}
&\text{\scalebox{0.65}{$\bullet\bullet\bullet$}}&
\begin{matrix}
\frac{p_{m}}{q_{1}}\overline{V^{\a}_{11}}V^{\a}_{m1}&\cdots&\frac{p_{m}}{q_{r}}\overline{V^{\a}_{11}}V^{\a}_{mr}\\
\vdots&&\vdots\\
\frac{p_{m}}{q_{1}}\overline{V^{\a}_{1r}}V^{\a}_{m1}&\cdots&\frac{p_{m}}{q_{r}}\overline{V^{\a}_{1r}}V^{\a}_{mr}\\
\end{matrix}\\
{\Huge\vdots}&&{\Huge\vdots}\\
\begin{matrix}
\frac{p_{1}}{q_{1}}\overline{V^{\a}_{m1}}V^{\a}_{11}&\cdots&\frac{p_{1}}{q_{r}}\overline{V^{\a}_{m1}}V^{\a}_{1r}\\
\vdots&&\vdots\\
\frac{p_{1}}{q_{1}}\overline{V^{\a}_{mr}}V^{\a}_{11}&\cdots&\frac{p_{1}}{q_{r}}\overline{V^{\a}_{mr}}V^{\a}_{1r}\\
\end{matrix}
&\text{\scalebox{0.65}{$\bullet\bullet\bullet$}}&
\begin{matrix}
\frac{p_{m}}{q_{1}}\overline{V^{\a}_{m1}}V^{\a}_{m1}&\cdots&\frac{p_{m}}{q_{r}}\overline{V^{\a}_{m1}}V^{\a}_{mr}\\
\vdots&&\vdots\\
\frac{p_{m}}{q_{1}}\overline{V^{\a}_{mr}}V^{\a}_{m1}&\cdots&\frac{p_{m}}{q_{r}}\overline{V^{\a}_{mr}}V^{\a}_{mr}\\
\end{matrix}
\end{bmatrix}
\ee
and ($\mathfrak{B}$ is obtained by a calculation similar to~(\ref{eq:ChoimatrixofBayesmatrixcasecompact}), but with $\mathds{1}_{r}$ and $0_{n-r}$ replaced by $0_{r}$ and $\mathds{1}_{n-r}$, respectively, and then all $0$'s are dropped)
\be
\uline{\mathfrak{B}}=
\sum_{\a}
\begin{bmatrix}
\begin{matrix}
\frac{p_{1}}{q_{1}}\overline{V^{\a}_{11}}V^{\a}_{1,r+1}&\cdots&\frac{p_{1}}{q_{1}}\overline{V^{\a}_{11}}V^{\a}_{1n}\\
\vdots&&\vdots\\
\frac{p_{1}}{q_{r}}\overline{V^{\a}_{1r}}V^{\a}_{1,r+1}&\cdots&\frac{p_{1}}{q_{r}}\overline{V^{\a}_{1r}}V^{\a}_{1n}\\
\end{matrix}
&\text{\scalebox{0.65}{$\bullet\bullet\bullet$}}&
\begin{matrix}
\frac{p_{1}}{q_{1}}\overline{V^{\a}_{11}}V^{\a}_{m,r+1}&\cdots&\frac{p_{1}}{q_{1}}\overline{V^{\a}_{11}}V^{\a}_{mn}\\
\vdots&&\vdots\\
\frac{p_{1}}{q_{r}}\overline{V^{\a}_{1r}}V^{\a}_{m,r+1}&\cdots&\frac{p_{1}}{q_{r}}\overline{V^{\a}_{1r}}V^{\a}_{mn}\\
\end{matrix}\\
{\Huge\vdots}&&{\Huge\vdots}\\
\begin{matrix}
\frac{p_{m}}{q_{1}}\overline{V^{\a}_{m1}}V^{\a}_{1,r+1}&\cdots&\frac{p_{m}}{q_{1}}\overline{V^{\a}_{m1}}V^{\a}_{1n}\\
\vdots&&\vdots\\
\frac{p_{m}}{q_{r}}\overline{V^{\a}_{mr}}V^{\a}_{1,r+1}&\cdots&\frac{p_{m}}{q_{r}}\overline{V^{\a}_{mr}}V^{\a}_{1n}\\
\end{matrix}
&\text{\scalebox{0.65}{$\bullet\bullet\bullet$}}&
\begin{matrix}
\frac{p_{m}}{q_{1}}\overline{V^{\a}_{m1}}V^{\a}_{m,r+1}&\cdots&\frac{p_{m}}{q_{1}}\overline{V^{\a}_{m1}}V^{\a}_{mn}\\
\vdots&&\vdots\\
\frac{p_{m}}{q_{r}}\overline{V^{\a}_{mr}}V^{\a}_{m,r+1}&\cdots&\frac{p_{m}}{q_{r}}\overline{V^{\a}_{mr}}V^{\a}_{mn}\\
\end{matrix}
\end{bmatrix}
.
\ee
Now, suppose that the vector 
\be
w=\begin{bmatrix}w_{11}&\cdots&w_{1r}&\cdots&w_{m1}&\cdots&w_{mr}\end{bmatrix}^{T}
\ee
is in the kernel of $\uline{\mathfrak{A}}$. Setting $\W$ to be the number of Kraus operators used in $F=\sum_{\a}\mathrm{Ad}_{V_{\a}}$, let $\vec{u}\in\C^{\W}$ be the vector whose $\a$-th component is 
\be
u_{\a}:=\sum_{i=1}^{m}\sum_{k=1}^{r}\frac{p_{i}}{q_{k}}V_{ik}^{\a}w_{ik}.
\ee
Using this, 
\be
\uline{\mathfrak{A}}w=\sum_{\a}\begin{bmatrix}\sum_{i=1}^{m}\sum_{k=1}^{r}\frac{p_{i}}{q_{k}}\overline{V^{\a}_{11}}V^{\a}_{ik}w_{ik}\\
\vdots\\
\sum_{i=1}^{m}\sum_{k=1}^{r}\frac{p_{i}}{q_{k}}\overline{V^{\a}_{mr}}V^{\a}_{ik}w_{ik}
\end{bmatrix}
=\sum_{\a}\begin{bmatrix}\overline{V^{\a}_{11}}u_{\a}\\\vdots\\\overline{V^{\a}_{mr}}u_{\a}
\end{bmatrix}
\ee
shows that $w\in\ker(\uline{\mathfrak{A}})$ is equivalent to
\be
\big\<\vec{V}_{jl},\vec{u}\big\>=0\qquad\forall\;j\in\{1,\dots,m\},\;\;l\in\{1,\dots,r\},
\ee
where $\vec{V}_{jl}$ has $\a$-th component given by $V_{jl}^{\a}$ and we assume the inner product is conjugate-linear in the left entry. However, $\vec{u}$ is itself a linear combination of the $\vec{V}_{jl}$ since 
\be
\vec{u}=\sum_{i=1}^{m}\sum_{k=1}^{r}\frac{p_{i}}{q_{k}}w_{ik}\vec{V}_{ik}. 
\ee
Hence, $\vec{u}=0$ since it is orthogonal to, and in the span of, a collection of vectors. Now, computing $\uline{\mathfrak{B}}^{\dag}w$ gives 
\be
\uline{\mathfrak{B}}^{\dag}w=
\sum_{\a}
\begin{bmatrix}
\sum_{i=1}^{m}\sum_{k=1}^{r}\frac{p_{i}}{q_{k}}\ov{V_{1,r+1}^{\a}}V_{ik}^{\a}w_{ik}\\
\vdots\\
\sum_{i=1}^{m}\sum_{k=1}^{r}\frac{p_{i}}{q_{k}}\ov{V_{1n}^{\a}}V_{ik}^{\a}w_{ik}\\
\vdots\\
\sum_{i=1}^{m}\sum_{k=1}^{r}\frac{p_{i}}{q_{k}}\ov{V_{m,r+1}^{\a}}V_{ik}^{\a}w_{ik}\\
\vdots\\
\sum_{i=1}^{m}\sum_{k=1}^{r}\frac{p_{i}}{q_{k}}\ov{V_{mn}^{\a}}V_{ik}^{\a}w_{ik}\\
\end{bmatrix}
=
\begin{bmatrix}
\<\vec{V}_{1,r+1},\vec{u}\>\\
\vdots\\
\<\vec{V}_{1n},\vec{u}\>\\
\vdots\\
\<\vec{V}_{m,r+1},\vec{u}\>\\
\vdots\\
\<\vec{V}_{mn},\vec{u}\>\\
\end{bmatrix}
=0.
\ee
This proves that the first two conditions of Lemma~\ref{lem:positivitySchur} hold (cf.\ Remark~\ref{rmk:positivitySchur}).
By Lemma~\ref{eq:extendingC}, $\mathfrak{C}\ge\mathfrak{B}^{\dag}\mathfrak{A}\mathfrak{B}$, which implies $\Choi(G)\ge0$, and therefore corresponds to a CP map $G$. 

The constructed $G$ is also unital for the following reasons. First, recall that $G(\mathds{1}_{m})=\mathds{1}_{n}$ if and only if $\tr_{\mathcal{M}_{m}(\C)}\big(\Choi(G)\big)=\mathds{1}_{n}$ because 
\be
G(\mathds{1}_{m})
=\sum_{i=1}^{m}G\big(E_{ii}^{(m)}\big)
=\tr_{\mathcal{M}_{m}(\C)}\big(\Choi(G)\big).
\ee
Second, 
\be
\sum_{i=1}^{m}P_{\xi}\hat{\s}F^{*}\big(\rho E_{ii}^{(m)}\big)
=P_{\xi}\hat{\s}F^{*}(\rho)
=P_{\xi}\hat{\s}\s
=P_{\xi}
\quad\implies\quad
\sum_{i=1}^{m}P_{\xi}\hat{\s}F^{*}\big(\rho E_{ii}^{(m)}\big)P_{\xi}^{\perp}
=0.
\ee
Hence, $\mathfrak{B}$ provides zero contribution to $\tr_{\mathcal{M}_{m}(\C)}\big(\Choi(G)\big)$, i.e.\ $\tr_{\mathcal{M}_{m}(\C)}\big(\mathfrak{A}+\mathfrak{B}+\mathfrak{B}^{\dag}\big)=P_{\xi}$. 
Thus, 
\be
G(\mathds{1}_{n})
%=\sum_{i=1}^{m}G(E_{ii}^{(m)})
=\tr_{\mathcal{M}_{m}(\C)}\big(\Choi(G)\big)
=\tr_{\mathcal{M}_{m}(\C)}(\mathfrak{A})+\tr_{\mathcal{M}_{m}(\C)}(\mathfrak{C})
=P_{\xi}+P_{\xi}^{\perp}
=\mathds{1}_{n}. 
\ee
The equality $\tr_{\mathcal{M}_{m}(\C)}(\mathfrak{C})=P_{\xi}^{\perp}$ holds by the definition of $\mathfrak{C}$ in~(\ref{eq:definingCforChoi}) and Lemma~\ref{eq:extendingC}. Notice that if $\tr_{\mathcal{M}_{m}(\C)}\big(\mathfrak{B}^{\dag}\hat{\mathfrak{A}}\mathfrak{B}\big)=P_{\xi}^{\perp}$, then $\mathfrak{C}$ necessarily equals $\mathfrak{B}^{\dag}\hat{\mathfrak{A}}\mathfrak{B}$ and the Bayesian inverse is unique. 
\eprf

The second condition in Theorem~\ref{thm:Bayesianinversionmatrixalgebracase} is not automatically satisfied even if the first one holds. The following example illustrates a situation where a CP Bayesian inverse exists on the supported corner but not the full algebra. 

\bx
\label{ex:problemwithsupport}
Fix $q\in(0,1)$, 
\be
\s:=\begin{bmatrix}q&0&0\\0&1-q&0\\0&0&0\end{bmatrix}
,\;\;
\rho:=\begin{bmatrix}1&0\\0&0\end{bmatrix}
,\;\;
V_{1}:=\begin{bmatrix}\sqrt{q}&0&0\\0&0&\sqrt{1-q}\end{bmatrix}
,\;\;
V_{2}:=\begin{bmatrix}0&\sqrt{1-q}&0\\0&0&\sqrt{q}\end{bmatrix}
,
\ee
and let $\mathcal{M}_{3}(\C)\xstoch{F}\mathcal{M}_{2}(\C)$ be given by $F:=\mathrm{Ad}_{V_{1}}+\mathrm{Ad}_{V_{2}}$. If $\w:=\tr(\rho\;\cdot\;)$ and $\xi:=\tr(\s\;\cdot\;)$ then $\xi=\w\circ F$. Furthermore, $F$ is CPU. 
The Choi matrix of the Bayesian inverse, where we know how it is defined, is given by 
\be
\label{eq:Choicannotbecompleted}
\Choi(G)=\begin{bmatrix}1&0&0&0&0&\sqrt{\frac{1-q}{q}}\\0&1&0&0&0&\sqrt{\frac{q}{1-q}}\\0&0&\vf(E_{11})&0&0&\vf(E_{12})\\0&0&0&0&0&0\\0&0&0&0&0&0\\\sqrt{\frac{1-q}{q}}&\sqrt{\frac{q}{1-q}}&\vf(E_{21})&0&0&\vf(E_{22})\end{bmatrix}
\ee
where the $\vf$ terms correspond to $\sum_{i,j}E_{ij}^{(2)}\otimes P_{\xi}^{\perp}G\big(E_{ij}^{(2)}\big)P_{\xi}^{\perp}$ and are as of yet unknown. The first condition of Theorem~\ref{thm:Bayesianinversionmatrixalgebracase} is clearly satisfied. However, upon permuting this Choi matrix (as in Appendix~\ref{appendix:permutation}) and setting (these matrices were underlined in the proof of Theorem~\ref{thm:Bayesianinversionmatrixalgebracase}) 
\be
\mathfrak{A}:=\begin{bmatrix}1&0&0&0\\0&1&0&0\\0&0&0&0\\0&0&0&0\end{bmatrix}
\quad\text{ and }\quad
\mathfrak{B}:=\begin{bmatrix}0&\sqrt{\frac{1-q}{q}}\\0&\sqrt{\frac{q}{1-q}}\\0&0\\0&0\end{bmatrix},
\ee
we see that
\be
\mathfrak{B}^{\dag}\hat{\mathfrak{A}}\mathfrak{B}=\begin{bmatrix}0&0\\0&\frac{q}{1-q}+\frac{1-q}{q}\end{bmatrix}.
\ee
Taking the partial trace of this over $\mathcal{M}_{2}(\C)$ gives the bottom-right entry. Since $\frac{q}{1-q}+\frac{1-q}{q}\ge2>1$ for all $q\in(0,1)$, the Choi matrix $\Choi(G)$ above can never be completed to a positive matrix whose partial trace is $\mathds{1}_{2}$. 
\ex

%%%%%%%%%%%%%%%%%%%%%%%%%%%%%%%%%%%%%
\subsection{Special cases of Bayesian inversion}
\label{sec:specialcases}
%%%%%%%%%%%%%%%%%%%%%%%%%%%%%%%%%%%%%

Here, we analyze Kraus decompositions for Bayesian inverses in certain special cases. We also describe the relationship between joint distributions and conditionals. Finally, we analyze Bayesian inversion for wave collapse and reconstruct the disintegration formula for Bayesian inverses of deterministic maps. 
The issue presented in Example~\ref{ex:problemwithsupport} never occurs if at least one of $\rho$ or $\sigma$ is invertible, as the following corollary explains.

\bc
Let $\mA,\mB,F,\w,\rho,\xi,\s,$ and $P_{\xi}$ be as in Proposition~\ref{prop:qBayesianinverseformula}. If either $\rho$ or $\sigma$ is invertible, then a Bayesian inverse for $(F,\w)$ exists if and only if 
\be
\label{eq:commutingconditionforbayes}
P_{\xi} F^*(\rho A)\sigma=\sigma F^*(A\rho)P_{\xi}
\qquad\forall\;A\in\mathcal{M}_{m}(\C). 
\ee
When $\s$ is invertible and~(\ref{eq:commutingconditionforbayes}) holds, the Bayesian inverse equals
\be
\label{eq:Leifersformula}
G=\sum_{\a}\mathrm{Ad}_{\sqrt{\s^{-1}}V_{\a}^{\dag}\sqrt{\rho}}. 
\ee
When $\rho$ is invertible and~(\ref{eq:commutingconditionforbayes}) holds, a Bayesian inverse is of the form
\be
\label{eq:Bayeswhenrhoinvertible}
G=\sum_{\a}\mathrm{Ad}_{\sqrt{\hat{\s}}V_{\a}^{\dag}\sqrt{\rho}}+\mathrm{Ad}_{P_{\xi}^{\perp}}, 
\ee
where $P_{\xi}^{\perp}=\mathds{1}_{n}-P_{\xi}$. 
\ec

\bprf
When $\s$ is invertible, this is a special case of Corollary~\ref{cor:KrausdecompBayesmatrixcase}. 
If $\rho$ is invertible, then the matrix $\mathfrak{B}$ in Theorem~\ref{thm:Bayesianinversionmatrixalgebracase} vanishes due to the condition $\s=F^*(\rho)$. In more detail, suppose that $\rho$ and $\s$ have been diagonalized (the vanishing of $\mathfrak{B}$ will not depend on such a diagonalization). Using the notation of the proof of (\ref{item:sa1}$\Rightarrow$\ref{item:sa6}) in Proposition~\ref{prop:selfadjointimpliesCP} by writing $F=\sum_{\a}\mathrm{Ad}_{V_{\a}}$, the condition $\s=F^*(\rho)$ can be expressed as 
\be
\hspace{-1mm}
\begin{bmatrix}
q_{1}&&0&0&\cdots&0\\
&\ddots&&\vdots&&\vdots\\
0&&q_{r}&0&\cdots&0\\
0&\cdots&0&0&\cdots&0\\
\vdots&&\vdots&\vdots&&\vdots\\
0&\cdots&0&0&\cdots&0\\
\end{bmatrix}
\!=\!
\sum_{\a}\!\sum_{i=1}^{m}p_{i}\!
\begin{bmatrix}
|V^{\a}_{i1}|^2&\cdots&\overline{V^{\a}_{i1}}V^{\a}_{ir}&\overline{V^{\a}_{i1}}V^{\a}_{i,r+1}&\cdots&\overline{V^{\a}_{i1}}V^{\a}_{in}\\
\vdots&&\vdots&\vdots&&\vdots\\
\overline{V^{\a}_{ir}}V^{\a}_{i1}&\cdots&|V^{\a}_{ir}|^{2}&\overline{V^{\a}_{ir}}V_{i,r+1}&\cdots&\overline{V^{\a}_{ir}}V^{\a}_{in}\\
\overline{V^{\a}_{i,r+1}}V^{\a}_{i1}&\cdots&\overline{V^{\a}_{i,r+1}}V^{\a}_{ir}&|V^{\a}_{i,r+1}|^2&\cdots&\overline{V^{\a}_{i,r+1}}V^{\a}_{in}\\
\vdots&&\vdots&\vdots&&\vdots\\
\overline{V^{\a}_{in}}V^{\a}_{i1}&\cdots&|V^{\a}_{in}|^{2}&\overline{V^{\a}_{in}}V^{\a}_{i,r+1}&\cdots&|V^{\a}_{in}|^2\\
\end{bmatrix}
\!\!\!
\ee
Because $p_{i}>0$ for all $i$ by assumption, this implies
\be
V_{ik}^{\a}=0
\qquad\forall\;k\in\{r+1,\dots,n\},
\;\;i\in\{1,\dots,m\},
\;\;\a\in\{1,\dots,\Omega\}. 
\ee
Since $\mathfrak{B}$ is defined only in terms of these components of $V^{\a}$, the matrix $\mathfrak{B}$ vanishes. Hence, both conditions in Theorem~\ref{thm:Bayesianinversionmatrixalgebracase} hold. The fact that $\mathfrak{B}$ vanishes means that $P_{\xi}G(A)P_{\xi}^{\perp}=0$ and  $P_{\xi}^{\perp}G(A)P_{\xi}=0$ for all $A\in\mM_{m}(\C)$. Therefore, $\mathfrak{C}=\frac{1}{m}\mathds{1}_{m}\otimes P_{\xi}^{\perp}$ by~(\ref{eq:definingCforChoi}), which, upon going from the Choi matrix to the corresponding linear map, reproduces~(\ref{eq:Bayeswhenrhoinvertible}). 
\eprf

\br
\label{rmk:whyBayesquantum}
Remark~\ref{rmk:whyBayesclassical} sets the background for the following discussion. 
The formulation of Theorem~\ref{thm:Bayesianinversionmatrixalgebracase} describes how to go from conditional to conditional bypassing the intermediate stage of constructing a joint distribution. Unlike in the classical setting, the joint state on $\mA\otimes\mB$ from the Bayes condition will rarely be a state in quantum theory. Even for the simplest case where $\mA=\mB=\mathcal{M}_{n}(\C)$ (with $n>1$), $F:=\id$, and an arbitrary state $\w:\mA\stoch\C$, one cannot construct a joint state via $\omega\circ\mu_{\mA}\circ(\id_{\mA}\otimes F)$, even though a Bayesian inverse clearly exists (it is $G=\id$). The reason is due to the non-commutativity of multiplication. In order for 
\be
\label{eq:AFBfunctional}
\mA\otimes\mB\ni A\otimes B\mapsto\w\big(AF(B)\big)
\ee
to be a state, one needs additional restrictions which are not generally equivalent to a Bayesian inverse existing. For example, if $\w$ is tracial, then the functional (\ref{eq:AFBfunctional}) is indeed a state (this is true even for arbitrary positive unital $F$). The proof of this is relatively straightforward because
\be
\w\big(a^*a F(b^*b)\big)=\w(aF(b^*b)a^*)=(\omega\circ\mathrm{Ad}_{a}\circ F)(b^*b)\ge0. 
\ee
Another sufficient condition for (\ref{eq:AFBfunctional}) to be a state is that the image of $F$ lands in the commutant $\mA'\subseteq\mA$. This is also a drastic condition because if $\mA=\mathcal{M}_{n}(\C)$, then the commutant consists only of scalar multiples of the identity. This condition is only sufficient but not necessary. There are intermediate cases where $\w$ is neither tracial and $F$ does not land in the commutant $\mA'\subseteq\mA$. When $\mA$ and $\mB$ are matrix factors and $\w=\tr(\rho\;\cdot\;)$, then Proposition~\ref{thm:jointstate} below covers these cases simultaneously. 
\er

\bn
\label{thm:jointstate}
Set $\mA:=\mathcal{M}_{m}(\C)$ and $\mB:=\mathcal{M}_{n}(\C)$. Fix a state $\mA\xstoch{\w=\tr(\rho\;\cdot\;)}\C$ and a PU map $F:\mB\stoch\mA$. If $\rho\in F(\mathcal{B})'\subseteq\mA$, then the linear functional uniquely determined by the assignment
$
\mA\otimes\mB\ni A\otimes B\mapsto\w\big(AF(B)\big)
$
is a state.
\en

\bprf
Let $A=a^*a\in\mA$ and $B\in\mB$ be positive. 
Since $F$ is positive, $F(B)$ is positive. 
Since $\mA$ is finite dimensional, 
the algebra generated by $F(\mathcal{B})$ is equal to $F(\mathcal{B})''\subseteq\mA$ by the double commutant theorem (cf.~\cite[Section~2]{To71}). 
Hence, by the functional calculus, 
the positive square root $c:=\sqrt{F(B)}$ is in $\mathrm{Alg}\big(F(\mB)\big)$, and therefore $c\in F(\mB)''$. Therefore, 
\be
\tr\big(\rho A F(B)\big)
=\tr(\rho a^*a c^* c)
=\tr(c\rho a^*a c^*)
=\tr(\rho ca^*ac^*)
=\w\big( (ac^*)^*(ac^*)\big)
\ge0.
\ee
Unitality of $\omega$ follows from unitality of $F$. 
\eprf

We are now in a position to prove Proposition~\ref{thm:wavecollapse} for wave collapse, which we will accomplish by illustrating more explicitly how one can obtain Bayesian inverses via matrix completions. However, to do so, and to also help us prove several other results of interest, we state a lemma that will allow us to compute pseudoinverses of Choi matrices. 

\blem
\label{lem:pseudoinverseofChoi}
Let $\mathcal{M}_{n}(\C)\xstoch{F}\mathcal{M}_{m}(\C)$ be a CP map. Then there exist $\Omega$-many Kraus operators $\{V_{\a}\}$ for $F$ and strictly positive numbers $\{\L_{\a}\}$ such that 
\be
\label{eq:orthogonalkrausdecomp}
\tr(V_{\a}^{\dag}V_{\b})=\L_{\a}\de_{\a\b}\qquad\forall\;\a,\b\in\{1,\dots,\W\}. 
\ee
Furthermore, in terms of such a set of Kraus operators for $F=\sum_{\a}\mathrm{Ad}_{V_{\a}}$, 
\be
\mathrm{support}\big(\Choi(F)\big)=\Choi\left(\sum_{\a}\mathrm{Ad}_{\frac{1}{\sqrt{\L_{\a}}}V_{\a}}\right)
\quad\text{ and }\quad
\reallywidehat{\Choi(F)}=\Choi\left(\sum_{\a}\mathrm{Ad}_{\frac{1}{\L_{\a}}V_{\a}}\right).
\ee
\elem

\bprf
The fact that Kraus operators satisfying~(\ref{eq:orthogonalkrausdecomp}) can be found for any CP map is a simple consequence of the existence of an orthogonal basis of eigenvectors of a positive matrix. Although this is a standard result, its proof provides useful techniques that also prove the remaining part of the lemma. Indeed, since the Choi matrix is positive, it can be expressed as $\Choi(F)=\sum_{\a}v_{\a}v_{\a}^{\dag}$, where the set of $v_{\a}\in\C^{n}\otimes\C^{m}\cong\C^{mn}$ is orthogonal (and each $v_{\a}$ can be assumed to be non-zero). One then sets $\L_{\a}:=v_{\a}^{\dag}v_{\a}$ and $V_{\a}$ to be the unique matrix satisfying
\be
v_{\a}=\begin{bmatrix}V_{\a}e_{1}\\\vdots\\V_{\a}e_{n}\end{bmatrix},
\ee
where $\{e_{1},\dots,e_{n}\}$ is the standard basis for $\C^{n}$. 
Orthogonality of these vectors is then precisely the statement
\be
\L_{\a}\de_{\a\b}=\<v_{\a},v_{\b}\>
=\sum_{j=1}^{n}(V_{\a}e_{j})^{\dag}V_{\b}e_{j}
=\sum_{j=1}^{n}\<e_{j},V_{\a}^{\dag}V_{\b}e_{j}\>
=\tr(V_{\a}^{\dag}V_{\b}). 
\ee
Therefore, one can set $u_{\a}:=\L_{\a}^{-1/2}v_{\a}$. With this definition, the set $\{u_{\a}\}$ is an orthonormal set of vectors in $\C^{mn}$. Hence, $\sum_{\a}u_{\a}u_{\a}^{\dag}$ is an orthogonal projection. Thus, 
\be
\Choi\left(\sum_{\a}\mathrm{Ad}_{\frac{1}{\sqrt{\L_{\a}}}V_{\a}}\right)
=\sum_{i,j}E_{ij}^{(n)}\otimes\sum_{\a}\frac{1}{\L_{\a}}(V_{\a}e_{i})(V_{\a}e_{j})^{\dag}
=\sum_{\a}\frac{1}{\L_{\a}}v_{\a}v_{\a}^{\dag}=\sum_{\a}u_{\a}u_{\a}^{\dag}
\ee
is indeed a projection, and in fact the support of $\Choi(F)$. 
Finally, 
\be
\Choi(F)\Choi\left(\sum_{\a}\mathrm{Ad}_{\frac{1}{\L_{\a}}V_{\a}}\right)
=\left(\sum_{\a}\L_{\a}u_{\a}u_{\a}^{\dag}\right)\left(\sum_{\b}\frac{1}{\L_{\b}}u_{\b}u_{\b}^{\dag}\right)
=\sum_{\a}u_{\a}u_{\a}^{\dag},
\ee
which coincides with the support of $\Choi(F)$.
\eprf

\bd
\label{defn:orthogonalkraus}
A Kraus decomposition satisfying the conditions of Lemma~\ref{lem:pseudoinverseofChoi} will be called an \define{orthogonal Kraus decomposition}. 
\ed

\bprf
[Proof of Proposition~\ref{thm:wavecollapse}]
We briefly recall some notation and set up some basic facts. Let $F=\sum_{\l}\mathrm{Ad}_{P_{\l}}$ with $\{P_{\l}\}$ a family of orthogonal projections such that $\sum_{\l}P_{\l}=\mathds{1}_{m}$ (here, the projections are associated with the eigenspaces of a self-adjoint matrix $H$ whose spectrum is denoted by $\s(H)$). Let $\s:=\sum_{\l}P_{\l}\rho P_{\l}$ with $\rho$ a density matrix and let $P_{\xi}$ denote the support of $\s$, $S_{\l}$ the support of $P_{\l}\rho P_{\l}$, and set
\be
\L_{\l}:=\tr(S_{\l})\equiv\mathrm{rank}\left(P_{\l}\rho P_{\l}\right). 
\ee
Note that $S_{\l}S_{\mu}=\de_{\l\mu}S_{\l}$ and $P_{\xi}=\sum_{\l}S_{\l}$ because  $P_{\l}P_{\mu}=\de_{\l\mu}P_{\l}$ and $S_{\l}\le P_{\l}$. 
We will frequently make use of identities such as 
\be
\label{eq:wavecollapsePSidentities}
P_{\l}\s=S_{\l}\s=S_{\l}\s S_{\l}=P_{\l}\s P_{\l}=\s S_{\l}=\s P_{\l}
\ee
and so on (and similarly for functions of $\s$ by the functional calculus). 
As a result of the orthogonality of the $P_{\l}$, 
\be
\label{eq:wavecollapsePShatidentities}
\hat{\s}=%\widehat{\sum_{\l}P_{\l}\rho P_{\l}}
\reallywidehat{\sum_{\l}P_{\l}\rho P_{\l}}
=\sum_{\l}\widehat{P_{\l}\rho P_{\l}}%=\sum_{\l}\widehat{\s_{\l}}.
\qquad\text{ and therefore }\qquad
S_{\l}\hat{\s}=P_{\l}\hat{\s}=\widehat{P_{\l}\rho P_{\l}}=\hat{\s}P_{\l}=\hat{\s}S_{\l}.
\ee

We now move on to the proof and first suppose that $\rho=\s$. Then it is straightforward to check $G=\sum_{\l}\mathrm{Ad}_{P_{\l}}$ is a Bayesian inverse of $(F,\w)$. 

Conversely, suppose that $(F,\w)$ has a Bayesian inverse $G$. We will break the proof into three parts. In the first part, we derive some relationships that will be useful in the latter two parts. In the second part, we compute the pseudo-inverse of the Choi matrix of $\mathrm{Ad}_{P_{\xi}}\circ G$. In the third part, we prove the conditions of Theorem~\ref{thm:Bayesianinversionmatrixalgebracase} are satisfied and explicitly find the Bayesian inverse to conclude the proof.

\begin{enumerate}[i.]
\itemsep0pt
\item 
First, 
\be
\label{eq:selfadjointforwavecollapse}
\xy0;/r.34pc/:
(0,11)*+{P_{\xi}G(A)P_{\xi}}="T";
(-24,6)*+{\sum_{\l}P_{\xi}\hat{\s}P_{\l}\rho AP_{\l}P_{\xi}}="L1";
(24,6)*+{\sum_{\l}P_{\xi}P_{\l}A\rho P_{\l}\hat{\s}P_{\xi}}="R1";
(-34,-3)*+{\sum_{\l,\mu,\nu}S_{\mu}\hat{\s}P_{\l}\rho AP_{\l}S_{\nu}}="L2";
(34,-3)*+{\sum_{\l,\mu,\nu}S_{\mu}P_{\l}A\rho P_{\l}\hat{\s}S_{\nu}}="R2";
(-29,-12)*+{\sum_{\l}P_{\l}\hat{\s}\rho AP_{\l}}="L3";
(29,-12)*+{\sum_{\l}P_{\l}A\rho\hat{\s}P_{\l}}="R3";
(0,0)*+{\sum_{\l}\mathrm{Ad}_{\sqrt{\hat{\s}}P_{\l}\sqrt{\rho}}(A)}="0";
(0,-12)*+{\sum_{\l}\mathrm{Ad}_{P_{\l}\sqrt{\hat{\s}}\sqrt{\rho}}(A)}="B";
{\ar@{=}@/^0.65pc/"T";"R1"^(0.7){\text{Prop~\ref{prop:selfadjointimpliesCP}}}};
{\ar@{=}@/^0.45pc/"R1";"R2"};
{\ar@{=}@/^0.55pc/"R2";"R3"};
{\ar@{=}@/_0.65pc/"T";"L1"_(0.6){\text{(\ref{eq:qBayesianinversepartialsupport})}}};
{\ar@{=}@/_0.45pc/"L1";"L2"};
{\ar@{=}@/_0.55pc/"L2";"L3"};
{\ar@{=}"T";"0"^{\text{Cor~\ref{cor:KrausdecompBayesmatrixcase}}}};
{\ar@{=}"0";"B"^{\text{(\ref{eq:wavecollapsePShatidentities})}}};
\endxy
,
\ee
which implies 
\be
\label{eq:wavecollapseusefulbreakup}
P_{\l}\hat{\s}\rho AP_{\l}
%=P_{\l}\sqrt{\hat{\s}}\sqrt{\rho}A\sqrt{\rho}\sqrt{\hat{\s}}P_{\l}
=\mathrm{Ad}_{P_{\l}\sqrt{\hat{\s}}\sqrt{\rho}}(A)
=P_{\l}A\rho\hat{\s}P_{\l}\qquad\forall\;A\in\mathcal{M}_{m}(\C),\;\;\l\in\s(H).
\ee
Taking the trace of the two ends of this for a fixed $\l$ and using cyclicity of trace gives
\be
\tr(P_{\l}\hat{\s}\rho A)=\tr(\rho\hat{\s}P_{\l}A)\qquad\forall\;A\in\mathcal{M}_{m}(\C)
\quad\implies\quad
P_{\l}\hat{\s}\rho=\rho\hat{\s}P_{\l}. 
\ee
Multiplying on the left by $P_{\mu}$ with $\mu\ne\l$ and multiplying on the right by $\s$ gives
\be
\label{eq:wavecollapseassumption}
0=P_{\mu}\rho\hat{\s}P_{\l}\s=P_{\mu}\rho S_{\l}
\qquad\forall\;(\mu,\l)\in\big(\s(H)\times\s(H)\big)\setminus\Delta\big(\s(H)\big).
\ee

\item
The pseudoinverse of $\mathfrak{A}:=\Choi(\mathrm{Ad}_{P_{\xi}}\circ G)$ is given by
\be
\label{eq:pseudoinverseChoiwavecollapse}
\hat{\mathfrak{A}}=%\Choi\left(\sum_{\l}\frac{1}{\left(\sum_{k}\lVert P_{\l}\sqrt{\hat{\s}}\sqrt{\rho}e_{k}\rVert^{2}\right)^{2}}\mathrm{Ad}_{\L_{\l}^{-1}P_{\l}\sqrt{\hat{\s}}\sqrt{\rho}}\right)
\Choi\left(\sum_{\l}\mathrm{Ad}_{\L_{\l}^{-1}P_{\l}\sqrt{\hat{\s}}\sqrt{\rho}}\right)
\ee
by Lemma~\ref{lem:pseudoinverseofChoi}, which holds since the Kraus decomposition along the middle list of equalities in ~(\ref{eq:selfadjointforwavecollapse}) is orthogonal, and since 
\be
\label{eq:wavecollapserankidentity}
\sum_{k}\left\lVert P_{\l}\sqrt{\hat{\s}}\sqrt{\rho}e_{k}\right\rVert^{2}
=\tr\left(\sqrt{\rho}\sqrt{\hat{\s}}P_{\l}P_{\l}\sqrt{\hat{\s}}\sqrt{\rho}\right)
=\tr(P_{\l}\hat{\s}P_{\l}\rho P_{\l})
=\tr(S_{\l})
=\L_{\l}.
\ee

\item
Set
\be
\mathfrak{B}:=(\mathds{1}_{m}\otimes P_{\xi})\left(\sum_{\l}\sum_{i,j}E_{ij}\otimes\hat{\s} P_{\l}\rho E_{ij}P_{\l}\right)(\mathds{1}_{m}\otimes P_{\xi}^{\perp}).
\ee 
The following calculation of $\mathfrak{B}^{\dag}\hat{\mathfrak{A}}\mathfrak{B}$ will be done in two steps. First, (dummy indices are frequently relabelled)
\be
\begin{split}
\hat{\mathfrak{A}}\mathfrak{B}
&\overset{\text{(\ref{eq:wavecollapseusefulbreakup})~\&~(\ref{eq:pseudoinverseChoiwavecollapse})}}{=\joinrel=\joinrel=\joinrel=\joinrel=\joinrel=\joinrel=\joinrel=\joinrel=}
\left(\sum_{i,j,k,l}\sum_{\l,\mu}\underbrace{E_{ij}E_{kl}}_{\de_{jk}E_{il}}\otimes\frac{1}{\L_{\l}^{2}}P_{\l}\hat{\s}\rho E_{ij}\!\!\!\underbrace{P_{\l}\hat{\s}P_{\mu}}_{\de_{\l\mu}P_{\l}\hat{\s}P_{\l}}\!\!\!\rho E_{kl}P_{\mu}\right)\left(\mathds{1}_{m}\otimes P_{\xi}^{\perp}\right)\\
&=\left(\sum_{i,j}\sum_{\l}E_{ij}\otimes\frac{\tr(P_{\l}\hat{\s}P_{\l}\rho)}{\L_{\l}^{2}}P_{\l}\hat{\s}\rho E_{ij}P_{\l}\right)\left(\mathds{1}_{m}\otimes P_{\xi}^{\perp}\right)\\
&\overset{\text{(\ref{eq:wavecollapserankidentity})}}{=\joinrel=\joinrel=}
\left(\sum_{i,j}\sum_{\l}E_{ij}\otimes\frac{1}{\L_{\l}}P_{\l}\hat{\s}\rho E_{ij}P_{\l}\right)\left(\mathds{1}_{m}\otimes P_{\xi}^{\perp}\right).
\end{split}
\ee
Multiplying $\mathfrak{B}^{\dag}$ on the left gives
\be
\begin{split}
\mathfrak{B}^{\dag}\hat{\mathfrak{A}}\mathfrak{B}&=\left(\mathds{1}_{m}\otimes P_{\xi}^{\perp}\right)\left(\sum_{i,j,k,l}\sum_{\l,\mu}E_{ij}E_{kl}\otimes\frac{1}{\L_{\mu}}P_{\l}E_{ij}\rho P_{\l}\hat{\s}P_{\mu}\hat{\s}\rho E_{kl}P_{\mu}\right)\left(\mathds{1}_{m}\otimes P_{\xi}^{\perp}\right)\\
&=\left(\mathds{1}_{m}\otimes P_{\xi}^{\perp}\right)\left(\sum_{i,j}\sum_{\l}E_{ij}\otimes\frac{\tr(\rho P_{\l}\hat{\s}P_{\l}\hat{\s}\rho)}{\L_{\l}}P_{\l}E_{ij}P_{\l}\right)\left(\mathds{1}_{m}\otimes P_{\xi}^{\perp}\right).
\end{split}
\ee
Now, notice that 
\be
\begin{split}
\tr(\rho P_{\l}\hat{\s}P_{\l}\hat{\s}\rho)
&=\sum_{\mu}\tr(P_{\mu}\rho P_{\l}\hat{\s}P_{\l}\hat{\s}\rho P_{\mu})
=\sum_{\mu}\tr(P_{\mu}\rho S_{\l}\hat{\s}S_{\l}\hat{\s}S_{\l}\rho P_{\mu})\\
&\overset{\text{(\ref{eq:wavecollapseassumption})}}{=\joinrel=\joinrel=}
\tr(P_{\l}\rho P_{\l}\hat{\s}P_{\l}\hat{\s}\rho P_{\l})
=\tr(S_{\l}^{2})
=\tr(S_{\l})
=\L_{\l},
\end{split}
\ee
where the first equation holds by cyclicity of trace and because $\sum_{\mu}P_{\mu}=\mathds{1}_{m}$. 
Hence, 
\be
\mathfrak{B}^{\dag}\hat{\mathfrak{A}}\mathfrak{B}
=\left(\mathds{1}_{m}\otimes P_{\xi}^{\perp}\right)\left(\sum_{i,j}E_{ij}\otimes\sum_{\l}P_{\l}E_{ij}P_{\l}\right)\left(\mathds{1}_{m}\otimes P_{\xi}^{\perp}\right)
=\Choi\big(\mathrm{Ad}_{P_{\xi}^{\perp}}\circ F\big).
\ee
Taking the partial trace of $\mathfrak{B}^{\dag}\hat{\mathfrak{A}}\mathfrak{B}$ gives
\be
\tr_{\mathcal{M}_{m}(\C)}\left(\mathfrak{B}^{\dag}\hat{\mathfrak{A}}\mathfrak{B}\right)
=P_{\xi}^{\perp}\sum_{i,j}\sum_{\l}\underbrace{\tr(E_{ij})}_{\de_{ij}}P_{\l}E_{ij}P_{\l}P_{\xi}^{\perp}
=P_{\xi}^{\perp}\sum_{\l}P_{\l}\sum_{i}E_{ii}P_{\l}P_{\xi}^{\perp}
=P_{\xi}^{\perp}. 
\ee
Hence, the last condition of Theorem~\ref{thm:Bayesianinversionmatrixalgebracase} is satisfied, and so the Bayesian inverse of the wave collapse is uniquely specified. In fact, the Bayesian inverse can be computed explicitly since $\Choi(G)$ is now completely known. But first, it helps to simplify the expression by using Lemma~\ref{lem:positivitySchur} to factor the Choi matrix of $G$. For this, note that 
\be
\hat{\mathfrak{A}}^{1/2}=\Choi\left(\sum_{\l}\mathrm{Ad}_{\L_{\l}^{-3/4}P_{\l}\sqrt{\hat{\s}}\sqrt{\rho}}\right).
\ee
Then, one can show 
\be
\hat{\mathfrak{A}}^{1/2}+\mathfrak{B}^{\dag}\hat{\mathfrak{A}}^{1/2}=\sum_{i,j}E_{ij}\otimes\sum_{\l}\frac{1}{\L_{\l}^{1/2}}P_{\l}E_{ij}\rho\hat{\s} P_{\l}
\ee
and  
\be
\Choi(G)=\left(\hat{\mathfrak{A}}^{1/2}+\mathfrak{B}^{\dag}\hat{\mathfrak{A}}^{1/2}\right)\left(\hat{\mathfrak{A}}^{1/2}+\mathfrak{B}^{\dag}\hat{\mathfrak{A}}^{1/2}\right)^{\dag}=\Choi(F)
\ee
using similar calculations to the ones that have already been done. This proves that the Bayesian inverse is $G=F$.
\end{enumerate} 
Hence, $\rho=G^*(\s)=\sum_{\l}P_{\l}\rho P_{\l}=\s$. 
%\AJP{all of this work for such a simple result!?}
\eprf

In the following example, we will work out Bayesian inverses for $^*$-homomorphisms $\mB\xrightarrow{F}\mA$ where $\mA$ and $\mB$ are matrix algebras. Although we know from~\cite{PaBayes} that Bayesian inverses of $^*$-homomorphisms are automatically disintegrations, and we already know the solution to the disintegration problem~\cite{PaRu19}, it is helpful to explicitly work out the solution from our matrix completion perspective. 

\bx
Every $^*$-homomorphism of matrix algebras is of the form 
\be
\begin{split}
\mB:=\mathcal{M}_{n}(\C)&\xrightarrow{F}\mathcal{M}_{p}(\C)\otimes\mathcal{M}_{n}(\C)=:\mA\\
B&\mapsto\mathds{1}_{p}\otimes B
\end{split}
\ee
up to a unitary conjugation on the codomain of $F$ (cf.\ Lecture~10 in~\cite{Werner17}). Let $\mB\xstoch{\w=\tr(\rho\;\cdot\;)}\C$ be a state on $\mB$ and write $\xi=\tr(\rho\;\cdot\;):=\w\circ F$. Note that $F^*=\tr_{\mathcal{M}_{p}(\C)}$ in this case. An arbitrary density matrix $\rho$ can be expressed as a linear combination 
\be
\rho=\sum_{\b}\t_{\b}\otimes\s_{\b},
\ee
where $\t_{\b}\in\mathcal{M}_{p}(\C)$, $\s_{\b}\in\mathcal{M}_{n}(\C)$, and such that $\{\t_{\b}\}$ is a linearly independent set of matrices (coefficients can been absorbed into the $\s_{\b}$ matrices).%
%footnote
\footnote{
One can take $\{\t_{\b}\}$ to be $\{E_{ij}^{(p)}\}$, which we will later. In this case, $\s_{ij}$ is precisely the $ij$-th block of $\rho$.
}
%end footnote
The fact that $\rho$ is a density matrix means $\rho\ge0$ and $\sum_{\b}\tr(\t_{\b})\tr(\s_{\b})=1$.  
The fact that $F^*(\rho)=\s$ entails 
\be
\s=\tr_{\mathcal{M}_{p}(\C)}\left(\sum_{\b}\t_{\b}\otimes\s_{\b}\right)
=\sum_{\b}\tr(\t_{\b})\s_{\b}.
\ee
By Proposition~\ref{prop:qBayesianinverseformula}, a Bayesian inverse $G$ of $(F,\w)$ must satisfy 
\be
\label{eq:bayesstarhomo}
P_{\xi}G(C\otimes B)=\hat{\s}\tr_{\mathcal{M}_{p}(\C)}\left(\sum_{\b}(\t_{\b}\otimes\s_{\b})(C\otimes B)\right)
=\sum_{\b}\tr(\t_{\b}C)\hat{\s}\s_{\b}B
\ee
for all $C\in\mathcal{M}_{p}(\C)$ and $B\in\mathcal{M}_{n}(\C).$
Hence, the Choi matrix of $\mathrm{Ad}_{P_{\xi}}\circ G$, with $G$ a Bayesian inverse of $F$, must be of the form 
\be
\begin{split}
\Choi\big(\mathrm{Ad}_{P_{\xi}}\circ G\big)
&=\sum_{i,j,k,l}E_{ij}^{(p)}\otimes E_{kl}^{(n)}\otimes\left(\sum_{\b}\tr\big(\t_{\b}E_{ij}^{(p)}\big)\hat{\s}\s_{\b}E_{kl}^{(n)}P_{\xi}\right)\\
&=\sum_{\b}\sum_{k,l}\t_{\b}^{T}\otimes E_{kl}^{(n)}\otimes\hat{\s}\s_{\b}E_{kl}^{(n)}P_{\xi}
\end{split}
\ee
Self-adjointness of $\mathrm{Ad}_{P_{\xi}}\circ G$ entails 
\be
\sum_{\b}\sum_{k,l}\t_{\b}^{T}\otimes E_{kl}^{(n)}\otimes\Big(\hat{\s}\s_{\b}E_{kl}^{(n)}P_{\xi}-P_{\xi}E_{kl}^{(n)}\s_{\b}\hat{\s}\Big)=0.
\ee
Since $\t_{\b}^{T}\otimes E_{kl}^{(n)}$ are linearly independent for all $\b,k,$ and $l$, we obtain 
\be
\hat{\s}\s_{\b}BP_{\xi}=P_{\xi}B\s_{\b}\hat{\s}
\qquad\forall\;B\in\mathcal{M}_{n}(\C)
\;\text{ and }\;\forall\;\b.
\ee
Taking the trace of this expression and taking the trace after multiplying both sides by $\s$ gives
\be
\label{eq:ssbcommute}
[\hat{\s},\s_{\b}]=0
\qquad\text{ and }\qquad
[\s,\s_{\b}]=0
\qquad\forall\;\b, 
\ee
respectively. Now, from Equation~(\ref{eq:bayesstarhomo}) and Proposition~\ref{prop:selfadjointimpliesCP}, we obtain 
\be
\sum_{\b}\tr(\t_{\b}C)\hat{\s}\s_{\b}BP_{\xi}=\sum_{\b}\tr(\t_{\b}C)P_{\xi}B\s_{\b}\hat{\s}
\qquad\forall\;C\in\mathcal{M}_{p}(\C),\;B\in\mathcal{M}_{n}(\C).
\ee
Since $\hat{\s}$, $\s_{\b}$, and $P_{\xi}$ all commute by~(\ref{eq:ssbcommute}), this implies
\be
\left[\sum_{\b}\tr(\t_{\b}C)\s_{\b}\hat{\s},P_{\xi}BP_{\xi}\right]=0
\qquad\forall\;B\in\mathcal{M}_{n}(\C),
\ee
which says $\sum_{\b}\tr(\t_{\b}C)\s_{\b}\hat{\s}$ is in the commutant of $P_{\xi}\mathcal{M}_{n}(\C)P_{\xi}$ inside $\mathcal{M}_{n}(\C)$. Since $\sum_{\b}\tr(\t_{\b}C)\s_{\b}\hat{\s}$ is also in $P_{\xi}\mathcal{M}_{n}(\C)P_{\xi}$, this gives 
\be
\sum_{\b}\tr(\t_{\b}C)\s_{\b}\hat{\s}\propto P_{\xi}
\qquad\forall\;C\in\mathcal{M}_{p}(\C), 
\ee
where $\propto$ means ``proportional to.'' At this point, it is convenient to specialize to a particular choice of linearly independent set $\{\t_{\b}\}$ by choosing the elementary matrices $\{E_{ij}^{(p)}\}$. The proportionality constraint entails $\s_{ij}\hat{\s}\propto P_{\xi}$ for all $i,j$ (choose $C=E^{(p)}_{ij}$). Multiplying on the right by $\s$ gives $\s_{ij}P_{\xi}\propto\s$. Taking the adjoint and using the fact that this holds for all $i,j$ also shows $P_{\xi}\s_{ij}\propto\s$. This proportionality forbids $\s_{ij}$ to have nonzero $P_{\xi}\s_{ij}P_{\xi}^{\perp}$ and $P_{\xi}^{\perp}\s_{ij}P_{\xi}$ contributions. Hence, 
\be
\s_{ij}=c_{ij}\s+P_{\xi}^{\perp}\s_{ij}P_{\xi}^{\perp}
\ee
for some constants $c_{ij}\in\C$ with $c_{ji}=\overline{c_{ij}}$. Thus, 
\be
\rho=\sum_{i,j}c_{ij}E_{ij}^{(p)}\otimes\s+\sum_{i,j}E_{ij}^{(p)}\otimes P_{\xi}^{\perp}\s_{ij}P_{\xi}^{\perp}. 
\ee
Let $\t:=\sum_{i,j}c_{ij}E_{ij}^{(p)}$. 
Then $\t$ is positive because
\be
\t=\tr_{\mathcal{M}_{n}(\C)}\left(\mathrm{Ad}_{\mathds{1}_{p}\otimes P_{\xi}}(\rho)\right)\ge0.
\ee
Furthermore, $\t$ is a density matrix because
\be
\s=F^*(\rho)=\tr_{\mathcal{M}_{p}(\C)}(\rho)
=\sum_{i}c_{ii}\s+\sum_{i}P_{\xi}^{\perp}\s_{ii}P_{\xi}^{\perp}
=\sum_{i}c_{ii}\s+\underbrace{P_{\xi}^{\perp}\s P_{\xi}^{\perp}}_{=0}
=\tr(\t)\s,
\ee
which entails $\tr(\t)=1$. 
Since
\be
\rho=\t\otimes\s+\sum_{i,j}E_{ij}^{(p)}\otimes P_{\xi}^{\perp}\s_{ij}P_{\xi}^{\perp}
\ee
is the sum of a density matrix ($\t\otimes\s$) and another positive matrix, the latter must therefore vanish in order for $\rho$ to also be a density matrix. 

Now that $\rho=\t\otimes\s$ has been established, the Choi matrix of $\mathrm{Ad}_{P_{\xi}}\circ G$ becomes 
\be
\mathfrak{A}:=\Choi\big(\mathrm{Ad}_{P_{\xi}}\circ G\big)=\tau^{T}\otimes\Choi\big(\mathrm{Ad}_{P_{\xi}}\big)
\ee
and the other known part of the Choi matrix due to the expression~(\ref{eq:qBayesianinversepartialsupport}) is
\be
%\begin{split}
\mathfrak{B}:=(\mathds{1}_{p}\otimes\mathds{1}_{n}\otimes P_{\xi})\big(\Choi(G)\big)(\mathds{1}_{p}\otimes\mathds{1}_{n}\otimes P_{\xi}^{\perp})%\\
%&
=\tau^{T}\otimes\sum_{k,l}E_{kl}^{(n)}\otimes P_{\xi}E_{kl}^{(n)}P_{\xi}^{\perp}. 
%\end{split}
\ee
Since 
\be
\Choi(\mathrm{Ad}_{P_{\xi}})=\begin{bmatrix}P_{\xi}e_{1}\\\vdots\\P_{\xi}e_{n}\end{bmatrix}\begin{bmatrix}(P_{\xi}e_{1})^{\dag}&\cdots&(P_{\xi}e_{n})^{\dag}\end{bmatrix}
\ee 
is a rank one matrix, it can be normalized to a rank 1 projection and thus its pseudo-inverse can be easily calculated. It is given by (cf.\ Lemma~\ref{lem:pseudoinverseofChoi}) 
\be
\reallywidehat{\Choi(\mathrm{Ad}_{P_{\xi}})}=\left(\sum_{i=1}^{n}\lVert P_{\xi}e_{i}\rVert^{2}\right)^{-2}\Choi(\mathrm{Ad}_{P_{\xi}})
=\frac{1}{r^2}\Choi(\mathrm{Ad}_{P_{\xi}}),  
\ee
where $r=\sum_{i=1}^{n}\lVert P_{\xi}e_{i}\rVert^2=\tr(P_{\xi})$ is the rank of $P_{\xi}$. 
Hence, 
\be
\begin{split}
\mathfrak{B}^{\dag}\hat{\mathfrak{A}}\mathfrak{B}
&=\frac{1}{r^2}
\sum_{\substack{i,j,k,\\l,m,o}}
\left(\tau^{T}\otimes E_{ij}\otimes P_{\xi}^{\perp}E_{ij}P_{\xi}\right)\left(\widehat{\t^{T}}\otimes E_{kl}\otimes P_{\xi}E_{kl}P_{\xi}\right)\left(\tau^{T}\otimes E_{mo}\otimes P_{\xi}E_{mo}P_{\xi}^{\perp}\right)\\
&=\frac{1}{r^2}
\sum_{i,j,k,l}\t^{T}\otimes E_{il}\otimes P_{\xi}^{\perp}E_{ij}P_{\xi}E_{jk}P_{\xi}E_{kl}P_{\xi}^{\perp}\\
&=\frac{1}{r^2}
\sum_{i,j,k,l}\t^{T}\otimes E_{il}\otimes (P_{\xi}^{\perp}e_{i})(P_{\xi}e_{j})^{\dag}(P_{\xi}e_{j})(P_{\xi}e_{k})^{\dag}(P_{\xi}e_{k})(P_{\xi}^{\perp}e_{l})^{\dag}\\
&=\sum_{i,j}\t^{T}\otimes E_{ij}\otimes (P_{\xi}^{\perp}e_{i})(P_{\xi}^{\perp}e_{j})^{\dag}
=\sum_{i,j}\t^{T}\otimes E_{ij}\otimes P_{\xi}^{\perp}E_{ij}P_{\xi}^{\perp}
=\t^{T}\otimes\Choi\big(\mathrm{Ad}_{P_{\xi}^{\perp}}\big)
\end{split}
\ee
Now taking the partial trace gives
\be
\tr_{\mathcal{M}_{p}(\C)\otimes\mathcal{M}_{n}(\C)}\left(\mathfrak{B}^{\dag}\hat{\mathfrak{A}}\mathfrak{B}\right)
=\sum_{i,j}\tr(\t^{T})\tr(E_{ij})P_{\xi}^{\perp}E_{ij}P_{\xi}^{\perp}
=\sum_{i}P_{\xi}^{\perp}E_{ii}P_{\xi}^{\perp}
=P_{\xi}^{\perp}. 
\ee
Thus, applying Theorem~\ref{thm:Bayesianinversionmatrixalgebracase} gives a unique Bayesian inverse. In fact, these calculations show
\be
G\big(E_{ij}^{(p)}\otimes E_{kl}^{(n)}\big)=\t_{ji}E_{kl}^{(n)}
\qquad\implies\qquad
G=\tr_{\mathcal{M}_{p}(\C)}\circ\mathrm{Ad}_{\sqrt{\t}\otimes\mathds{1}_{n}},
\ee
which agrees with the disintegration formula from~\cite{PaRu19}. 
\ex

%%%%%%%%%%%%%%%%%%%%%%%%%%%%%%%%%%%%%
\section{A quantum Bayes' theorem for finite-dimensional $C^*$-algebras}
\label{sec:quantumBayesgeneral}
%%%%%%%%%%%%%%%%%%%%%%%%%%%%%%%%%%%%%

Theorem~\ref{thm:Bayesianinversionmatrixalgebracase} can be relatively easily generalized to the finite-dimensional $C^*$-algebra setting. Most of the work was already done, and the main challenge is to set up all the notation. We refer the reader to Section~5.2 in~\cite{PaRu19} for more background. The following notation will be used throughout this section. 

\begin{notation}
\label{not:directsums}
The general form of a finite-dimensional $C^*$-algebra is a direct sum of matrix algebras. 
Hence, we set 
\be
\mA:=\bigoplus_{x\in X}\mathcal{M}_{m_{x}}(\C)
\quad\text{ and }\quad
\mB:=\bigoplus_{y\in Y}\mathcal{M}_{n_{y}}(\C),
\ee
where $X$ and $Y$ are finite sets. Since every element $A\in\mA$ can be expressed as $\bigoplus_{x\in X}A_{x},$ 
a state $\mA\xstoch{\w}\C$ on $\mA$ can be expressed as 
\be
\mA\ni A\mapsto\w(A)=:\sum_{x\in X}p_{x}\tr(\rho_{x}A_{x}),
\ee
where $\{p_{x}\}$ defines a probability measure on $X$ and $\rho_{x}$ is a density matrix on $\mathcal{M}_{m_{x}}(\C)$ for every $x\in X$. Similarly, we let $\mB\xstoch{\xi}\C$ be written (somewhat abusively) as $\xi=:\sum_{y\in Y}q_{y}\tr(\s_{y}\;\cdot\;)$ with the understanding that given $B=\bigoplus_{y\in Y}B_{y}$, one plugs in $B_{y}$ into the term $\tr(\s_{y}\;\cdot\;)$. 
Let $N_{p}:=\{x\in X\;:\;p_{x}=0\}$ and $N_{q}:=\{y\in Y\;:\;q_{y}=0\}$.
Let $P_{x}$ and $Q_{y}$ be the supports of $\rho_{x}$ and $\s_{y}$, respectively. Let $\mathfrak{P}_{x}$ and $\mathfrak{Q}_{y}$ be the supports of $p_{x}\rho_{x}$ and $q_{y}\s_{y}$, respectively. In many of the arguments that follow, we will often use the identities $p_{x}P_{x}=p_{x}\mathfrak{P}_{x}$ and $q_{y}Q_{y}=q_{y}\mathfrak{Q}_{y}$. 
Note that $P_{\w}$ and $P_{\xi}$, the supports of $\w$ and $\xi$, are equal to $P_{\w}=\bigoplus_{x\in X}\mathfrak{P}_{x}$ and $P_{\xi}=\bigoplus_{y\in Y}\mathfrak{Q}_{y}$.
Hence, 
\be
\label{eq:Nomeganullspacedecomp}
\mathcal{N}_{\w}=\mA P_{\w}^{\perp}=\bigoplus_{x\in X}\Big(\mathcal{M}_{m_{x}}(\C)\mathfrak{P}_{x}^{\perp}\Big)\cong\bigoplus_{x\in X\setminus N_{p}}\Big(\mathcal{M}_{m_{x}}(\C){P}_{x}^{\perp}\Big)
\ee
and similarly
\be
\label{eq:Nxinullspacedecomp}
\mathcal{N}_{\xi}=\mB P_{\xi}^{\perp}=\bigoplus_{y\in Y}\Big(\mathcal{M}_{n_{y}}(\C)\mathfrak{Q}_{y}^{\perp}\Big)
\cong\bigoplus_{y\in Y\setminus N_{q}}\Big(\mathcal{M}_{n_{y}}(\C){Q}_{y}^{\perp}\Big). 
\ee
Every linear map $\mB\xstoch{F}\mA$ has a ``matrix decomposition'' with $yx$-th entry given by a linear map $\mathcal{M}_{m_{x}}(\C)\xstoch{F_{yx}}\mathcal{M}_{n_{y}}(\C)$, which is CP for all $x\in X$ and $y\in Y$ if and only if $F$ is CP. 
\end{notation}

If a lemma is written without proof, it is because the proof is straightforward or similar to proofs we have already been exposed to in earlier sections. 

\blem
\label{lem:positivecorner}
Let $\mathcal{M}_{n}(\C)\xstoch{\vf}\mathcal{M}_{m}(\C)$ be a positive map and let $P\in\mathcal{M}_{m}(\C)$ be an orthogonal projection. If $P\vf(A)P=0$ for all $A\ge0$, then $\vf\big(\mathcal{M}_{n}(\C)\big)\subseteq P^{\perp}\mathcal{M}_{m}(\C)P^{\perp}$. 
\elem

\bprf
This is proved in Lecture 7 of~\cite{Werner17}. Briefly, 
$P\vf(A)P=\big(\sqrt{\vf(A)}P\big)^{\dag}\big(\sqrt{\vf(A)}P\big)=0$ implies $\sqrt{\vf(A)}P=0$ and $P\sqrt{\vf(A)}=0$ for all $A\ge0$ (the latter follows from the first by taking the adjoint). Hence $\vf(A)\in P^{\perp}\mathcal{M}_{m}(\C)P^{\perp}$ for all $A\ge0$. The rest follows from the fact that every element of $\mathcal{M}_{n}(\C)$ is a linear combination of four positive matrices. 
\eprf

\blem
\label{lem:statepreservingdirectsum}
In terms of Notation~\ref{not:directsums}, 
the condition $\w\circ F=\xi$ is equivalent to%
%footnote
\footnote{$F^{*}_{xy}$ denotes the Hilbert--Schmidt adjoint of $F_{xy}$.}
%end footnote
\be
\label{eq:statepreservingconditiongeneralcase}
q_{y}\s_{y}=\sum_{x\in X}p_{x}F^*_{xy}(\rho_{x})\qquad\forall\;y\in Y. 
\ee
If $F$ is positive, then
\be
\label{eq:positivedirectsumnull}
p_{x}F_{xy}^{*}(\rho_{x})=0
\quad\text{ and }\quad
F_{xy}\big(\mathcal{M}_{n_{y}}(\C)\big)\subseteq\mathfrak{P}_{x}^{\perp}
\mathcal{M}_{m_{x}}(\C)
\mathfrak{P}_{x}^{\perp}
\qquad\forall\;
(x,y)\in X\times N_{q}. 
\ee
If $F$ is 2-positive, then 
\be
\label{eq:nullspacepreservation}
F_{xy}\Big(\mathcal{M}_{n_{y}}(\C)\mathfrak{Q}_{y}^{\perp}\Big)\subseteq\mathcal{M}_{m_{x}}(\C)\mathfrak{P}_{x}^{\perp}\qquad\forall\;x\in X,\;y\in Y. 
\ee
\elem

\bprf
For any $B_{y}\in\mathcal{M}_{n_{y}}(\C)$, the state-preserving condition gives
\be
\sum_{x\in X}p_{x}\tr\big(\rho_{x}F_{xy}(B_{y})\big)=q_{y}\tr(\s_{y}B_{y}).
\ee
Equation~(\ref{eq:statepreservingconditiongeneralcase}) follows from this by calculations similar to those we have seen earlier. When $y\in N_{q}$, the equality $p_{x}F_{xy}^{*}(\rho_{x})=0$ in~(\ref{eq:positivedirectsumnull}) follows from~(\ref{eq:statepreservingconditiongeneralcase}) and the fact that $p_{x}F^*_{xy}(\rho_{x})\ge0$ for all $x\in X$ and the only way a sum of positive operators adds to zero is if each of them are zero. For the second condition in~(\ref{eq:positivedirectsumnull}), let $y\in N_{q}$ and $B_{y}\in\mathcal{M}_{n_{y}}(\C)$. Then, 
\be
0=\xi(B_{y})=\w\big(F(B_{y})\big)=\sum_{x\in X}p_{x}\tr\big(\rho_{x}F_{xy}(B_{y})\big)=\sum_{x\in X}p_{x}\tr\left(\rho_{x}^{1/2}F_{xy}(B_{y})\rho_{x}^{1/2}\right). 
\ee
If $B_{y}\ge0$, then 
\be
p_{x}\tr\left(\rho_{x}^{1/2}F_{xy}(B_{y})\rho_{x}^{1/2}\right)=0\qquad\forall\;x\in X
\ee
because $p_{x}\rho_{x}^{1/2}F_{xy}(B_{y})\rho_{x}^{1/2}\ge0$. 
Since the trace is faithful and $p_{x}\rho_{x}^{1/2}F_{xy}(B_{y})\rho_{x}^{1/2}\ge0$, this implies 
$p_{x}\rho_{x}^{1/2}F_{xy}(B_{y})\rho_{x}^{1/2}=0.$ 
Since every matrix can be expressed as a linear combination of at most four positive matrices, 
$p_{x}\rho_{x}^{1/2}F_{xy}\big(\mathcal{M}_{n_{y}}(\C)\big)\rho_{x}^{1/2}=0.$ 
Multiplying both sides by $\widehat{\rho_{x}}^{1/2}$ gives 
\be
\label{eq:PxFxyPx}
\mathfrak{P}_{x}F_{xy}\big(\mathcal{M}_{n_{y}}(\C)\big)\mathfrak{P}_{x}=0.
\ee
Since $F_{xy}$ is positive, the second claim in~(\ref{eq:positivedirectsumnull}) follows from Lemma~\ref{lem:positivecorner}. 
Finally, condition~(\ref{eq:nullspacepreservation}) follows from~(\ref{eq:Nomeganullspacedecomp}),~(\ref{eq:Nxinullspacedecomp}), and  
Lemma~\ref{lem:nullspaces}. 
\eprf

\bn
\label{prop:bayesnecessarydirectsum}
Suppose $F$ is positive and unital. 
A linear map $\mA\xstoch{G}\mB$ satisfies the Bayes condition for $(F,\w)$ if and only if (cf.\ Notation~\ref{not:directsums})
\be
\label{eq:bayesconditiondirectsumcase}
q_{y}\tr\big(\s_{y}G_{yx}(A_{x})B_{y}\big)=p_{x}\tr\big(\rho_{x}A_{x}F_{xy}(B_{y})\big)
\ee
for all $A_{x}\in\mathcal{M}_{m_{x}}(\C), B_{y}\in\mathcal{M}_{n_{y}}(\C)$ and for all $x\in X,y\in Y$. If $G$ satisfies the Bayes condition, then 
\be
\label{eq:bayesdirectsumqyn=0}
Q_{y}G_{yx}(A_{x})=\frac{p_{x}}{q_{y}}\widehat{\s_{y}}F^{*}_{xy}(\rho_{x}A_{x})\qquad\forall\;A_{x}\in\mathcal{M}_{m_{x}}(\C)
\quad\forall\;(x,y)\in X\times(Y\setminus N_{q})
.
\ee
Conversely, 
suppose there exists a linear map $G$ satisfying~(\ref{eq:bayesdirectsumqyn=0}). Then $G$ is a Bayes map for $(F,\w)$. 
\en

\bprf
Equation~(\ref{eq:bayesconditiondirectsumcase}) follows directly from the Bayes condition. 
Equation~(\ref{eq:bayesdirectsumqyn=0}) follows a similar analysis to the one in Proposition~\ref{prop:qBayesianinverseformula}. 
Conversely, suppose 
there exists 
a linear map $G$ satisfying~(\ref{eq:bayesdirectsumqyn=0}). In what follows, we will prove~(\ref{eq:bayesconditiondirectsumcase}). First, let $y\in N_{q}$. Then 
\be
p_{x}\tr\big(\rho_{x}A_{x}F_{xy}(B_{y})\big)
%=p_{x}\tr\big(\rho_{x}A_{x}F_{xy}(B_{y})P_{x}\big)
=p_{x}\tr\big(\rho_{x}A_{x}F_{xy}(B_{y})\mathfrak{P}_{x}\big)
=0
\ee
because $F_{xy}(B_{y})\in\mathfrak{P}_{x}^{\perp}\mathcal{M}_{m_{x}}(\C)\mathfrak{P}_{x}^{\perp}$ by~(\ref{eq:positivedirectsumnull}). Now, let $y\in Y\setminus N_{q}$ so that $q_{y}\ne0$. Then essentially the same argument as in~(\ref{eq:Bayesholdsfromotherassumptions}) goes through using~(\ref{eq:nullspacepreservation}) in place of~(\ref{eq:nullspacepreservation}). Note that one uses that fact that $\mathfrak{Q}_{y}=Q_{y}$ for $y\in Y\setminus N_{q}$ and $p_{x}P_{x}=p_{x}\mathfrak{P}_{x}$ for $x\in X$. This proves that $G$ is a Bayes map. 
\eprf

\blem
Suppose $\mA\xstoch{G}\mB$ is a Bayes map for $(F,\w)$ and
let $(x,y)\in X\times(Y\setminus N_{q})$. 
Then $\mathrm{Ad}_{Q_{y}}\circ G_{yx}:\mM_{m_{x}}(\C)\stoch\mM_{n_{y}}(\C)$ is $*$-preserving if and only if 
it is CP. 
\elem

\bprf
By Proposition~\ref{prop:bayesnecessarydirectsum}, $\mathrm{Ad}_{Q_{y}}\circ G_{yx}$ takes the form
\be
(\mathrm{Ad}_{Q_{y}}\circ G_{yx})(A_{x})=\frac{p_{x}}{q_{y}}\widehat{\s_{y}}F^{*}_{xy}(\rho_{x}A_{x})Q_{y}\qquad\forall\;A_{x}\in\mathcal{M}_{m_{x}}(\C).
\ee
If $p_{x}=0,$ then $\mathrm{Ad}_{Q_{y}}\circ G_{yx}$ is the zero map and is therefore CP. If $p_{x}\ne0$, by Proposition~\ref{prop:selfadjointimpliesCP}, all of the items listed in that proposition are equivalent to $\mathrm{Ad}_{Q_{y}}\circ G_{yx}$ being $*$-preserving. In particular, $\mathrm{Ad}_{Q_{y}}\circ G_{yx}$ is CP. 
\eprf

\blem
\label{eq:unitalitydirectsum}
The following conditions on a linear map $\mA\xstoch{G}\mB$ are equivalent.
\begin{enumerate}[i.]
\itemsep0pt
\item
$G$ is unital. 
\item
$G$ satisfies 
$\sum_{x\in X}G_{yx}(\mathds{1}_{m_{x}})=\mathds{1}_{n_{y}}$
for all $y\in Y$.
\item
$G$ satisfies 
$\sum_{x\in X}\tr_{\mathcal{M}_{m_{x}}(\C)}\big(\Choi(G_{yx})\big)=\mathds{1}_{n_{y}}$
for all $y\in Y$.
\end{enumerate}
\elem

\bt
[A Bayes' theorem for finite-dimensional $C^*$-algebras]
\label{thm:Bayesianinversiongeneralcase}
For each $x\in X$ and $y\in Y\setminus N_{q}$, set (cf.\ Notation~\ref{not:directsums})
\be
\mathfrak{A}_{yx}:=\frac{p_{x}}{q_{y}}\!\sum_{i,j=1}^{m_{x}}E_{ij}^{(m_{x})}\!\otimes\widehat{\s_{y}}F_{xy}^*\big(\rho_{x} E_{ij}^{(m_{x})}\big)Q_{y}
\;\text{ and }\;
\mathfrak{B}_{yx}:=\frac{p_{x}}{q_{y}}\!\sum_{i,j=1}^{m_{x}}E_{ij}^{(m_{x})}\!\otimes\widehat{\s_{y}}F_{xy}^*\big(\rho_{x} E_{ij}^{(m_{x})}\big)Q_{y}^{\perp},
\ee
which are matrices in $\mathcal{M}_{m_{x}}(\C)\otimes\mathcal{M}_{n_{y}}(\C)$. 
Then $(F,\w)$ has a (CPU) Bayesian inverse if and only if 
\be
\mathfrak{A}_{yx}^{\dag}=\mathfrak{A}_{yx}\qquad\forall\;(x,y)\in X\times(Y\setminus N_{q})
\ee
and
\be
\sum_{x\in X}\tr_{\mathcal{M}_{m_{x}}(\C)}\left(\mathfrak{B}_{yx}^{\dag}\widehat{\mathfrak{A}_{yx}}\mathfrak{B}_{yx}\right)\le Q_{y}^{\perp}
\qquad\forall\;y\in Y\setminus N_{q}
.
\ee
\et

The proof will be similar to the proof of Theorem~\ref{thm:Bayesianinversionmatrixalgebracase} with a few minor changes. We will therefore spell this out, but on occasion we will shorten arguments which would otherwise be repetitive. 

\bprf[Proof of Theorem~\ref{thm:Bayesianinversiongeneralcase}]
{\color{white}{you found me!}}

\noindent
($\Rightarrow$)
Suppose that $(F,\w)$ has a Bayesian inverse $G$. Then by Proposition~\ref{prop:bayesnecessarydirectsum}, 
\be
\label{eq:Choidirectsum}
(\mathds{1}_{m_{x}}\otimes Q_{y})\big(\Choi(G_{yx})\big)(\mathds{1}_{m_{x}}\otimes Q_{y})=\mathfrak{A}_{yx}
\quad\text{ and }\quad
(\mathds{1}_{m_{x}}\otimes Q_{y})\big(\Choi(G_{yx})\big)(\mathds{1}_{m_{x}}\otimes Q_{y}^{\perp})=\mathfrak{B}_{yx}
\ee
hold for all $(x,y)\in X\times(Y\setminus N_{q})$. 
The rest of the proof in this direction follows similar arguments as in the proof of Theorem~\ref{thm:Bayesianinversionmatrixalgebracase}.

\noindent
($\Leftarrow$)
Conversely, suppose the two conditions hold. For $y\in N_{q}$, set 
\be
\label{eq:Gyxoffsupport}
\mathcal{M}_{m_{x}}(\C)\ni A_{x}\mapsto G_{yx}(A_{x}):=\frac{\tr(A_{x})}{m_{x}|X|}\mathds{1}_{n_{y}}\qquad\forall\;x\in X. 
\ee
Note that $G_{yx}$ is completely positive and satisfies 
\be
\label{eq:Gunitaloffsupport}
\sum_{x\in X}G_{yx}(\mathds{1}_{m_{x}})=\mathds{1}_{n_{y}}\qquad\forall\;y\in N_{q}. 
\ee
For $y\in Y\setminus N_{q}$, most of the proof follows through as in the proof of Theorem~\ref{thm:Bayesianinversionmatrixalgebracase}. In particular, one builds $G_{yx}$ in terms of its Choi matrix by using~(\ref{eq:Choidirectsum}).
The first main difference occurs in defining $\mathfrak{C}_{yx}\equiv(\mathds{1}_{m_{x}}\otimes Q_{y}^{\perp})\big(\Choi(G_{yx})\big)(\mathds{1}_{m_{x}}\otimes Q_{y}^{\perp})$, which can be set as
\be
\mathfrak{C}_{yx}:=\mathfrak{B}_{yx}^{\dag}\widehat{\mathfrak{A}_{yx}}\mathfrak{B}_{yx}+\frac{1}{m_{x}|X|}\left(\mathds{1}_{m_{x}}\otimes\left(Q_{y}^{\perp}-\sum_{x'\in X}\tr_{\mathcal{M}_{m_{x'}}(\C)}\Big(\mathfrak{B}_{yx'}^{\dag}\widehat{\mathfrak{A}_{yx'}}\mathfrak{B}_{yx'}\Big)\right)\right)
.
\ee
Note that $\mathfrak{C}_{yx}\ge0$ and satisfies 
\be
\label{eq:partialtracedirectsum}
\sum_{x\in X}\tr_{\mathcal{M}_{m_{x}}(\C)}(\mathfrak{C}_{yx})=Q_{y}^{\perp}.
\ee
With these definitions, $G$ has been defined everywhere and automatically satisfies the Bayes condition by the assumptions made. Finally,~(\ref{eq:Gunitaloffsupport}), (\ref{eq:partialtracedirectsum}), 
\be
\sum_{x\in X}\tr_{\mathcal{M}_{m_{x}}(\C)}(\mathfrak{A}_{yx})=Q_{y},
\qquad\text{ and }\qquad
\sum_{x\in X}\tr_{\mathcal{M}_{m_{x}}(\C)}(\mathfrak{B}_{yx})=0
\qquad\forall\;y\in Y\setminus N_{q}
\ee
show $G$ is unital by Lemma~\ref{eq:unitalitydirectsum}. 
\eprf

In the following example, we will describe when Bayesian inverses exist for quantum instruments. 

\bx
An \define{instrument} on a Hilbert space $\mathcal{H}$ indexed by a finite set $Y$ is a CPU map $\mathcal{B}(\Hi)\otimes\C^{Y}\stoch\mathcal{B}(\Hi)$. In what follows, let $\Hi$ be finite-dimensional so that such a map specifies a family of CP maps $\mathcal{M}_{m}(\C)\xstoch{F_{y}}\mathcal{M}_{m}(\C)$, indexed by $y\in Y$, such that $\sum_{y\in Y}F_{y}$ is CPU. Given a state $\w=\tr(\rho\;\cdot\;)$ on $\mathcal{M}_{m}(\C),$ set $\xi:=\w\circ F$. Set $q_{y}:=\tr\big(F_{y}^{*}(\rho)\big)$. If $q_{y}=0,$ let $\s_{y}$ be any density matrix, while if $q_{y}\ne0$, set $\s_{y}:=\frac{1}{q_{y}}F_{y}^{*}(\rho)$. Then $\xi=\sum_{y\in Y}q_{y}\tr(\s_{y}\;\cdot\;)$. 
A Bayesian inverse of $(F,\w)$ would consist of a family $G_{y}:\mathcal{M}_{m}(\C)\xstoch{\;}\mathcal{M}_{m}(\C)$ of CPU maps (note the difference and that $F_{y}$ was not required to be unital). At present, we have not yet found a simpler description of when Bayesian inverses exist other than to demand the assumptions in Theorem~\ref{thm:Bayesianinversiongeneralcase}. Nevertheless, if one has an explicit instrument with known states, then it should not be too difficult to apply the criteria of Theorem~\ref{thm:Bayesianinversiongeneralcase} to see if a CPU Bayesian inverse exists. Furthermore, one may always compute the required pseudoinverses (provided they are not too large) to determine the formula for the Bayesian inverse. 
\ex

Finally, we end with the verification that Theorem~\ref{thm:Bayesianinversiongeneralcase} reproduces the standard classical Bayes' theorem as a corollary. 

\bx
Suppose $m_{x}=1$ and $n_{y}=1$ for all $(x,y)\in X\times Y$ in Theorem~\ref{thm:Bayesianinversiongeneralcase}. 
In this case, $F$ can be viewed as a matrix of (linear transformations given by multiplication by) non-negative numbers. In this case, $F_{xy}^*=F_{xy}$ as numbers. 
If $q_{y}\ne0$, then $Q_{y}=\mathds{1}_{n_{y}}$ so that $\mathfrak{A}_{yx}=\frac{p_{x}F_{xy}}{q_{y}}$, $\mathfrak{B}_{yx}=0$, and $\mathfrak{C}_{yx}=0$. The two conditions in Theorem~\ref{thm:Bayesianinversiongeneralcase} are automatically satisfied. 
Hence a Bayesian inverse exists. When $q_{y}\ne0$, one sets
$G_{yx}:=\frac{p_{x}F_{xy}}{q_{y}}$. 
When $q_{y}=0$, the assignment in~(\ref{eq:Gyxoffsupport}) defines $G_{yx}$ as $G_{yx}:=\frac{1}{|X|}$. Thus, our proof of our quantum Bayes' theorem reproduces classical Bayesian inference.
\ex

%%%%%%%%%%%%%%%%%%%%%%%%%%%%%%%%%%%%%
\appendix 

%%%%%%%%%%%%%%%%%%%%%%%%%%%%%%%%%%%%%
\section{Permutations of the Choi matrix}
\label{appendix:permutation}
%%%%%%%%%%%%%%%%%%%%%%%%%%%%%%%%%%%%%

Let $\mathcal{M}_{m}(\C)\xstoch{G}\mathcal{M}_{n}(\C)$ be a linear map. 
Decompose the Choi matrix of $G$ as%
%the code for coloring rows and columns is taken from Gonzalo Medina's answer to https://tex.stackexchange.com/questions/69713/matrix-change-row-or-column-background
%footnote
\footnote{$t=$ top, $b=$ bottom, $r=$ right, $l=$ left. The matrix rows and columns have been colored to illustrate how one slides them to obtain the permutation discussed afterwards.}
%end footnote
\be
\Choi(G)=
  \left[
  \begin{matrix}[0.91]\\ \\ \\ \\ \\ \\ \\ \end{matrix}\!%%need this to get spacing right for brackets
  \begin{array}{>{\columncolor{red!30}}c>{\columncolor{blue!30}}cccc>{\columncolor{yellow!30}}c>{\columncolor{green!30}}c}
    \rowcolor{red!30}
    G\big(E_{11}^{(m)}\big)^{\mathrm{tl}}  & \cellcolor{blue!30!red!30}{G\big(E_{11}^{(m)}\big)^{\mathrm{tr}}}  & & \raisebox{-6pt}{\scriptsize$\bullet\bullet\bullet$} & & \cellcolor{yellow!30!red!30}{G\big(E_{1m}^{(m)}\big)^{\mathrm{tl}}} & \cellcolor{green!30!red!30}{G(E_{1m}^{(m)})^{\mathrm{tr}}} \\
    \rowcolor{blue!30}
    \cellcolor{red!30!blue!30}{G\big(E_{11}^{(m)}\big)^{\mathrm{bl}}}   & G\big(E_{11}^{(m)}\big)^{\mathrm{br}}  & & & & \cellcolor{yellow!30!blue!30}{G\big(E_{1m}^{(m)}\big)^{\mathrm{bl}}} & \cellcolor{green!30!blue!30}{G\big(E_{1m}^{(m)}\big)^{\mathrm{br}}} \\
       \shifttext{63pt}{\Huge\vdots}&\shifttext{-62pt}{\Huge\vdots}    & & & &   \shifttext{63pt}{\Huge\vdots}&\shifttext{-63pt}{\Huge\vdots}  \\
    \rowcolor{yellow!30}
    \cellcolor{red!30!yellow!30}{G\big(E_{m1}^{(m)}\big)^{\mathrm{tl}}}   & \cellcolor{blue!30!yellow!30}{G\big(E_{m1}^{(m)}\big)^{\mathrm{tr}}}   & & \raisebox{-6pt}{\scriptsize$\bullet\bullet\bullet$} & & G\big(E_{mm}^{(m)}\big)^{\mathrm{tl}}  & \cellcolor{green!30!yellow!30}{G\big(E_{mm}^{(m)}\big)^{\mathrm{tr}}} \\
    \rowcolor{green!30}
    \cellcolor{red!30!green!30}{G\big(E_{m1}^{(m)}\big)^{\mathrm{bl}}}  &  \cellcolor{blue!30!green!30}{G\big(E_{m1}^{(m)}\big)^{\mathrm{br}}}  & & & & \cellcolor{yellow!30!green!30}{G\big(E_{mm}^{(m)}\big)^{\mathrm{bl}}} &  G\big(E_{mm}^{(m)}\big)^{\mathrm{br}}\\
  \end{array}
  \!%to get the brackets to look nicely
  \right]
\ee
so that 
\be
G\big(E_{ij}^{(m)}\big)=
\begin{bmatrix}
G\big(E_{ij}^{(m)}\big)^{\mathrm{tl}}&G\big(E_{ij}^{(m)}\big)^{\mathrm{tr}}\\
G\big(E_{ij}^{(m)}\big)^{\mathrm{bl}}&G\big(E_{ij}^{(m)}\big)^{\mathrm{br}} 
\end{bmatrix}
.
\ee
Note that the size of the square matrices $G\big(E_{ij}^{(m)}\big)^{\mathrm{tl}}$ and $G\big(E_{ij}^{(m)}\big)^{\mathrm{br}}$ need not be the same, but we assume that the size of 
$G\big(E_{ij}^{(m)}\big)^{\mathrm{tl}}$ equals the size of  $G\big(E_{kl}^{(m)}\big)^{\mathrm{tl}}$ for all $i,j,k,l\in\{1,\dots,m\}$, and similarly for $G\big(E_{ij}^{(m)}\big)^{\mathrm{br}}$ and $G\big(E_{kl}^{(m)}\big)^{\mathrm{br}}$. 
Then, there exists a permutation matrix $U$ such that 
\be
U\Choi(G)U^{\dag}=
  \left[
  \begin{matrix}[0.91]\\ \\ \\ \\ \\ \\ \\ \\ \end{matrix}
  \!
  \begin{array}{>{\columncolor{red!30}}cc>{\columncolor{yellow!30}}c>{\columncolor{blue!30}}cc>{\columncolor{green!30}}c}
    \rowcolor{red!30}
    G\big(E_{11}^{(m)}\big)^{\mathrm{tl}}  &\cdots& \cellcolor{yellow!30!red!30}{G\big(E_{1m}^{(m)}\big)^{\mathrm{tl}}} & \cellcolor{blue!30!red!30}{G\big(E_{11}^{(m)}\big)^{\mathrm{tr}}} &\cdots& \cellcolor{green!30!red!30}{G(E_{1m}^{(m)})^{\mathrm{tr}}} \\
       \vdots&&\vdots  &  \vdots&&\vdots  \\
    \rowcolor{yellow!30}
    \cellcolor{red!30!yellow!30}{G\big(E_{m1}^{(m)}\big)^{\mathrm{tl}}}   &\cdots& G\big(E_{mm}^{(m)}\big)^{\mathrm{tl}} & \cellcolor{blue!30!yellow!30}{G\big(E_{m1}^{(m)}\big)^{\mathrm{tr}}} &\cdots& \cellcolor{green!30!yellow!30}{G\big(E_{mm}^{(m)}\big)^{\mathrm{tr}}} \\
    \rowcolor{blue!30}
    \cellcolor{red!30!blue!30}{G\big(E_{11}^{(m)}\big)^{\mathrm{bl}}}   &\cdots& \cellcolor{yellow!30!blue!30}{G\big(E_{1m}^{(m)}\big)^{\mathrm{bl}}}   & G\big(E_{11}^{(m)}\big)^{\mathrm{br}}  &\cdots& \cellcolor{green!30!blue!30}{G(E_{1m}^{(m)})^{\mathrm{br}}} \\
       \vdots&&\vdots   &   \vdots&&\vdots  \\
    \rowcolor{green!30}
    \cellcolor{red!30!green!30}{G\big(E_{m1}^{(m)}\big)^{\mathrm{bl}}}  &\cdots&  \cellcolor{yellow!30!green!30}{G\big(E_{mm}^{(m)}\big)^{\mathrm{bl}}} & \cellcolor{blue!30!green!30}{G\big(E_{m1}^{(m)}\big)^{\mathrm{br}}} &\cdots&  G\big(E_{mm}^{(m)}\big)^{\mathrm{br}}\\
  \end{array}
  \!
  \right]
\ee
(which corresponds to the matrix $\left[\begin{smallmatrix}\uline{\mathfrak{A}}&\uline{\mathfrak{B}}\\\uline{\mathfrak{B}}^{\dag}&\uline{\mathfrak{C}}\end{smallmatrix}\right]$ in the proof of Theorem~\ref{thm:Bayesianinversionmatrixalgebracase}).
This is just a special case of what is discussed in Remark~\ref{rmk:positivitySchur}, where the projection $P$ is of the form $\mathds{1}_{m}\otimes\left[\begin{smallmatrix}\mathds{1}_{r}&0\\0&0_{n-r}\end{smallmatrix}\right]$ with $0\le r\le n$.
The permutation matrix $U$ is the matrix associated to the permutation 
\be
\left(\begin{matrix}1&2&3&4&\cdots&2m-3&2m-2&2m-1&2m\\
1&m+1&2&m+2&\cdots&m-1&2m-1&m&2m\end{matrix}\right)
\ee
where $\begin{smallmatrix}i\\j\end{smallmatrix}$ means that the $i$-th row is sent to the $j$-th row. 
Note that this is a permutation for the blocks of the appropriate sizes.
If $G(E_{ij}^{(m)})^{\mathrm{tl}}$ has size $r\times r$, then $G(E_{ij}^{(m)})^{\mathrm{br}}$ has size $(n-r)\times(n-r)$, and the general form of $U$ is
\be
U=\begin{bmatrix}
\mathds{1}_{r}&0&0&0&&0&0\\
0&0&\mathds{1}_{r}&0&&0&0\\
&&&&\ddots&&\\
0&0&0&0&&\mathds{1}_{r}&0\\
0&\mathds{1}_{n-r}&0&0&&0&0\\
0&0&0&\mathds{1}_{n-r}&&0&0\\
&&&&\ddots&&\\
0&0&0&0&&0&\mathds{1}_{n-r}
\end{bmatrix}
.
\ee 
For example, the matrix $U$ that achieves this for~(\ref{eq:Choicannotbecompleted}) is 
\be
U=\begin{bmatrix}
\mathds{1}_{2}&0&0&0\\
0&0&\mathds{1}_{2}&0\\
0&\mathds{1}_{1}&0&0\\
0&0&0&\mathds{1}_{1}
\end{bmatrix}
\ee
and corresponds to the permutation 
\be
\left(\begin{matrix}1&2&3&4\\
1&3&2&4
\end{matrix}\right)
,
\ee
which swaps the second and third blocks but leaves the first and fourth blocks alone. 

\section*{Acknowledgements}

We gratefully acknowledge support from the Simons Center for Geometry and Physics, Stony Brook University and for the opportunity to participate in the workshop ``Operator Algebras and Quantum Physics'' in June 2019.
We thank Kenta Cho, Chris Heunen, and Bart Jacobs for sharing their code for the string diagrams used in this paper. 
We thank Luca Giorgetti, Alessio Ranallo, and Fidel I.~Schaposnik for discussions. 
A part of this project was done while AJP was an Assistant Research Professor at the University of Connecticut.

%%%%%%%%BIBLIOGRAPHY%%%%%%%%%%%%
\addcontentsline{toc}{section}{\numberline{}Bibliography}
\bibliographystyle{plain}
\bibliography{Bayesbib}

\Addresses
%%based on egreg's code (see after title and authors in preamble)

\end{document}

%% file: tikzdefs.tex
\newlength\stateheight
\newlength\minimumstatewidth
\tikzset{width/.initial=\minimummorphismwidth}
\tikzset{colour/.initial=white}

\newif\ifblack\pgfkeys{/tikz/black/.is if=black}
\newif\ifwedge\pgfkeys{/tikz/wedge/.is if=wedge}
\newif\ifvflip\pgfkeys{/tikz/vflip/.is if=vflip}
\newif\ifhflip\pgfkeys{/tikz/hflip/.is if=hflip}
\newif\ifhvflip\pgfkeys{/tikz/hvflip/.is if=hvflip}

\pgfdeclarelayer{foreground}
\pgfdeclarelayer{background}
\pgfsetlayers{background,main,foreground}

\def\thickness{0.4pt}

\makeatletter
\pgfkeys{%
  /tikz/on layer/.code={
    \pgfonlayer{#1}\begingroup
    \aftergroup\endpgfonlayer
    \aftergroup\endgroup
  },
  /tikz/node on layer/.code={
    \gdef\node@@on@layer{%
      \setbox\tikz@tempbox=\hbox\bgroup\pgfonlayer{#1}\unhbox\tikz@tempbox\endpgfonlayer\pgfsetlinewidth{\thickness}\egroup}
    \aftergroup\node@on@layer
  },
  /tikz/end node on layer/.code={
    \endpgfonlayer\endgroup\endgroup
  }
}
\def\node@on@layer{\aftergroup\node@@on@layer}
\makeatother

\makeatletter
\pgfdeclareshape{state}
{
    \savedanchor\centerpoint
    {
        \pgf@x=0pt
        \pgf@y=0pt
    }
    \anchor{center}{\centerpoint}
    \anchorborder{\centerpoint}
    \saveddimen\overallwidth
    {
        \pgf@x=3\wd\pgfnodeparttextbox
        \ifdim\pgf@x<\minimumstatewidth
            \pgf@x=\minimumstatewidth
        \fi
    }
    \savedanchor{\upperrightcorner}
    {
        \pgf@x=.5\wd\pgfnodeparttextbox
        \pgf@y=.5\ht\pgfnodeparttextbox
        \advance\pgf@y by -.5\dp\pgfnodeparttextbox
    }
    \anchor{A}
    {
        \pgf@x=-\overallwidth
        \divide\pgf@x by 4
        \pgf@y=0pt
    }
    \anchor{B}
    {
        \pgf@x=\overallwidth
        \divide\pgf@x by 4
        \pgf@y=0pt
    }
    \anchor{text}
    {
        \upperrightcorner
        \pgf@x=-\pgf@x
        \ifhflip
            \pgf@y=-\pgf@y
            \advance\pgf@y by 0.4\stateheight
        \else
            \pgf@y=-\pgf@y
            \advance\pgf@y by -0.4\stateheight
        \fi
    }
    \backgroundpath
    {
       \begin{pgfonlayer}{foreground}
        \pgfsetstrokecolor{black}
        \pgfsetlinewidth{\thickness}
        \pgfpathmoveto{\pgfpoint{-0.5*\overallwidth}{0}}
        \pgfpathlineto{\pgfpoint{0.5*\overallwidth}{0}}
        \ifhflip
            \pgfpathlineto{\pgfpoint{0}{\stateheight}}
        \else
            \pgfpathlineto{\pgfpoint{0}{-\stateheight}}
        \fi
        \pgfpathclose
        \ifblack
            \pgfsetfillcolor{black!50}
            \pgfusepath{fill,stroke}
        \else
            \pgfusepath{stroke}
        \fi
       \end{pgfonlayer}
    }
}

\pgfdeclareshape{my ground}
{
  \inheritsavedanchors[from=rectangle ee]
  \inheritanchor[from=rectangle ee]{center}
  \inheritanchor[from=rectangle ee]{north}
  \inheritanchor[from=rectangle ee]{south}
  \inheritanchor[from=rectangle ee]{east}
  \inheritanchor[from=rectangle ee]{west}
  \inheritanchor[from=rectangle ee]{north east}
  \inheritanchor[from=rectangle ee]{north west}
  \inheritanchor[from=rectangle ee]{south east}
  \inheritanchor[from=rectangle ee]{south west}
  \inheritanchor[from=rectangle ee]{input}
  \inheritanchor[from=rectangle ee]{output}
  \anchorborder{\csname pgf@anchor@my ground@input\endcsname}

  \backgroundpath{
    \pgf@process{\pgfpointadd{\southwest}{\pgfpoint{\pgfkeysvalueof{/pgf/outer xsep}}{\pgfkeysvalueof{/pgf/outer ysep}}}}
    \pgf@xa=\pgf@x \pgf@ya=\pgf@y
    \pgf@process{\pgfpointadd{\northeast}{\pgfpointscale{-1}{\pgfpoint{\pgfkeysvalueof{/pgf/outer xsep}}{\pgfkeysvalueof{/pgf/outer ysep}}}}}
    \pgf@xb=\pgf@x \pgf@yb=\pgf@y
    \pgfpathmoveto{\pgfqpoint{\pgf@xa}{\pgf@ya}}
    \pgfpathlineto{\pgfqpoint{\pgf@xa}{\pgf@yb}}
    \pgfmathsetlength\pgf@xa{.5\pgf@xa+.5\pgf@xb}
    \pgfmathsetlength\pgf@yc{.16666\pgf@yb-.16666\pgf@ya}
    \advance\pgf@ya by\pgf@yc
    \advance\pgf@yb by-\pgf@yc
    \pgfpathmoveto{\pgfqpoint{\pgf@xa}{\pgf@ya}}
    \pgfpathlineto{\pgfqpoint{\pgf@xa}{\pgf@yb}}
    \advance\pgf@ya by\pgf@yc
    \advance\pgf@yb by-\pgf@yc
    \pgfpathmoveto{\pgfqpoint{\pgf@xb}{\pgf@ya}}
    \pgfpathlineto{\pgfqpoint{\pgf@xb}{\pgf@yb}}
  }
}
\makeatother

\tikzset{inline text/.style =
  {text height=1.2ex,text depth=0.25ex,yshift=0.5mm}}
\tikzset{arrow box/.style =
  {rectangle,inline text,fill=white,draw,
    minimum height=5mm,yshift=-0.5mm,minimum width=5mm}}
\tikzset{bubble/.style =
  {inner sep=0mm,minimum width=3mm,minimum height=3mm,
    draw,shape=circle,fill=white}}
\tikzset{dot/.style =
  {inner sep=0mm,minimum width=1mm,minimum height=1mm,
    draw,shape=circle}}
\tikzset{white dot/.style = {dot,fill=white,text depth=-0.2mm}}
\tikzset{scalar/.style = {diamond,draw,inner sep=1pt}}
\tikzset{square/.style =
  {inner sep=0mm,minimum width=2mm,minimum height=2mm,
    draw,shape=rectangle}}%%me trying to make a fakecopier

\tikzset{star/.style = {dot,fill=white,text depth=-0.2mm}}
\tikzset{copier/.style = {dot,fill,text depth=-0.2mm}}
\tikzset{fakecopier/.style = {square,fill,text depth=-0.2mm}}
\tikzset{discarder/.style = {my ground,draw,inner sep=0pt,
    minimum width=4.2pt,minimum height=11.2pt,anchor=input,rotate=90}}

\tikzset{xshiftu/.style = {shift = {(#1, 0)}}}
\tikzset{yshiftu/.style = {shift = {(0, #1)}}}
\tikzset{scriptstyle/.style={font=\everymath\expandafter{\the\everymath\scriptstyle}}}